\documentclass[12pt]{article}
\usepackage[utf8]{inputenc}
\usepackage{graphicx} 
\usepackage{csquotes} 
\usepackage{hyperref}   

\usepackage{amssymb}
\usepackage{amsmath}
\usepackage{authblk}
\usepackage[numbers, sort&compress]{natbib}
\usepackage{lineno}
\usepackage{caption}
\usepackage{geometry}
\usepackage{parskip}
\usepackage{placeins}

\usepackage{booktabs}

\begin{document}
	
		

	\title{\textbf{Predicting co-segregation in alloys with solute-solute interactions}}
	\author{Zuoyong Zhang and Chuang Deng\footnote{Corresponding author: Chuang.Deng@umanitoba.ca}}
	\affil{\small \textit{Department of Mechanical Engineering, University of Manitoba, Winnipeg, R3T 5V6 Manitoba, Canada}}
	\date{}
	\maketitle

	\begin{abstract}
		The co-segregation of impurities in systems with multiple solute species has been widely recognized as an effective strategy for tailoring material properties. However, reliable predictions of co-segregation behavior remain a significant challenge for alloy design in these systems. In this work, we develop an extended dual-solute (DS) segregation framework to semi-quantitatively predict co-segregation behavior with solute-solute interactions, including both homoatomic and heteroatomic contributions. A machine-learning workflow is first established to predict the pairwise segregation energy to construct the DS segregation energy spectra that intrinsically include both types of solute-solute interactions. The resulting spectral information is then used to determine the upper and lower bounds of segregation for individual solute species. When applied to magnesium-based ternary systems constructed by alloying Mg with any two of the 11 candidate solute species (Ag, Al, Ca, Co, Cu, Gd, Nd, Ni, Pb, Pd, and Zn), the extended DS segregation framework is successfully validated by hybrid molecular dynamics/Monte Carlo simulations and experimental results available in the literature. Furthermore, we introduce a design strategy to promote co-segregation by incorporating additional solute species that exhibit attractive interactions with existing solutes, thereby enabling enhanced co-segregation even in the presence of strong site competition. These results underscore the critical role of solute-solute interactions in governing co-segregation behavior and provide a predictive pathway for the design and optimization of metallic alloys.
	\end{abstract}
	\textbf{Keywords: }Grain boundary co-segregation; Machine learning; Atomistic simulations; Segregation energy spectrum; Solute-solute interactions

	\section{Introduction}
	\label{intro}
	In alloys, co-segregation refers to the concurrent enrichment of multiple solute species at grain boundaries (GBs), which arises from their coupled thermodynamic and kinetic interactions rather than independent segregation behavior \cite{guttmannEquilibriumSegregationTernary1975,guttmannThermodynamicsInteractiveCosegregation1982,xingSoluteInteractionEffects2018,wangCALPHADIntegratedGrain2023}. Early experimental observations of this phenomenon can be traced back to the 1970s, where deviations from single-solute segregation models were reported \cite{seahGrainBoundarySegregation1973,mulfordTemperEmbrittlementNiCr1976,hondrosSegregationInterfaces1977}. Since then, co-segregation has evolved from a largely descriptive concept into a powerful strategy for tailoring interfacial chemistry and bulk material properties.
	
	Recently, solute co-segregation strategy has been widely employed in the design and optimization of advanced metallic systems. In particular, the simultaneous segregation of multiple solute species with substantial atomic size mismatch can significantly reduce the GB excess energy and promote complexion transitions. This often leads to the formation of amorphous intergranular films, which are structurally disordered, glass-like layers confined at GBs \cite{dillonComplexionNewConcept2007,schulerMaterialsSelectionRules2017,cantwellGrainBoundaryComplexion2020}. At high temperatures, the presence of amorphous intergranular films plays a critical role in stabilizing the GB network by reducing their mobility and suppressing abnormal grain growth \cite{grigorianThickAmorphousComplexion2019,leiBulkNanocrystallineAlloys2021,hessongThickerAmorphousGrain}. Upon rapid quenching, such disordered interfacial structures can be kinetically frozen and retained at lower temperatures, thereby preventing their reversion to more ordered GB structures \cite{grigorianCriticalCoolingRates2021,hessongModulationStructuralShortrange2026,leiBinaryNanocrystallineAlloys2023}. Recent progress demonstrates that stable GB complexions can even be achieved under relatively slow cooling rates, attributed to proper solute selection that promotes strong co-segregation and favorable interfacial thermodynamics \cite{leiBulkNanocrystallineMg2025}.
	 
	Beyond promoting complexion transition, solute co-segregation also provides a versatile route to enhance mechanical performance, such as strength, ductility, and formability, e.g., in magnesium alloys \cite{basuRoleAtomicScale2016,zhangDesigningHighMechanical2024,xiaoCurrentProgressSolute2026}. Strategic co-alloying with multiple solute species can effectively modify the local atomic environments (LAEs) at regular GBs and twin boundaries (TBs) in Mg-based alloys, thereby facilitating the activation of non-basal slip systems and enhancing their plasticity. For instance, the co-segregation of Nd-Ag \cite{zhaoDirectObservationImpact2019} and Gd-Zn \cite{niePeriodicSegregationSolute2013} has been shown to stabilize magnesium TBs by reducing local elastic strain energy. Likewise, co-segregation in systems such as Mg-Ca-Zn \cite{qianInfluenceAlloyingElement2022,qianImprovedTensileForming2026,zengTextureEvolutionStatic2016} and Mg-Al-Ca-Zn \cite{mengAchievingExtraordinaryThermal2022,peiGrainBoundaryCosegregation2021,peiSoluteCosegregationMechanisms2026,peiSynergisticEffectCa2022} can stabilize non-basal orientations, decrease GB mobility, and suppress grain coarsening. These findings highlight that co-segregation is not merely a passive outcome of multicomponent alloying, but rather an active design strategy for engineering interfacial structure and properties. Consequently, the quantitative prediction of co-segregation behavior in systems with multiple solute species is essential for guiding next-generation alloy design and performance optimization.
	
	The classical McLean model \cite{mcleanGrainBoundariesMetalsM1957} often fails to accurately predict GB segregation under non-dilute conditions, primarily because it neglects the spectral nature of LAEs within the GB network. In reality, GB sites exhibit a wide distribution of energetics due to structural disorder and varying atomic coordination, which cannot be captured by a single average binding energy. To address this limitation, a spectral approach has been developed to explicitly account for site-wise variations in segregation energetics, enabling a more realistic description of GB segregation behavior \cite{wagihSpectrumGrainBoundary2019,wagihLearningGrainBoundary2020,wagihLearningGrainBoundarySegregation2022}. Building on this framework, it has been widely applied to investigate the roles of solute-solute interactions \cite{wagihGrainBoundarySegregation2020,matsonBondFocusedLocalAtomic2024,zhangGrainBoundarySegregation2024}, grain size effects \cite{tuchindaGrainSizeDependencies2022}, vibrational entropy contributions \cite{tuchindaVibrationalEntropySpectra2023}, hydrostatic pressure effects \cite{zhangHydrostaticPressureinducedTransition2023}, segregation-induced GB phase transitions \cite{zhangGrainBoundaryInterstitial2025,zhangDisconnectionFormationSegregationinduced2026}, and solute clustering \cite{nenningerSoluteClusteringPolycrystals2025,nenningerLocalAtomicEnvironment2023,nikitinSizedependentAttractionCu2025,} in a wide range of binary systems. 
	
	More recently, this approach has been extended to predict co-segregation behavior in ternary systems. In contrast to classical co-segregation models \cite{guttmannEquilibriumSegregationTernary1975,xingSoluteInteractionEffects2018}, which typically rely on simplified interaction terms, the spectrum-based method \cite{wagihDesigningCooperativeGrain2025} is primarily grounded in a site-competition mechanism. Within this approach, the tendency of co-segregation is related to the linear correlation, referred to as the site competition parameter, between segregation energies of different solute species obtained from the dataset in Ref. \cite{wagihLearningGrainBoundarySegregation2022}. Although this treatment captures the competition for available sites, it does not fully account for solute-solute interactions, particularly those arising from short-range chemical bonding and local strain coupling at GBs. Therefore, its predictive capability may be limited in systems where, although site competition exists, co-segregation behavior is predominantly governed by strong synergistic interactions between solute species.
	
	In parallel, machine-learning (ML) models have been developed to predict co-segregation trends in chemically complex alloys \cite{huDatadrivenPredictionGrain2022,dasBayesianOptimizationGrainBoundary2025,aksoyMachineLearningFramework2024}, offering improved scalability and efficiency compared to physics-based models. However, most existing ML models remain largely phenomenological, making it challenging to extract physically meaningful insights or ensure reliable extrapolation beyond the training domain. Meanwhile, substantial progress has been made in modeling site-specific segregation energetics \cite{wagihSpectrumGrainBoundary2019}, solute-solute interactions \cite{wagihGrainBoundarySegregation2020,matsonBondFocusedLocalAtomic2024,matsonAtomisticAssessmentSoluteSolute2021}, and multiple solute effects \cite{scheiberImpactSolutesoluteInteractions2018c,maiSegregationTransitionMetals2022b}. Nevertheless, these aspects are often addressed separately within different modeling frameworks, or their simultaneous treatment requires substantially greater computational effort, making their integration into a unified and quantitatively predictive framework challenging. Addressing this challenge is essential for guiding the design of advanced alloys through controlled co-segregation and interfacial engineering.
	
	In this study, we extend the original dual-solute (DS) segregation framework for binaries \cite{zhangGrainBoundarySegregation2024}, which is built upon the spectral framework developed by Wagih and Schuh \cite{wagihSpectrumGrainBoundary2019}, to enable semi-quantitative prediction of co-segregation behavior in ternary alloys. By incorporating both homoatomic (same-species) and heteroatomic (different-species) pairwise interactions, the extended DS segregation framework captures interaction-induced spectral shifts associated with co-segregation and constructs physically meaningful lower and upper bounds for the segregation behavior of individual solute species in the interested systems. Using magnesium-based binaries and ternaries as representative systems, we establish an ML workflow to predict the dual-solute (DS) segregation energy spectra. These spectra enable identification of systems with strong heteroatomic attraction, where co-segregation is most favorable, and allow determination of upper and lower segregation bounds employing the DS model \cite{zhangGrainBoundarySegregation2024}. The prediction results are validated using hybrid molecular dynamics (MD)/Monte Carlo (MC) simulations at finite temperatures and experimental data available in existing literature. For ternary systems exhibiting strong site competition, as assessed by the single-solute (SS) segregation energy correlation matrix \cite{wagihDesigningCooperativeGrain2025}, together with heteroatomic repulsive or weakly attractive interactions, solutes with stronger segregation tendencies are expected to preferentially occupy favorable segregation sites, thereby reducing the available segregation sites and suppressing the segregation of competing solutes. To mitigate this effect and promote co-segregation, we introduce an additional solute as a co-segregation mediator that is simultaneously attractive to others. We demonstrate that the proposed framework retains its predictive capability in such higher-order systems. Our findings reveal the significance of solute-solute interactions in predicting co-segregation behavior in alloys, and provide a scalable, physically grounded strategy for alloy design with tailored GB chemistry.

	\section{GB segregation thermodynamics}\label{thermodynamics}
	\subsection{SS and DS segregation framework for binary systems}\label{binary}
	
	Before introducing the ML workflow, it is necessary to clarify the thermodynamic frameworks employed throughout this study, especially the DS model. This includes a rigorous definition of the segregation energy, together with a detailed formulation of how solute-solute interactions are incorporated into the segregation energy spectrum.
	
	The SS model, proposed by Wagih and Schuh \cite{wagihSpectrumGrainBoundary2019}, has been widely applied to dilute conditions. Its framework is straightforward: in each iteration, a single solute atom is used to substitute a target GB site, such that only solute-GB interactions are captured, while solute-solute interactions are neglected. After sampling all GB sites, the resulting SS segregation energy dataset is described using the skew-normal distribution \cite{wagihSpectrumGrainBoundary2019}. The spectrum is then used to numerically solve the SS model, yielding the GB segregation profile for the given solute species, expressed as the GB solute concentration ($X^{gb}$) as a function of the global solute concentration ($X^{tot}$).
	
	Regarding the DS model, further details can be found in our previous study \cite{zhangGrainBoundarySegregation2024}. Here, we highlight several key aspects of this framework. In contrast to the SS model, the DS framework considers the simultaneous substitution of two solute atoms at neighboring sites, with at least one site located at the GB. Therefore, a nearest neighbor ($i-j$) pair list is required, in contrast to the GB site list used in the SS segregation framework. In this pair list, the left-hand sites ($i$) are restricted to GB sites, while the right-hand sites ($j$) may correspond to either GB or near-GB bulk sites. 
	
	Within this framework, the pairwise segregation energy for an A-X binary system, $\Delta E_{i(j)}^{A(X-X)}$, is defined as \cite{zhangGrainBoundarySegregation2024}: 
	\begin{equation}\label{eq1}
		\Delta E_{i(j)}^{A(X-X)}=\Delta H_{ij}^{A(X-X)}-\Delta E_j^{A(X)}
	\end{equation}
	where $\Delta H_{ij}^{A(X-X)}$ denotes the segregation energy of two solute atoms occupying site $i$ and $j$, which is given by:
	
	\begin{equation}\label{eq2}
		\Delta H_{ij}^{A(X-X)}= E_{ij}^{A(X-X)}-2\Delta E_{ref}^{A(X)}-E_{0}
	\end{equation}
	
	Here, $H$ is used in $\Delta H_{ij}^{A(X-X)}$ purposely to avoid possible confusion and highlight its difference from the pairwise segregation energy $\Delta E_{i(j)}^{A(X-X)}$; the former represents the segregation energy of two solute atoms while the latter represents only one. $E_{ij}^{A(X-X)}$ represents the system energy with solute atoms at GB site $i$ and its neighboring site $j$, and $E_{0}$ is the system energy without any solute atoms of the pristine system. The reference energy $\Delta E_{ref}^{A(X)}$ is defined as:
		
	\begin{equation}\label{eq3}
		\Delta E_{ref}^{A(X)}=E_{bulk}^{A(X)}-E_0
	\end{equation}
	where $E_{bulk}^{A(X)}$ is the system energy with a solute atom X located at a bulk reference site, which is far away from the GB network \cite{wagihGrainBoundarySegregation2020}. In Eq. \ref{eq1}, the SS segregation energy, $E_j^{A(X)}$,which defines the segregation energy of an isolated solute atom occupying a site $j$, is given by:
		
	\begin{equation}\label{eq4}
		\Delta E_j^{A(X)}=E_j^{A(X)}-E_0-\Delta E_{ref}^{A(X)}
	\end{equation}
	
	It is important to clarify that, the SS segregation energy in Eq. \ref{eq4} includes contributions from both GB and near-GB bulk sites, which are necessary to determine the complete DS segregation energy spectra. After computing all pairwise and SS segregation energies, the two datasets are combined to construct the DS segregation energy dataset, $\Delta E_i^{DS,XX}$. This combined dataset therefore captures both dilute (without solute-solute interactions) and beyond-dilute (with solute-solute interactions) conditions. The resulting distribution is subsequently fitted using a skew-normal function \cite{wagihSpectrumGrainBoundary2019}:
	\begin{equation}\label{eq5}
		P_i^{gb}=\frac{1}{\sqrt {2\pi }\sigma}exp\left[-\frac{(\Delta E_i^{DS}-\mu )^2}{2\sigma ^2}\right]erfc\left[-\frac{\alpha (\Delta E_i^{DS})-\mu }{\sqrt{2} \sigma }\right]
	\end{equation}
	where $P_i^{gb}$ describes the probability density of corresponding segregation energy, and $\alpha$, $\mu$, and $\sigma$ are three fitting parameters representing the skewness, characteristic segregation energy, and width of the skew-normal distribution, respectively. Then, integrating these parameters into the DS model \cite{zhangGrainBoundarySegregation2024}:
	\begin{equation}\label{eq6}
		X^{tot}=(1-f^{gb})X^{bulk}+f^{gb}\int_{-\infty}^{+\infty} P_i^{gb}\left[1+\frac{1-X^{bulk}}{X^{bulk}}exp\left(\frac{\Delta E_i^{DS}}{k_BT}\right)\right]^{-1}d\Delta E_i^{DS}
	\end{equation}
	where $X^{tot}$ and $X^{bulk}$ denote the global and bulk solute concentrations, respectively. The term $f^{gb}$ refers to the GB volume fraction, which is a function of grain size and GB thickness \cite{trelewiczGrainBoundarySegregation2009} and is introduced to account for the finite-size effects. The average GB solute concentration, $\bar{X}^{gb}$, can be derived from \cite{wagihSpectrumGrainBoundary2019}:
	\begin{equation}\label{eq7}
		X^{tot}=(1-f^{gb})X^{bulk}+f^{gb}\bar{X}^{gb}
	\end{equation}
	
	By numerically solving the DS model, the GB segregation profile is obtained, which inherently captures solute–solute interactions in the binary A–X system. 
	
	It should be noted that only nearest neighbor pairwise interactions are considered in this DS segregation framework, while long-range interactions are neglected. This simplification approximates the complex nature of solute-solute interactions using only LAEs, thereby omitting more intricate configurations that may arise in realistic materials. Nevertheless, despite this approximation, the DS model has demonstrated a remarkable agreement with hybrid MD/MC simulations at finite temperatures across various alloy systems \cite{zhangGrainBoundarySegregation2024}. Thus, the DS model is a good strategy that considers both accuracy and computational efficiency.
	
\begin{figure}
		\centering
		\includegraphics[width=0.9\textwidth]{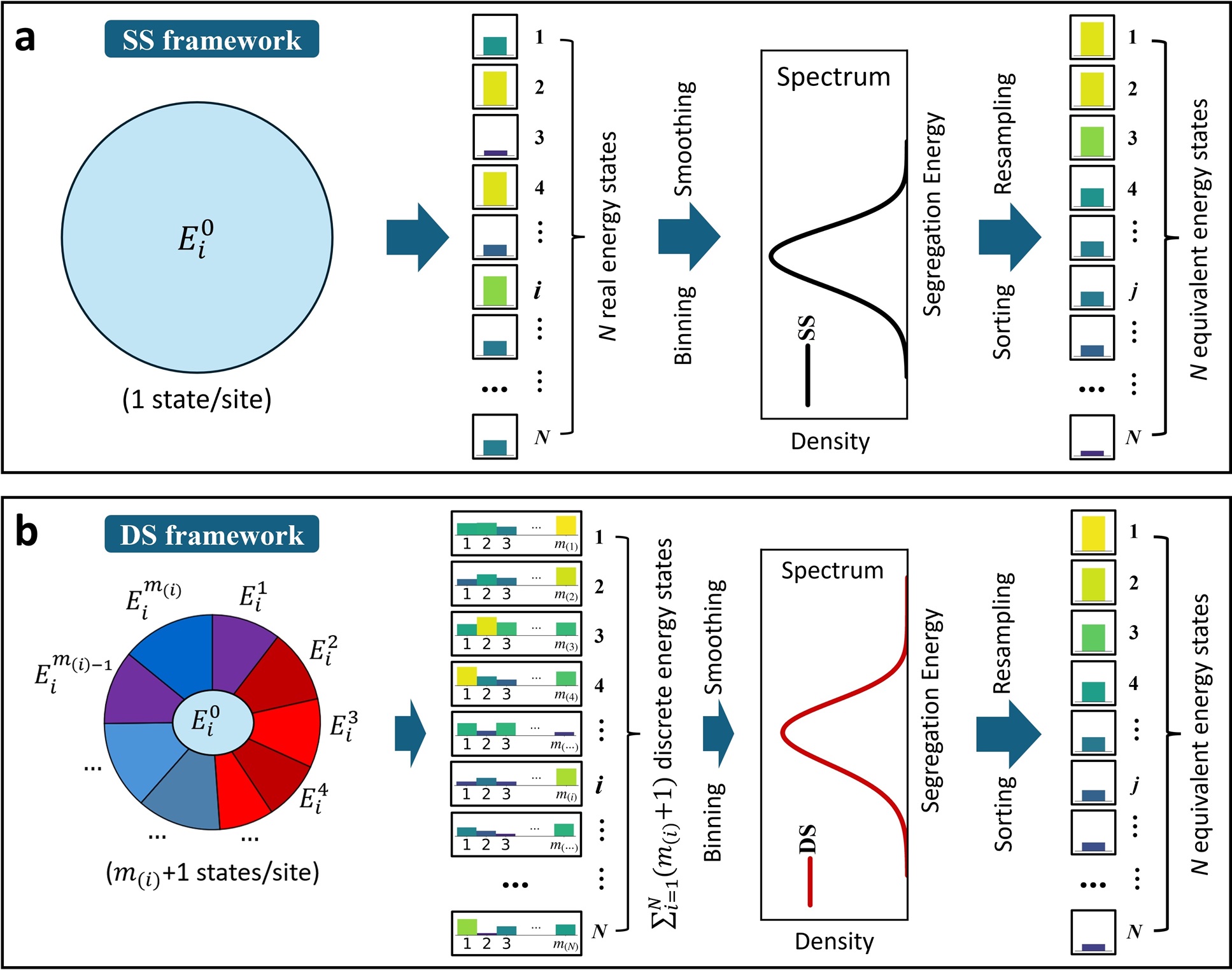}
		\caption{Schematic comparison of the (a) SS and (b) DS segregation frameworks.}
		\label{Fig1}
	\end{figure}

	A schematic comparison of the SS and DS frameworks is presented in Fig. \ref{Fig1}. In the SS model, each GB site is assigned a segregation energy determined by the interaction of a single solute atom with its neighboring host atoms at either GB or bulk sites, as illustrated in Fig. \ref{Fig1}(a). Therefore, a closed system containing $N$ GB sites has $N$ discrete segregation energy states. These discrete states are mathematically sorted, binned, and smoothed to construct a continuous segregation energy spectrum, which is represented by a skew-normal distribution. The resulting fitting parameters provide a compact spectral description of the dilute-limit segregation energy distribution and are subsequently used in the SS isotherm to predict the statistical average segregation behavior. Alternatively, the fitted spectrum can then be reversibly resampled into $N$ equivalent energy states that preserve the same spectral characteristics. These equivalent states provide a discrete representation of the continuous spectrum and produce identical statistical predictions, demonstrating that the spectral representation and its equivalent discrete representation are interchangeable.
	
	To account for solute-solute interactions, we extend the SS framework by adapting the concept of diffusive molecular dynamics (DMD) \cite{liDMD2011,najafabadiFiniteTemperatureStructure1990a,dontsovaSoluteSegregationKinetics2015}. In DMD, each atomic site is described by a continuous occupancy probability rather than being treated as either fully occupied or vacant. Inspired by this concept, we associate each GB site with multiple segregation configurations instead of a single segregation energy. Specifically, a GB site $i$, which has $m_{(i)}$ individual nearest neighbors, is assigned $m_{(i)}+1$ segregation energy states: one SS energy state corresponding to an isolated solute at site $i$, and $m_{(i)}$ pairwise energy states, ranging from $E_i^1$ to $E_i^{m_{(i)}}$, representing interacting two-solute configurations, as illustrated in Fig. \ref{Fig1}(b). By exhausting all isolated sites and neighboring pairs, the DS framework produces a dataset containing $\Sigma_{i=1}^N(m_{(i)}+1)$ discrete energy states. Compared to the SS framework, this expanded dataset samples a substantially diverse segregation energy space by explicitly incorporating local pairwise interaction environments.

	The DS energy dataset is then processed in the same manner as the SS dataset by sorting, binning, and smoothing to construct a continuous segregation energy spectrum, as shown in Fig. \ref{Fig1}(b). As in the SS framework, the fitted spectrum may optionally be resampled into $N$ equivalent energy states that preserve the statistical characteristics of the spectrum. These equivalent states serve the same conceptual role as the $N$ discrete segregation energy states in the SS framework and enable a direct comparison between the two approaches. It should be emphasized that this resampling procedure is not an essential part of either framework. It is introduced solely as a conceptual tool to illustrate that both frameworks can ultimately be represented by an equivalent set of $N$ discrete energy states, thereby allowing the DS dataset to be treated conceptually in the same manner as the SS dataset and enabling the same spectral isotherm to be used for predicting segregation behavior.

	The fundamental distinction between the two models lies not in the representation of the spectrum, but in the construction of the underlying energy space. Whereas the SS framework assigns one segregation energy to each GB site, the DS framework enriches the energy space by explicitly incorporating interacting two-solute configurations before constructing the spectrum. Consequently, the DS spectrum contains significantly more information regarding the local segregation thermodynamics, allowing the statistical model to more accurately describe systems in which solute-solute interactions are important.

	\subsection{DS segregation framework for ternary systems}\label{ternary}
	In this study, we extend the original DS model, which considers only homoatomic interactions in binary systems, to ternary systems by incorporating both homoatomic interactions between identical species and heteroatomic interactions among different species. These additional interactions represent new factors that must be accounted for in the extended framework.
	
	In an A-(B,C) ternary system, where B and C are two solute species in A matrix, two distinct scenarios should be considered: (i) B segregating in the presence of C, and (ii) C segregating in the presence of B. In this study, we adopt the notation A-B-C to indicate that B segregates with the presence of C, where C is a pre-existing solute atom occupying a right-hand site within the pair list, and vice versa. According to Section \ref{binary}, the SS and DS segregation energy datasets can be independently obtained for the A-B and A-C binary systems. 
	
	In this section, we derive the pairwise segregation energies between dissimilar solute species by extending the DS segregation framework for binaries. Specifically, the pairwise segregation energy for a B atom occupying a GB site $i$, in the presence of a neighboring site $j$ already occupied by a C atom, denoted as, denoted as $\Delta E_{i(j)}^{A(B-C)}$, is defined as:
	\begin{equation}\label{eq8}
		\Delta E_{i(j)}^{A(B-C)}=\Delta H_{ij}^{A(B-C)}-\Delta E_j^{A(C)}
	\end{equation}
	where $\Delta H_{ij}^{A(B-C)}$ defines the two-atom segregation energy corresponding to the configuration in which GB site $i$ and its neighbor site $j$ are occupied by B and C atoms, respectively, and is given by:
	\begin{equation}\label{eq9}
		\Delta H_{ij}^{A(B-C)}=E_{ij}^{A(B-C)}-\Delta E_{ref}^{A(B)}-\Delta E_{ref}^{A(C)}-E_0
	\end{equation}
	
	For the A-B-C system, the pairwise segregation energy dataset involving heteroatomic interactions, $\Delta E_{i(j)}^{A(B-C)}$, is combined with the DS segregation energy dataset for B, $\Delta E_i^{DS,BB}$, to construct the ternary DS segregation energy dataset, $\Delta E_i^{DS,BC}$. Thus, this dataset includes more complex interacting conditions, including dilute B, homoatomic (B-B) interactions, and heteroatomic (B-C) interactions. Accordingly, corresponding datasets can also be obtained for the A-C-B configuration.
	
	It should be noted that the SS spectrum for an A-B binary system represents the dilute-limit segregation behavior \cite{wagihGrainBoundarySegregation2020}, whereas the corresponding DS spectrum captures segregation under beyond-dilute conditions \cite{zhangGrainBoundarySegregation2024}. In the absence of, or under very weak, homoatomic interactions, the DS spectrum is expected to coincide with or closely follow the SS spectrum. In contrast, strong repulsive B-B interactions induce a rightward shift of the DS spectrum, while pronounced attractive interactions result in a leftward shift, which may originate from preferences in chemical bonding, complementary local strain fields, or relaxation of local lattice distortion through neighboring solute atoms. Similarly, attractive heteroatomic B-C interactions lead to a leftward shift of its ternary DS spectrum relative to that of the A-B binary, whereas repulsive interactions cause a rightward shift, which may arise from unfavorable chemical interactions or local strain accumulation. Accordingly, the DS prediction curve for A-B binary obtained from solving Eq. \ref{eq6} can be regarded as the limiting case where the concentration of solute C approaches zero, or equivalently, where B-C interactions are negligible. Conversely, the DS prediction curve for the A-B-C ternary represents the limiting case that B-C interactions are strong during co-segregation. Within this framework, the segregation behavior of solute B is expected to fall within the bounds defined by Eq. \ref{eq6}, evaluated using the DS spectra associated with B-B and B-C interactions, corresponding to the lower and upper limits, respectively. The lower and upper bounds are expected to carry clear physical meanings: (i) the lower bound, derived from the binary DS spectrum, corresponds to the case where heteroatomic contributions are absent and segregation is governed solely by homoatomic interactions; and (ii) the upper bound, derived from the ternary DS spectrum at a 1:1 solute ratio, represents the scenario in which heteroatomic attractive interactions are maximized.
	
	In systems where heteroatomic interactions dominate the co-segregation behavior, increasing the concentration of the partner solute is expected to result in a higher GB concentration of the solute of interest. Nevertheless, the extended DS framework does not account for significant evolution of the GB atomic structure or complexion. Consequently, this segregation-enhancing effect may no longer hold when the partner species induces substantial changes in the GB structure or promotes complexion transitions.
	
	Within the applicability of the present framework, the practical width of the predicted bounds is determined by the spectral shift between the heteroatomic and homoatomic DS spectra. A larger spectral shift indicates stronger heteroatomic interactions and therefore produces a wider separation between the lower and upper bounds. Conversely, weaker heteroatomic interactions lead to a smaller spectral shift and correspondingly narrower bounds.
	
	\section{Learning segregation energy spectra with solute-solute interactions}\label{MLflow}
	ML, as a rapidly evolving branch of artificial intelligence, has emerged as a powerful and versatile tool for accelerating materials discovery and design \cite{liuMaterialsDiscoveryDesign2017,raccugliaMachinelearningassistedMaterialsDiscovery2016,fang2022machine}. In particular, the application of ML methods enables rapid prediction and generation of large datasets, such as segregation energy spectra \cite{wagihGrainBoundarySegregation2020,wagihLearningGrainBoundarySegregation2022,matsonBondFocusedLocalAtomic2024} and vibrational entropy contributions \cite{tuchindaComputedEntropySpectra2024} with much lower computational cost compared to traditional approaches.
	
	Prior to predicting segregation energy spectra in the presence of solute-solute interactions, two key steps are required. First, a comprehensive list extraction is performed, which includes both the unique pair list and the dilute list. This step ensures that all GB atomic arrangements, spanning from isolated solutes to interacted solute pairs, are systematically included. Second, a subset of representative nearest-neighbor pairs is selected from the unique pair list to compute their DS segregation energy. This selection focuses on the most physically significant interactions, thereby reducing computational cost while retaining the essential features governing co-segregation behavior. Together, these steps establish a structured and efficient foundation for accurately capturing both homoatomic and heteroatomic interaction effects in segregation predictions.	
	
	\begin{figure}[htbp]
		\centering
		\includegraphics[width=1\textwidth]{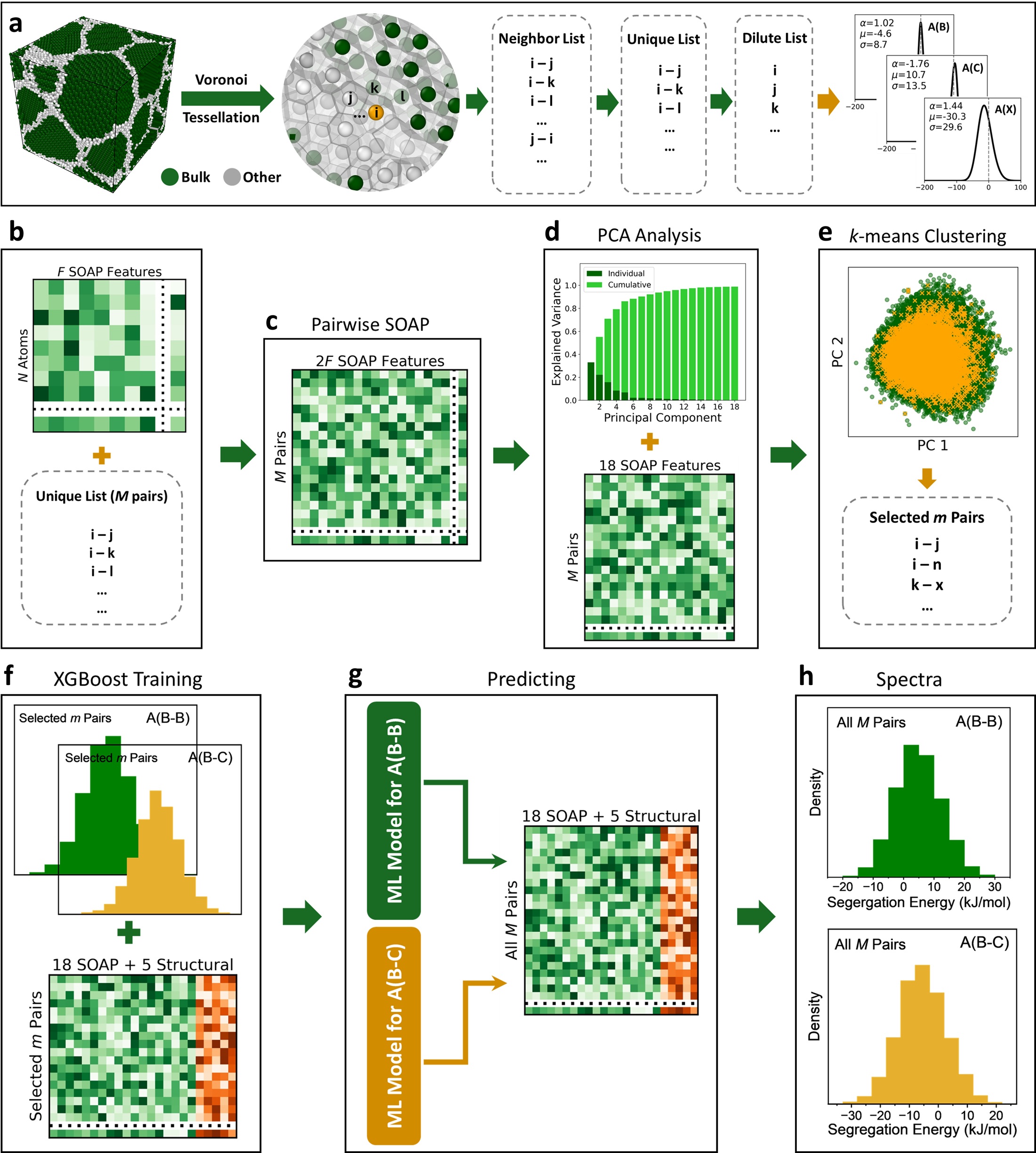}
		\caption{Schematic illustration of ML workflow for segregation energy spectrum prediction. (a) Extraction of unique pairwise list and dilute list, and determination of single-solute segregation energy spectra. (b)-(e) Typical nearest pair selection based on the SOAP features and the unique pairwise list. (f)-(h) ML model training and DS spectrum prediction for both binaries and ternaries.}
		\label{Fig2}
	\end{figure}
	
	As shown in Fig. \ref{Fig2}(a), Voronoi polyhedra are constructed for each atom in the annealed polycrystal sample employing the Voronoi tessellation. The nearest neighbor list is then extracted from the resulting dataset, where two atoms sharing a common face are defined as nearest neighbors regardless of their separation distance. Duplicate pairs ($i-j$ and $j-i$) are removed to generate a unique pair list. Furthermore, a dilute list containing all individual atomic sites is constructed from the unique pair list. Subsequently, the dilute (i.e., SS) segregation energy is calculated for each site in the dilute list using Eq. \ref{eq4}, from which the corresponding GB SS segregation energy spectrum is determined.
	
	Then, the smooth overlap of atomic positions (SOAP) \cite{bartok2013mlprb} is employed as a descriptor to represent the LAEs. This method has been widely used in predicting segregation energy \cite{wagihLearningGrainBoundary2020,wagihLearningGrainBoundarySegregation2022,matsonBondFocusedLocalAtomic2024,zhangGrainBoundaryInterstitial2025} and entropy spectra \cite{tuchindaComputedEntropySpectra2024}.Full SOAP descriptors are first generated for all N atoms in the annealed structure using the DScribe Python package \cite{laakso2023dscribe}. The hyperparameters are set to $n_{max}=12$, $l_{max}=6$, and $r_{cut}=6.0$ \AA, which yield 546 SOAP features for each atom. Based on the unique pair list, pairwise SOAP descriptors are then constructed by concatenating atomic descriptors, resulting in 1092 features for each of the $M$ pairs, as illustrated in Fig. \ref{Fig2}(b) and (c). To reduce dimensionality, principal component analysis (PCA)  \cite{pedregosa2011scikit} is applied, compressing the feature space to 18 components while retaining over 98$\%$ of the explained variance, as shown in Fig. \ref{Fig2}(d). Then, a \textit{k}-means clustering algorithm \cite{pedregosa2011scikit} is employed to identify $m$ representative atomic pairs, as shown in \ref{Fig2}(e). 
	
	After that, full pairwise descriptors are constructed, comprising 18 SOAP features and 5 structural features. The five structural descriptors, shown in Fig. \ref{Fig1}(f) and (g), include the free volume and hydrostatic stress at each site within the pair, as well as the inter-site distance between the two neighboring sites. These descriptors are selected based on the fundamental physical mechanisms governing segregation behavior \cite{zhangHydrostaticPressureinducedTransition2023,xie2025predsegmg,pal2021spectrumvol}. Specifically, the free volume describes the local geometric environment and the ability of a GB site to accommodate solute atoms with different atomic sizes, while the hydrostatic stress reflects the local elastic state and its interaction with local lattice distortion at/around GB sites. The inter-site distance further characterizes the spatial configuration between neighboring sites and influences solute-solute interactions during co-segregation. These structural factors therefore provide direct physical insights into the origin of segregation energy variations. For example, previous studies have shown that some rare-earth elements, such as Gd and Nd, preferentially occupy sites with larger free volume, whereas certain transition metal solutes favor compressive environments due to their distinct atomic size and bonding characteristics \cite{niePeriodicSegregationSolute2013}. Therefore, the combination of representative SOAP descriptors and physically motivated structural features provides a comprehensive representation of the local co-segregation environment, enabling the ML models to capture the underlying relationships between atomic structure, solute interactions, and DS segregation energies.
	
	At this stage, the pairwise segregation energies are evaluated for the binary systems A-B and A-C, as well as the ternary systems A-B-C and A-C-B, for the selected $m$ pairs according to the Section \ref{thermodynamics}. Combined with the full set of descriptors for these $m$ pairs, the XGBoost algorithm \cite{chen2016xgboost} is employed to train ML models for predicting the pairwise segregation energy for each binary and ternary systems of all $M$ pairs using a 90/10 training-to-testing split. Detailed hyperparameters are listed in Table S1. In the Supplementary Materials (SM) (i). The full process for training and predicting pairwise segregation energy is shown in Fig. \ref{Fig2}(e)-(h). Here, XGBoost is adopted due to its high predictive accuracy and great computational efficiency in dealing with GB segregation behavior \cite{yuan2025multitaskml}. 
	
	\FloatBarrier
	\section{Simulation methods}\label{method}
	\subsection{Sample preparation}\label{sec4.1}
	
	In this study, all MD and MS simulations are performed using the LAMMPS software package \cite{plimpton1995lammps,thompson2022lammps}, while structural visualization and analysis are carried out using OVITO \cite{stukowski2022ovito}. The additive-common neighbor analysis (a-CNA) method \cite{larsen2016acna}, as implemented in OVITO, is used to characterize local atomic structures and distinguish GB regions from the bulk. Periodic boundary conditions are applied to all three spatial directions of the simulation cell in all simulations. Throughout this study, interatomic interactions between Mg and 11 solute species (Ag, Al, Ca, Co, Cu, Gd, Nd, Ni, Pb, Pd, and Zn), as well as in additional testing systems such as Al-(Mg,Zn) and Ni-(Cu,Pd), are described using a recently developed high-accuracy neuroevolution potential (NEP89) \cite{liang2025nep89}. This potential is selected for its broad elemental coverage and its strong capability to reproduce density functional theory (DFT) results while maintaining good agreement with experimental data. 
	
	The nanocrystalline (NC) structures are constructed using the Atomsk \cite{hirel2015atomsk}, followed by a standard structural relaxation procedure to obtain annealed samples. First, energy minimization is performed using the conjugate gradient (CG) algorithm to optimize the initial structures, with the convergence criteria for energy and force set to $10^{-20}$ eV and $10^{-20}$ eV/Å, respectively. This is followed by thermal relaxation at ~0.5$T_{m}$ (i.e., half of the melting point) under zero-pressure conditions within the isothermal-isobaric (NPT) ensemble for 500 ps. Subsequently, the NC structures are quenched to 0 K at a constant cooling rate of 3 K/ps. During this process, the temperature is controlled using the canonical (NVT) ensemble, while the pressure is maintained at 0 bar using the Berendsen barostat \cite{berendsen1984mdcoupling}. Finally, the NC structures are further optimized using the CG method with the same settings with initial minimization, while the pressure is precisely controlled at 0 bar using the Parrinello-Rahman method \cite{parrinello1981poly}.
	
	\subsection{Determination of segregation energy}\label{sec4.2}

	According to Sections \ref{thermodynamics} and \ref{MLflow}, the accuracy of pairwise segregation energy predictions depends on the SS segregation energies of each binary system. Therefore, to ensure computational efficiency, a small NC Mg structure (NC13) is employed to determine the segregation energy spectra. After relaxation, the cubic sample contains 94209 magnesium atoms, with an edge length of approximately 13 nm, and consists of six randomly oriented grains with an average grain size of $\sim$6.9 nm. In the dilute list, 46370 individual sites are identified, including 26603 GB sites, while the remainders are near-GB bulk sites. In addition, 229650 pairs are identified in the unique list. Following the workflow illustrated in Fig. \ref{Fig2}(b)-(e), a subset of 10000 pairs (approximately 4.3$\%$ of the unique entries) is selected to determine their pairwise segregation energy, which intrinsically includes solute-solute interactions. MS simulations at 0 K are then performed using the CG method, during which the pressure is precisely controlled at 0 bar using the Parrinello-Rahman method \cite{parrinello1981poly}. To balance accuracy and computational efficiency, the energy and force convergence criteria are set to $10^{-12}$ eV and $10^{-12}$ eV/Å, respectively \cite{zhangHydrostaticPressureinducedTransition2023}. 
	
	From Eqs. \ref{eq3} and \ref{eq5}, the pairwise segregation energy is shown to depend on the SS segregation energy. Consequently, calculating SS segregation energies for all relevant binaries is necessary to ensure the accuracy of pairwise segregation energy predictions. This dependence also justifies our use of small NC structures for the segregation energy calculations. During the calculation of SS segregation energies, only one site from the dilute list is substituted with a specific solute atom in each iteration. In contrast, for pairwise segregation energies, a pair of sites from the selected unique list is substituted with two solute atoms to account for the homoatomic and heteroatomic solute-solute interactions. These two solute atoms are of the same species for binary systems and of different species for ternary systems. After each calculation, the sample structure is restored before proceeding to the next iteration until all the sites or pairs in the dilute or selected unique lists are exhausted. It is worth noting that the segregation energies obtained from MS simulations are enthalpy-based quantities, as the contributions from entropy and pressure-volume work are negligible \cite{wagihSpectrumGrainBoundary2019}.
	
	\subsection{Hybrid MD/MC simulations}\label{sec4.3}
	
	To better capture the structural diversity of GB environments and improve the statistical reliability of the results, a larger sample (denoted as NC16) with dimensions of 16×16×16 $nm^3$ is used for hybrid MD/MC simulations. The increased system size not only provides a broader range of GB characters but also helps reduce artificial correlations arising from periodic boundary conditions. After the standard annealing process, this NC16 sample contains a total of 175768 magnesium atoms and is composed of nine randomly oriented grains with the average grain size of $\sim$7.6 nm. 
	
	In this study, hybrid MD/MC simulations are conducted using the GPUMD \cite{fanGPUMDPackage2022,xu2025gpumd4.0}, which is a high-performance MD simulation package designed to run efficiently on GPUs, particularly for simulations employing NEP models. First, the NC16 structure is equilibrated at 300 K under zero pressure using the NPT ensemble for 100 ps. Then, hybrid MD/MC simulations are carried out using the variance-constrained semi-grand-canonical (VC-SGC) method \cite{sadigh2012vcsgc} at 300 K and zero-pressure, in which solute concentrations are precisely controlled by the corresponding chemical potentials. During the hybrid process, 1000 MC trials are attempted every 100 MD steps, resulting in a total of 10 million MC trials for each desired composition. Thereafter, the system is further relaxed at the same temperature and pressure conditions for another 100 ps. This is followed by a quenching process to 0 K with a constant cooling rate of 3 K/ps. Finally, the system energy is minimized by a FIRE algorithm \cite{bitzekStructuralRelaxationMade2006,guenole2020fire} with the force tolerance of  $10^{-6}$ eV/ Å.
	
	\section{Results}\label{results}
	\subsection{Segregation energy spectra}\label{sec5.1}
	
	\begin{figure}
		\centering
		\includegraphics[width=1\textwidth]{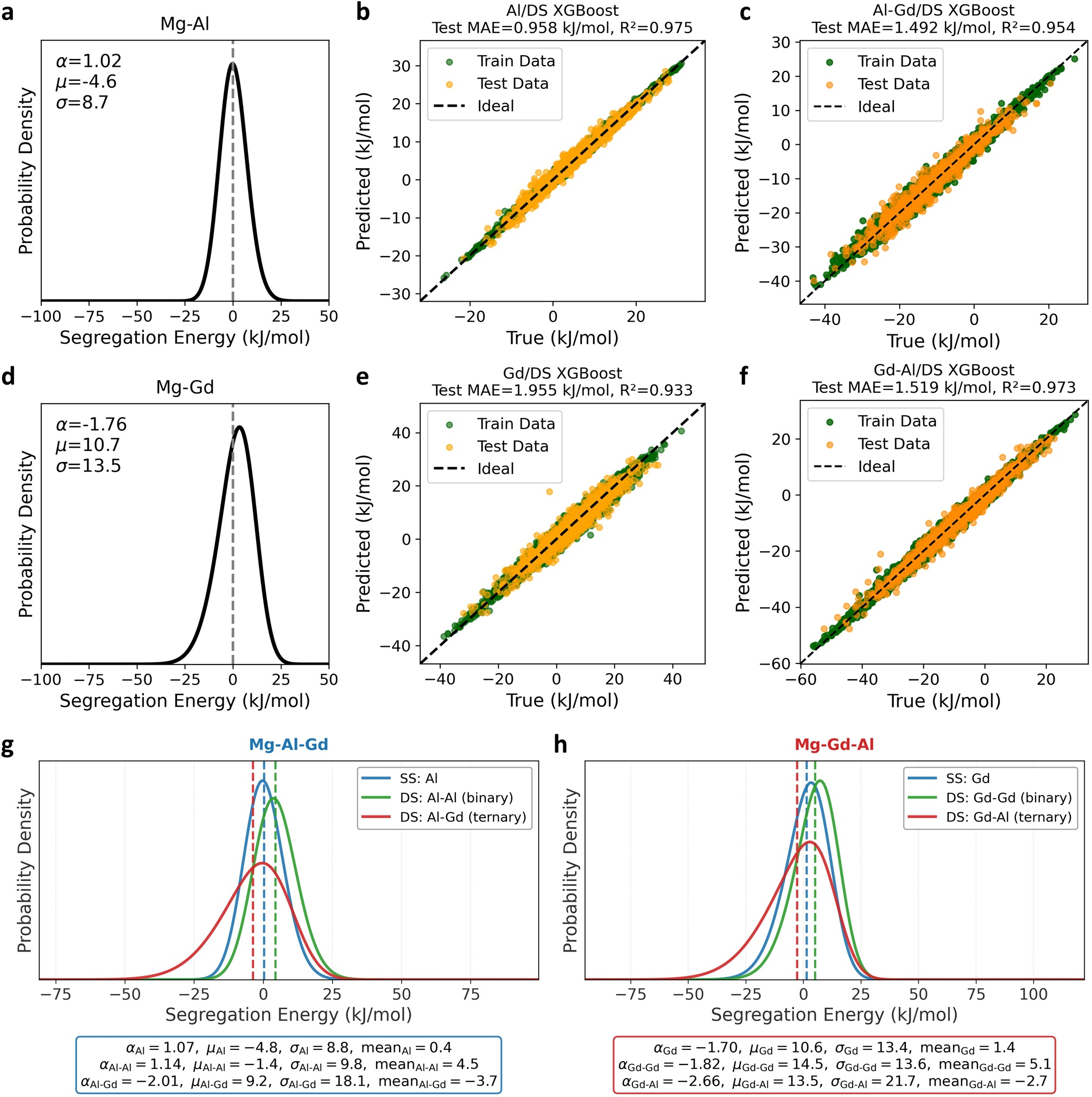}
		\caption{(a) and (d) show the SS GB segregation energy spectra for Mg-Al and Mg-Gd binary systems, respectively. (b) and (e) evaluate the prediction performance of the ML models for predicting pairwise segregation energies with homoatomic interactions, while (c) and (f) illustrate the corresponding performance of the ML models for predicting pairwise segregation energies with heteroatomic interactions. (g) and (h) compare all spectra for the Mg-Al-Gd and Mg-Gd-Al, respectively. The vertical lines indicate the mean values of the corresponding segregation energy spectra. Following the SS model \cite{wagihSpectrumGrainBoundary2019}, the SS spectra shown here only include the contributions from GB sites.}
		\label{Fig3}
	\end{figure}

	Here, we begin with the Mg-(Al,Gd) ternary alloy as a representative system to elucidate and validate the extended DS segregation framework. The calculated SS spectra for the Mg-Al and Mg-Gd systems indicate that both solute species span similar segregation energy ranges and exhibit a moderate segregation tendency under dilute conditions within the Mg GB network, as shown in Fig. \ref{Fig3}(a) and (d). The skew-normal distributions reveal that the two spectra possess opposite skewness: the Mg-Al system is left-skewed (positive $\alpha$ value), whereas the Mg-Gd system is right-skewed (negative $\alpha$ value). This contrast results in significant differences in the characteristic segregation energies, i.e., $\mu _{Al}^{SS}=-4.6\ kJ/mol$ and $\mu _{Gd}^{SS}=10.7\ kJ/mol$, despite the overall similarity in spectral positions. This observation indicates that $\mu$ is highly sensitive to the skewness of the distribution. Therefore, it is not an appropriate descriptor for comparing spectral positions across different systems. In this work, $mean$ values are adopted to enable a consistent comparison of spectral positions \cite{azzaliniClass1985}:
	\begin{equation}\label{eq10}
		mean=\mu + \sigma \cdot \sqrt{\frac{2}{\pi}} \cdot \frac{\alpha}{\sqrt{1+\alpha^2}}
	\end{equation}
	
	For positive $\alpha$, $mean>\mu$, and vice versa. Therefore, the mean is a better descriptor of the average spectral position than $\mu$. 
	
	In Fig. \ref{Fig3}, panels (b) and (e) present the performance of the ML models in predicting DS pairwise segregation energy for the Mg-Al and Mg-Gd binary systems. The results demonstrate that the XGBoost models achieve excellent predictive accuracy, with coefficients of determination ($R^2$) as high as 0.975 and corresponding mean absolute errors (MAE) as low as 0.958 $kJ/mol$. These results indicate that the combined SOAP-based structural descriptors effectively capture the underlying homoatomic interactions governing segregation behaviors. Furthermore, the high fidelity of the predictions suggests that the descriptor set provides a robust representation of the LAEs, enabling reliable modeling of interacting effects within the DS framework.
	
	Fig. \ref{Fig3}(c) and (f) demonstrate the performance of the ML models in predicting DS pairwise segregation energies for the Mg-Al-Gd and Mg-Gd-Al ternary configurations, respectively, which inherently incorporate heteroatomic interactions between different solute species. Specifically, Fig. \ref{Fig3}(c) corresponds to Al segregation at sites with pre-existing Gd atoms at their neighbors, while Fig. \ref{Fig3}(f) represents the reverse configuration, i.e., Gd segregation in the presence of Al occupying neighbor sites. The models exhibit excellent prediction accuracy, with $R^2$ values reaching up to 0.973 and MAE as low as 1.492 $kJ/mol$. These results confirm that the ML workflow is also capable of accurately capturing heteroatomic interactions, even in the presence of complex LAEs. The strong agreement between predicted and reference values further suggests that the combined descriptors effectively capture both chemical and structural contributions to segregation energetics.
	
	The predicted DS segregation energy spectra for the Mg-Al-Gd and Mg-Gd-Al systems are presented in Fig. \ref{Fig3}(g) and (h), respectively, with the corresponding SS spectra included as references. In particular, the SS spectra for Mg-Al and Mg-Gd are employed to compare with the DS spectra and characterize the homoatomic interaction effects, as they represent the dilute-limit behavior in the absence of solute-solute interactions \cite{zhangGrainBoundarySegregation2024}. These reference spectra therefore provide a baseline for isolating the contributions arising from additional solute interactions in the DS segregation framework.
	
	A clear deviation between the DS and SS spectra is observed for both systems. Specifically, in Fig. \ref{Fig3}(g), the DS spectrum for the Mg-Al system exhibits a notable shift toward higher segregation energies relative to its SS counterpart, as reflected by its greater mean value (4.5 $kJ/mol$, also denoted by the olive dashed vertical line) compared to that (0.4 $kJ/mol$) of the SS spectrum. This rightward shift indicates the repulsive Al-Al interactions as the solute concentration exceeds the dilute limit, thereby reducing the driving force for further Al segregation at the GBs in this binary system. A similar trend is evident in Fig. \ref{Fig2}(h) for the Mg-Gd system, where the DS spectrum also shifts toward higher energies compared to its SS spectrum. This behavior demonstrates the presence of repulsive Gd-Gd interactions at elevated solute concentrations, which act to suppress further Gd segregation in the Mg-Gd binary system.
	
	Furthermore, the DS spectra of the corresponding binary systems are taken as references to elucidate the influence of introducing an additional solute species. By directly comparing the positions of the ternary and binary DS spectra, specifically their mean values (indicated by corresponding dashed vertical lines), the effects of heteroatomic interactions can be isolated beyond the homoatomic contributions already embedded in the binary baseline. As shown in Fig. \ref{Fig3}(g), the DS spectrum for the Mg-Al-Gd ternary system exhibits a significant leftward shift with respect to that of the Mg-Al binary system. This shift toward lower segregation energies indicates the strong attractive interactions between Al and Gd atoms within the Mg GB network. Therefore, the Al-Gd bonding is expected to be energetically more favorable than Al-Al interactions in the presence of Gd solutes. 
	
	Similarly, in Fig. \ref{Fig3}(h), the Mg-Gd-Al ternary DS spectrum shows a significant leftward shift compared to the Mg-Gd binary DS spectrum. This behavior demonstrates that introducing Al atoms enhances the Gd segregation at GBs, again reflecting the strong Gd-Al attractive interactions. Consistent with the Mg-Al-Gd system, these results indicate that Gd-Al bonding is energetically preferred over Gd-Gd interactions. Accordingly, the strong heteroatomic interactions in the Mg-(Al,Gd) ternary systems suggest a pronounced tendency of Al and Gd co-segregation at Mg GB network.
	
	To assess the accuracy of the skew-normal approximation, the fitted distributions were compared with the corresponding discrete segregation energy spectra using quantitative goodness-of-fit metrics. As shown in the SM (i), the skew-normal fits exhibit excellent agreement with the discrete spectra, with $R^2$ values (Fig. S1)  close to unity and very low root mean square error ($RMSE$) values (Fig. S2) for all investigated systems. These results demonstrate that the skew-normal function successfully captures the characteristic features of the discrete segregation energy distributions with minimal deviations. Therefore, the extracted distribution parameters provide reliable statistical descriptors for quantifying the spectral shifts induced by solute-solute interactions in the systems considered.
	
	\subsection{Validation of predicted segregation behavior}\label{sec5.2}
	
\begin{figure}
		\centering
		\includegraphics[width=1\textwidth]{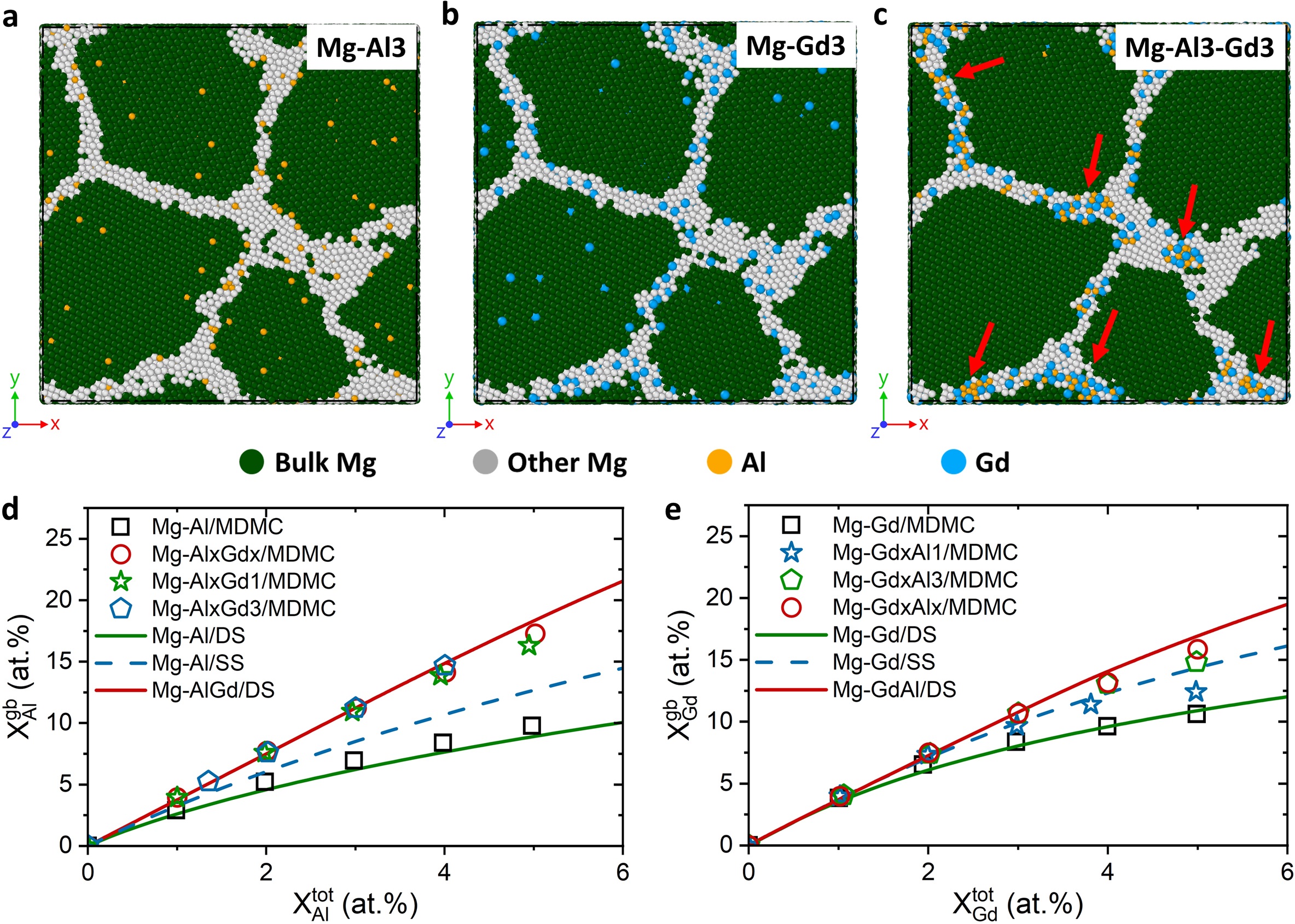}
		\caption{(a)-(c) Hybrid MD/MC results for the Mg-Al (3\ at.\%), Mg-Gd (3\ at.\%), and Mg-Al (3\ at.\%)-Gd (3\ at.\%) systems at 300 K, respectively. (d) and (e) Comparison between DS model predictions and hybrid MD/MC results for Al and Gd segregation, respectively.}
		\label{Fig4}
	\end{figure}

	In this study, hybrid MD/MC simulations performed at 300 K were employed to benchmark the prediction capability of the extended DS segregation framework. For the Mg-Al and Mg-Gd binary systems, a substantial fraction of solute atoms remains randomly distributed within the bulk regions, as illustrated in Fig. \ref{Fig4}(a) and (b), respectively. This behavior indicates that, although the segregation of solute atoms to magnesium GB network occurs, it is not sufficiently strong to fully deplete the solute atoms in bulk, even at a relatively low global solute concentration of 3 at.\%. These observations are consistent with the moderate segregation tendencies reflected in their corresponding SS and DS segregation energy spectra shown in Fig. \ref{Fig3}(g) and (h). Specifically, the relatively narrow energy distributions and the absence of pronounced low-energy tails suggest limited thermodynamic driving forces for strong segregation of either solute species in their binary systems.
	
	However, in the Mg-(Al,Gd) ternary system, most solute atoms, including both Al and Gd, segregate to the GBs after hybrid MD/MC simulations at 300 K, as shown in Fig. \ref{Fig4}(c). This observation not only confirms the occurrence of Al-Gd co-segregation within the Mg GBs but also demonstrates that the segregation of each solute species is significantly enhanced by the presence of the other, namely enhanced or synergistic co-segregation. Furthermore, Al-Gd clusters are observed within the GB network, as highlighted by red arrows in Fig. \ref{Fig4}(c), indicating a relatively strong heteroatomic attraction between Al and Gd. This observation is consistent with the leftward shifts of the corresponding DS spectra for the ternary system with respect to their corresponding binary DS spectra shown in Fig. \ref{Fig3}(g) and (h), further supporting the presence of energetically favorable Al-Gd interactions at the GBs.
	
	As discussed in Section \ref{thermodynamics}, the prediction curves obtained from solving Eq. \ref{eq6} for binary systems describe segregation behavior in the absence of a second solute species. In the Mg-Al and Mg-Gd binary systems, this assumption is validated by the strong agreement between binary DS prediction curves (solid olive) and the corresponding hybrid MD/MC sampling data points (open dark squares) in Fig. \ref{Fig4}(d) and (e), respectively. 
	
	The SS prediction curve (dashed blue) serves as a reference for assessing the influence of solute-solute interactions on GB segregation. In the Mg-Al system, repulsive Al-Al interactions suppress Al segregation, as evidenced in Fig. \ref{Fig4}(d), where the Mg-Al binary DS prediction curve lies below the corresponding SS curve, consistent with the sampled data points. A similar trend is observed in the Mg-Gd system, where repulsive homoatomic interactions likewise reduce the extent of Gd segregation in Mg, as shown in Fig. \ref{Fig3}(e). In contrast, both DS prediction curves for the Mg-Al-Gd and Mg-Gd-Al ternary systems lie above their respective SS curves, suggesting that heteroatomic interactions in both systems are attractive and enhance solute segregation. These trends are consistent with the spectral shifts observed in Fig. \ref{Fig3}(g) and (h), which further demonstrate that the $mean$ value provides a more reliable indicator of solute-solute interaction effects compared to the $\mu$ parameter.
	
	Moreover, the DS spectral predictive curves for ternary systems correspond to limiting scenarios in which heteroatomic interactions reach their maximum influence. As such, the prediction curves derived from ternary DS spectra represent the extreme GB segregation states of Al and Gd in the Mg-(Al, Gd) system. This interpretation is supported by the close agreement between the ternary prediction curves (solid red) and the highest GB solute concentrations obtained from hybrid MD/MC simulations shown in Fig. \ref{Fig4}(d) and (e). Additionally, the monotonic increase in Gd enrichment at GBs with increasing Al concentration supports the hypothesis that a higher concentration of the partner solute enhances the segregation of the solute of interest. Importantly, all other sampled data points fall between the binary and ternary DS prediction curves, confirming that these curves define the lower and upper bounds of segregation behavior in the Mg-(Al,Gd) ternary system. This consistency among the segregation spectra, the corresponding predictions, and the sampled data demonstrates that the extended DS segregation framework provides a robust and predictive approach for capturing solute co-segregation behavior in Mg-based alloys, owing to its explicit treatment of solute-solute interactions. 
	
	\section{Discussion}\label{sec6}
	\subsection{Site competition and solute-solute interactions}\label{sec6.1}
	
	We have shown that the extended DS segregation framework successfully predicts co-segregation behavior driven by strong heteroatomic interactions between Al and Gd in the Mg-(Al,Gd) ternary system. Apart from solute-solute interaction effects \cite{xingSoluteInteractionEffects2018}, site competition is another critical factor that governs co-segregation behavior in ternary alloys, where strong site competition can naturally suppress co-segregation \cite{hondrosSegregationInterfaces1977,leaSiteCompetitionSurface1975}. In contrast, the absence of site competition may facilitate co-segregation. For example, in the Mg-(Ag,Nd) system, Nd preferentially occupies extensive sites, whereas Ag favors compressive sites in Mg TBs. In other words, no site competition exists between these two solute species. As such, Nd and Ag exhibit cooperative co-segregation at Mg TBs, which significantly reduces the mobility of the Ag-Nd co-segregated TBs \cite{zhaoDirectObservationImpact2019}. Similarly, Zn and Gd co-segregation within Mg TBs, also occurring in the absence of site competition, leads to pronounced pinning effects \cite{niePeriodicSegregationSolute2013}. By assessing site-competition factors between different solute species, Wagih et al. \cite{wagihDesigningCooperativeGrain2025} successfully designed the co-segregation of Ni and Hf at Al GBs.
	
\begin{figure}
		\centering
		\includegraphics[width=1\textwidth]{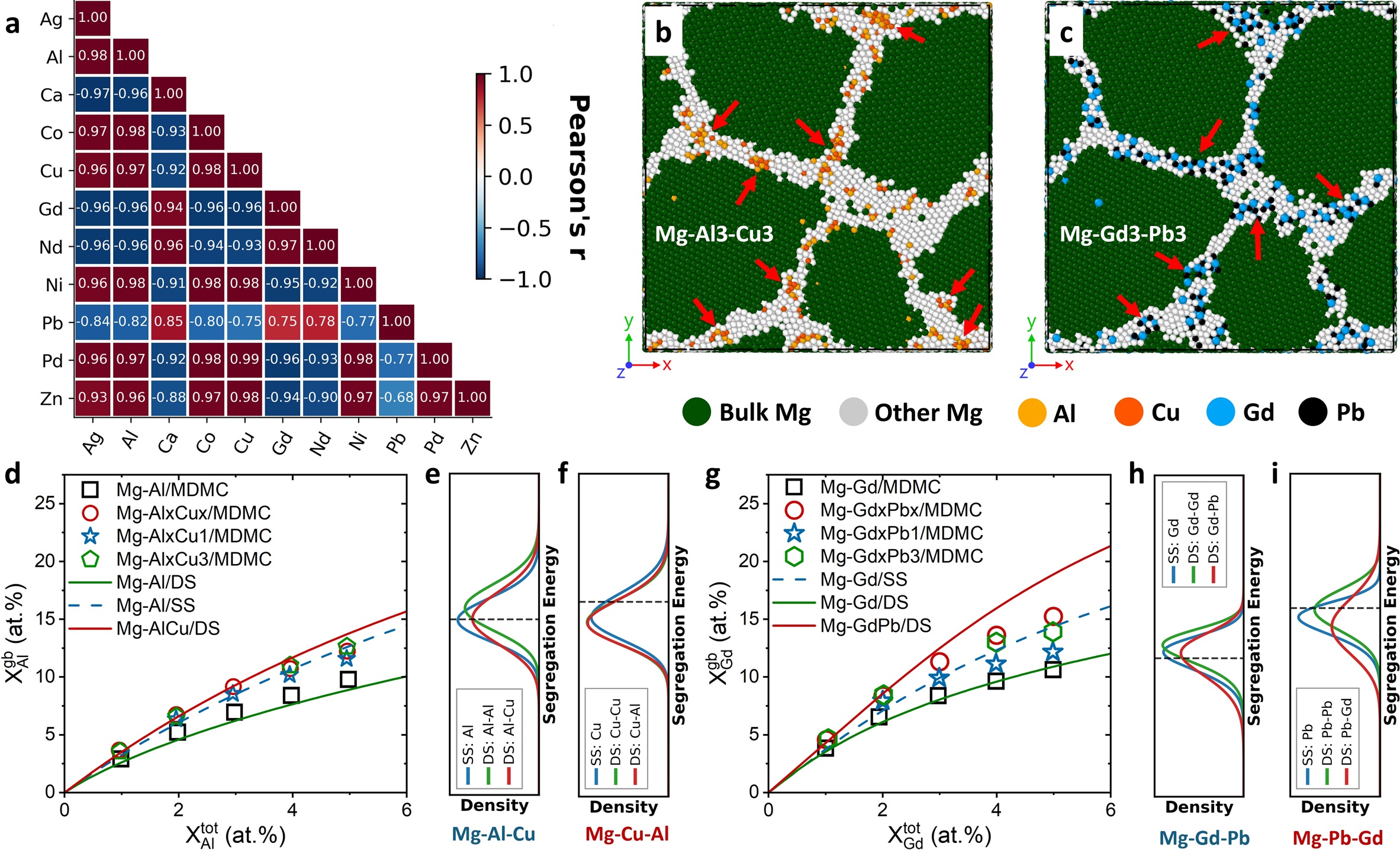}
		\caption{(a) Site-wise GB segregation energy correlation matrix between different solute species in Mg. (b) and (c) Hybrid MD/MC results at 300 K of the Mg-Al (3 at.\%)-Cu (3 at.\%) and Mg-Gd (3 at.\%)-Pb (3 at.\%), respectively. (d) and (d) demonstrate the prediction performance of the extended DS framework in the presence of strong site competition in the Mg-(Al,Cu) and Mg-(Gd,Pb) systems, respectively. (e)-(f) and (h)-(i) present the corresponding segregation energy spectra for comparison of solute-solute interactions. }
		\label{Fig5}
	\end{figure}

	Following the method proposed by Wagih et al. \cite{wagihDesigningCooperativeGrain2025}, site-wise correlations of SS GB segregation energies between different solute species in Mg are quantified, as shown in Fig. \ref{Fig5}(a). In this correlation matrix, positive values indicate that the two solute species preferentially occupy similar GB sites, signifying the presence of site competition. In contrast, negative values indicate that the two solute species favor dissimilar GB sites, suggesting the potential for cooperative co-segregation. 
	
	Surprisingly, even in the presence of ultra-strong site competition, the extended DS framework retains excellent predictive capability for co-segregation behavior in ternary systems with pronounced heteroatomic attraction. For example, in the Mg-(Al,Cu) and Mg-(Gd,Pb) ternary systems, the GB segregation energy correlations reach as high as 0.97 and 0.75 (Fig. \ref{Fig5}(a)), respectively, indicating strong site competition between Al and Cu, and between Gd and Pb. However, strong co-segregation is still observed in both ternary systems after hybrid MD/MC simulations, as shown in Fig. \ref{Fig5}(b) and (c). Specifically, most Al and Cu atoms segregate to Mg GBs and form small Al-Cu clusters rather than being randomly dispersed within the GB network, as highlighted by the red arrows in Fig. \ref{Fig5}(b). Similar heteroatomic clustering is also evidenced in the Mg-(Gd,Pb) system (Fig. \ref{Fig5}(c)). These observations demonstrate that the strong heteroatomic attraction dominates the co-segregation behavior, where the segregation of Al and Gd has been significantly enhanced by the presence of Cu and Pb, respectively, despite the high degree of site competition in both ternary systems.

	Cu exhibits a strong segregation tendency in Mg at dilute conditions, which is further enhanced by strong Cu-Cu attraction. This is demonstrated in the Mg-Cu-Al spectra (Fig. \ref{Fig5}(f)), where the ternary DS spectrum nearly overlaps with that of the Mg-Cu binary system, indicating that Al-Cu interactions are comparable to Cu-Cu bonding. As a result, Cu segregation remains strong regardless of Al addition and can be reliably predicted by the DS model, as shown in Fig. S3 (see SM (i)). In this ternary system, the primary focus therefore shifts to predicting Al segregation in the presence of Cu. The predictions show that all hybrid MD/MC data points fall on or between the predictive curves obtained from the Mg-Al and Mg-Al-Cu DS spectra, as shown in Fig. \ref{Fig5}(d). Moreover, increasing Cu concentration leads to a monotonic increase in Al segregation at GBs, indicating strong Al–Cu attraction. This trend is further supported by the significant leftward shift of the ternary Al-Cu DS spectrum relative to that of Mg-Al (Fig. \ref{Fig5}(e)), which reflects the enhanced contribution of heteroatomic interactions.
	
	When applied to the Mg-Gd-Pb ternary system, the extended DS framework also accurately captures Gd segregation behavior. As shown in Fig. \ref{Fig5}(g), all hybrid MD/MC data points fall on or between the corresponding DS prediction curves. Moreover, increasing Pb concentration also monotonically enhances Gd segregation at GBs, which is consistent with the trends in the Mg-Gd-Al (Fig. \ref{Fig4}(e)) and Mg-Al-Cu (Fig. \ref{Fig5}(d)) systems. Although Gd-Gd and Pb-Pb interactions are repulsive in their respective binary systems, the Gd-Pb interactions are strongly attractive. This contrast is evident from the Mg-Gd-Pb and Mg-Pb-Gd spectra shown in Fig. \ref{Fig5}(h) and (i), respectively, which clearly reflect the dominance of heteroatomic bonding in governing co-segregation in the ternary system. 
	
	It is worth noting that the spectra shown in this section are simplified for qualitatively interpreting the solute-solute interactions in these systems. Complete spectra together with the solute-pair GB SS segregation correlation plots for all 55 unique ternary systems are provided in the SM (ii). Overall, these findings demonstrate that strong heteroatomic attraction may be able to override the co-segregation trends predicted by the site-competition framework in systems that are otherwise unfavorable for co-segregation and further underscore the robustness of the extended DS segregation framework in accurately capturing co-segregation behavior even under conditions of extreme site competition. Here, we would like to clarify that neither the current framework nor Eq. (6) explicitly incorporates site competition. The discussion of site competition is intended to demonstrate the capability of the extended DS framework to predict co-segregation behavior even in the presence of competing segregation tendencies between different solute species.
	
	Additionally, among the 55 Mg-based ternary systems, 29 are characterized as candidates for synergistic co-segregation attributed to their pronounced heteroatomic attraction. For those systems with available experimental data, their co-segregation behavior predicted by the extended DS segregation framework shows good agreement with the corresponding experimental observations, as shown in Table 1. For systems without significant site competition, heteroatomic attraction readily induces synergistic co-segregation. However, despite significant site competition (\ref{Fig5}(a)), some systems still exhibit co-segregation, as confirmed by the corresponding hybrid MD/MC results shown in Fig. S4 in SM (i).

	\setcounter{table}{0}  
	\renewcommand{\thetable}{\arabic{table}}
	\renewcommand{\arraystretch}{1.3}
	\begin{table}
		
		\centering
		\caption{Synergistic co-segregation systems identified by the DS segregation framework with associated supporting information. MD/MC denotes that the supporting evidence is from hybrid MD/MC simulations for systems with significant site competition but no available experimental data.}\label{Tab1}
		\begin{tabular}{cccc}

			\toprule[1.5pt]
			\textbf{Alloy systems} & \textbf{Site Competition} & \textbf{Heteroatomic interactions} & \textbf{Support}\\
			\midrule[1.0pt]
			Mg-(Ag,Ca) & No & Attractive & \cite{bian2021solute}\\
			Mg-(Ag,Gd) & No & Attractive & \cite{yang2026age-hardening,yang2025toward,zhouEffectAgInterfacial2015}\\
			Mg-(Ag,Nd) & No & Attractive & \cite{zhaoDirectObservationImpact2019}\\
			Mg-(Al,Ca) & No & Attractive & \cite{peiGrainBoundaryCosegregation2021,peiSoluteCosegregationMechanisms2026,peiSynergisticEffectCa2022}\\
			Mg-(Al,Co) & Yes & Attractive & MD/MC\\
			Mg-(Al,Cu) & Yes & Attractive & \\
			Mg-(Al,Gd) & No & Attractive & \\
			Mg-(Al,Nd) & No & Attractive & \\
			Mg-(Al,Ni) & Yes & Attractive & \\
			Mg-(Al,Pd) & Yes & Attractive & MD/MC\\
			Mg-(Ca,Cu) & No & Attractive & \cite{bian2022Improving}\\
			Mg-(Ca,Ni) & No & Attractive & \\
			Mg-(Ca,Pb) & Yes & Attractive & MD/MC\\
			Mg-(Ca,Pd) & No & Attractive & \\
			Mg-(Ca,Zn) & No & Attractive & \cite{mengAchievingExtraordinaryThermal2022,zengTextureEvolutionStatic2016,qian2026jmst,zhang2022significantly}\\
			Mg-(Co,Gd) & No & Attractive & \\
			Mg-(Co,Nd) & No & Attractive & \\
			Mg-(Cu,Gd) & No & Attractive & \\
			Mg-(Cu,Nd) & No & Attractive & \\
			Mg-(Gd,Ni) & No & Attractive & \cite{maEffectExtrusionParameters2025}\\
			Mg-(Gd,Pb) & Yes & Attractive & MD/MC\\
			Mg-(Gd,Pd) & No & Attractive & \\
			Mg-(Gd,Zn) & No & Attractive & \cite{niePeriodicSegregationSolute2013,mouhib2022MSEA}\\
			Mg-(Nd,Ni) & No & Attractive & \\
			Mg-(Nd,Pb) & Yes & Attractive & MD/MC\\
			Mg-(Nd,Pd) & No & Attractive & \\
			Mg-(Nd,Zn) & No & Attractive & \cite{qianInfluenceAlloyingElement2022}\\
			Mg-(Ni,Zn) & Yes & Attractive & MD/MC\\
			Mg-(Pb,Pd) & No & Attractive & \\
			\bottomrule[1.5pt]
			
		\end{tabular}

	\end{table}

	\subsection{Mediating co-segregation by additional solute species}\label{sec6.2}
	
\begin{figure}
		\centering
		\includegraphics[width=1\textwidth]{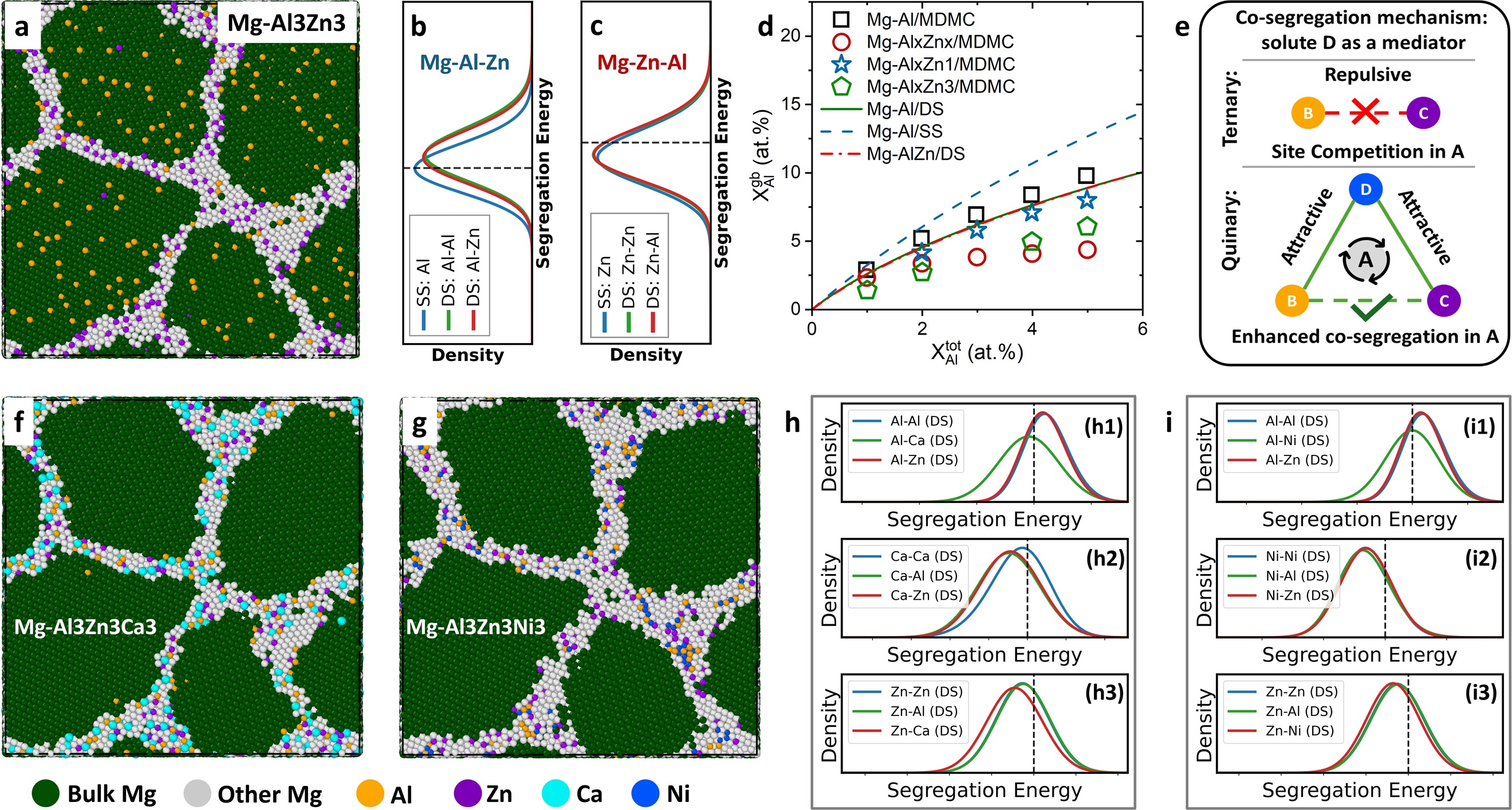}
		\caption{(a) Hybrid MD/MC results of Mg-Al (3 at.\%)-Zn (3 at.\%). (b) and (c) correspond to the spectra for the Mg-Al-Zn and Mg-Zn-Al systems. (d) Predictions of GB Al concentration versus hybrid MD/MC data points in the Mg-(Al,Zn) system. (e) Schematic illustration of additional solute species enhanced co-segregation. (f) and (g) demonstrate the hybrid MD/MC results of the Mg-Al (3 at.\%)-Zn (3 at.\%)-Ca (3 at.\%) and Mg-Al (3 at.\%)-Zn (3 at.\%)-Ni (3 at.\%) systems, respectively. (h) and (i) Simplified DS segregation energy spectra for each solute species in the Mg-(Al,Zn,Ca) and the Mg-(Al,Zn,Ni) systems, respectively.}
		\label{Fig6}
	\end{figure}

	In the Mg-(Al,Zn) system, the extended DS model fails to accurately predict the segregation behavior of Al. As shown in Fig. \ref{Fig5}(a), the Al-Zn pair exhibits a very high correlation coefficient (0.96) in site-wise GB segregation energies, indicating the strong site competition between Al and Zn. This is consistent with experimental and DFT data, where both Zn and Al tend to segregate to compressive sites \cite{peiSoluteCosegregationMechanisms2026,xue2025utilizing,jang2022acomparative,linSynergistic2023a}. Together with the very weak Al-Zn attractive interactions, as interpreted from the spectra positions in Fig. \ref{Fig5}(b), this strong site competition dominates the complex LAEs within the GB network, leading to most Al atoms being displaced from available GB sites and remaining in bulk regions, as shown in Fig. \ref{Fig6}(a). 
	
	Zn exhibits a strong intrinsic tendency to segregate to Mg GBs over Al \cite{zeng2026MSEA}, whereas the presence of Al has little influence on the DS segregation energy landscape of Zn, as shown in Fig. \ref{Fig6}(c). As a result, Zn preferentially occupies most available GB sites and effectively excludes Al from GBs. Increasing the Zn-to-Al ratio further amplifies this effect, significantly reducing Al segregation and thereby suppressing Al-Zn co-segregation, as evidenced by the hybrid MD/MC results in Fig. \ref{Fig6}(d), where de-segregation of Al is observed when the Zn-to-Al ratio is 1:1. Nevertheless, this does not necessarily preclude Al-Zn co-segregation in the Mg-(Al,Zn) system. At low Zn-to-Al ratios, Al can still co-segregate with Zn, as shown in Fig. \ref{Fig6}(d). For example, at a Zn-to-Al ratio of 1:3 (\textbf{\textit{mass ratio}}), the GB Al concentration is comparable to that in the Mg-Al binary system, consistent with experimental observations \cite{nakata2022scripta}.

	According to the analysis in Section \ref{sec6.1}, strong heteroatomic attraction can promote co-segregation even in the presence of extreme site competition. Following this, we propose a new mechanism to enhance the co-segregation in systems where pronounced site competition or strong heteroatomic repulsion induces de-segregation. This co-segregation mechanism is schematically illustrated in Fig. \ref{Fig6}(e). In an A-(B,C) ternary system, where B and C are repulsive to each other in A or compete for similar GB sites, the co-segregation of B and C is unfavorable. To overcome this limitation, we introduce an additional solute species D that exhibits attractive interactions with both B and C and has a strong tendency to segregate at GBs. The presence of D effectively bridges the interactions between B and C, enabling a cooperative segregation configuration. As a result, multiple co-segregation can emerge, thereby enhancing the overall co-segregation behavior despite the underlying site competition between B and C.
	
	Applying this strategy to the Mg-(Al,Zn) system, Ca is identified as a candidate for co-segregation mediator. Specifically, no significant site competition is observed between Al and Ca or between Zn and Ca, as Ca favors to segregate at extension sites due to its large atomic size \cite{linSynergistic2023a}. In addition, strong Al-Ca and Zn-Ca heteroatomic attractions are evidenced by the corresponding spectral shifts, as shown in Fig. \ref{Fig6}(h), and see also Fig. S5 in SM (i) for detailed spectral information. As a result, this Al-Ca-Zn co-segregation occurs in Mg, and the presence of Ca enhances Al-Zn co-segregation, as shown in Fig. \ref{Fig6}(f). These findings are supported by existing experimental data, where Al-Ca-Zn co-segregation in Mg GBs has been reported \cite{peiSynergisticEffectCa2022}, and Ca additions are shown to promote Al-Zn co-segregation \cite{peiGrainBoundaryCosegregation2021,peiSoluteCosegregationMechanisms2026,mengGrainBoundaryMisorientationdependent2026}.

	\begin{figure}
		\centering
		\includegraphics[width=0.5\textwidth]{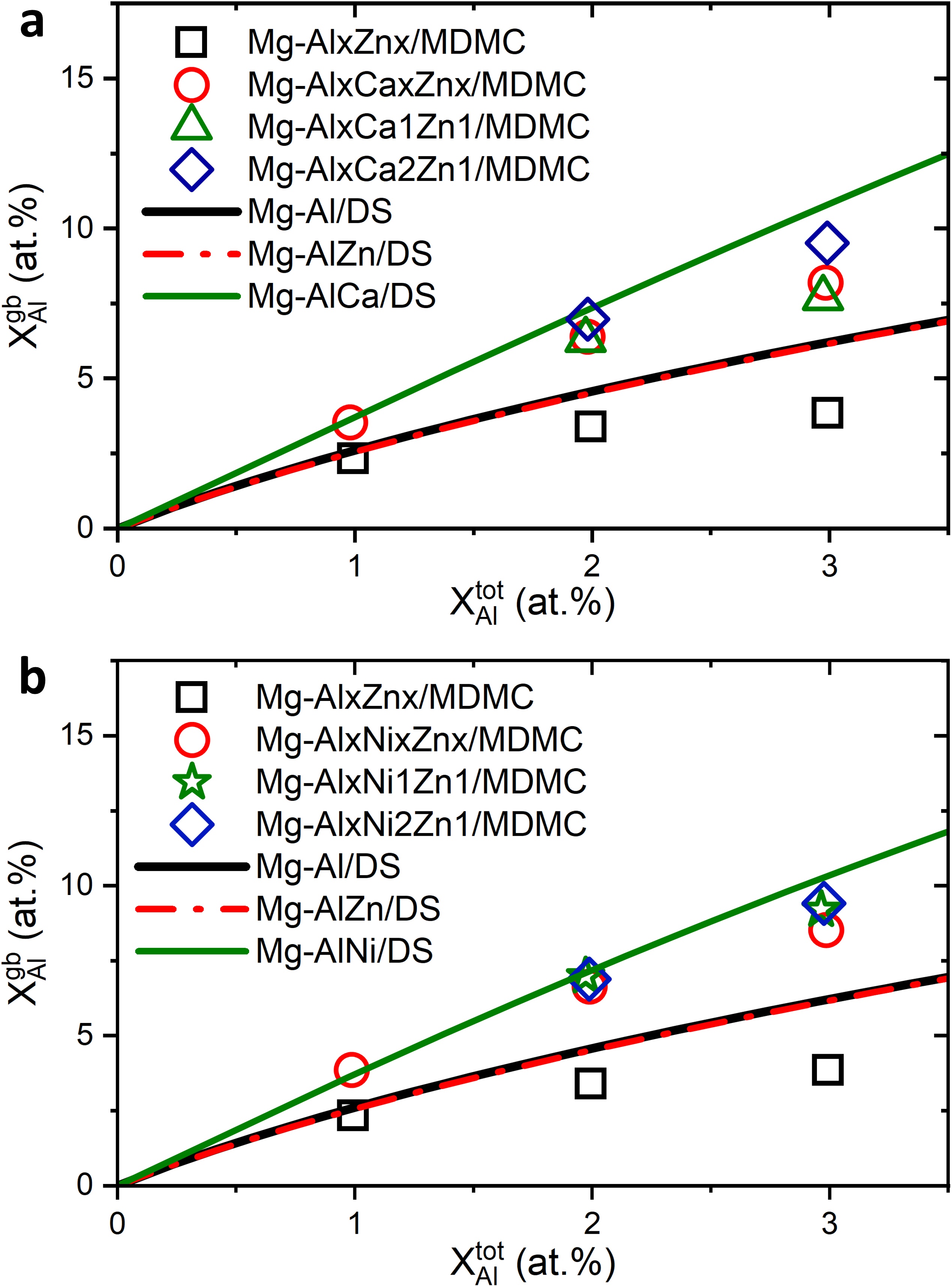}
		\caption{(a) Segregation prediction of Al in the Mg-(Al,Zn,Ca) system. (b) Segregation prediction of Al in the Mg-(Al,Zn,Ni) system.}
		\label{Fig7}
	\end{figure}
	
	In the Mg-(Al,Zn,Ni) system, all three solute species strongly compete for available GB sites, as each solute pair exhibits GB segregation energy correlations close to 1, as shown in Fig. \ref{Fig5}(a). Such site competition is further supported by DFT calculations \cite{linSynergistic2023a}. However, the addition of Ni enhances the Al-Zn co-segregation, as shown in Fig. \ref{Fig6}(g), where most solute atoms, including Al, Zn, and Ni, segregate to GBs. Therefore, this multiple co-segregation can only be attributed to the strong Al-Ni and Zn-Ni heteroatomic attractions, as evidenced by the left shifts of the corresponding DS spectra relative to those of other homoatomic and heteroatomic pairs shown in Fig. \ref{Fig6}(i), and see also Fig. S6 in SM (i) for detailed spectral information.

	As Ca, Ni, and Zn exhibit strong intrinsic segregation tendencies to magnesium GBs, the presence of additional solute species has negligible influence on their segregation behavior. Accordingly, their GB concentrations can be directly predicted from the DS spectra of their respective binary systems and are not the focus at this stage. Here, we instead focus on predicting the GB concentrations of Al in the presence of co-segregation mediators (i.e., Ca and Ni), when Al is chemically mixed with Zn in the Mg matrix. In doing so, we assess the capability of the extended DS framework to predict Al segregation in these two quinary systems.
	
	Following the prediction strategy established for ternary systems, we employ the left-most DS spectrum in Fig. \ref{Fig6}(h1), corresponding to Al-Ca, to define the upper bound of Al segregation in the Mg-(Al,Zn,Ca) system. This choice is justified as the Al-Ca spectrum represents the strongest heteroatomic interaction states. Accordingly, the right-most DS spectrum (Al-Al) defines the lower bound for Al segregation at GBs. As expected, all hybrid MD/MC data points, except those for the Mg-(Al,Zn) system, fall within these bounds, as shown in Fig. \ref{Fig7}(a). Applying to the Mg-(Al,Zn,Ni) system, it also successfully predicts the lower and upper bounds of Al segregation, as shown in Fig. \ref{Fig7}(b), employing the right-most and left-most DS spectra in Fig. \ref{Fig6}(i1). The data points for Mg-(Al,Zn) shown in Fig. \ref{Fig7}(a) and (b) are outside these bounds to highlight the enhancement of Al segregation due to the addition of co-segregation mediators. Moreover, increasing the concentration of mediators monotonically enhances the GB enrichment of the solute of interest, consistent with the expected role of heteroatomic interactions in promoting co-segregation. These results further demonstrate that the developed framework provides robust predictive capability for co-segregation behaviors and can be easily extended into systems with more solute species.
	
	\subsection{Application to other systems}\label{sec6.3}
	
\begin{figure}
		\centering
		\includegraphics[width=0.5\textwidth]{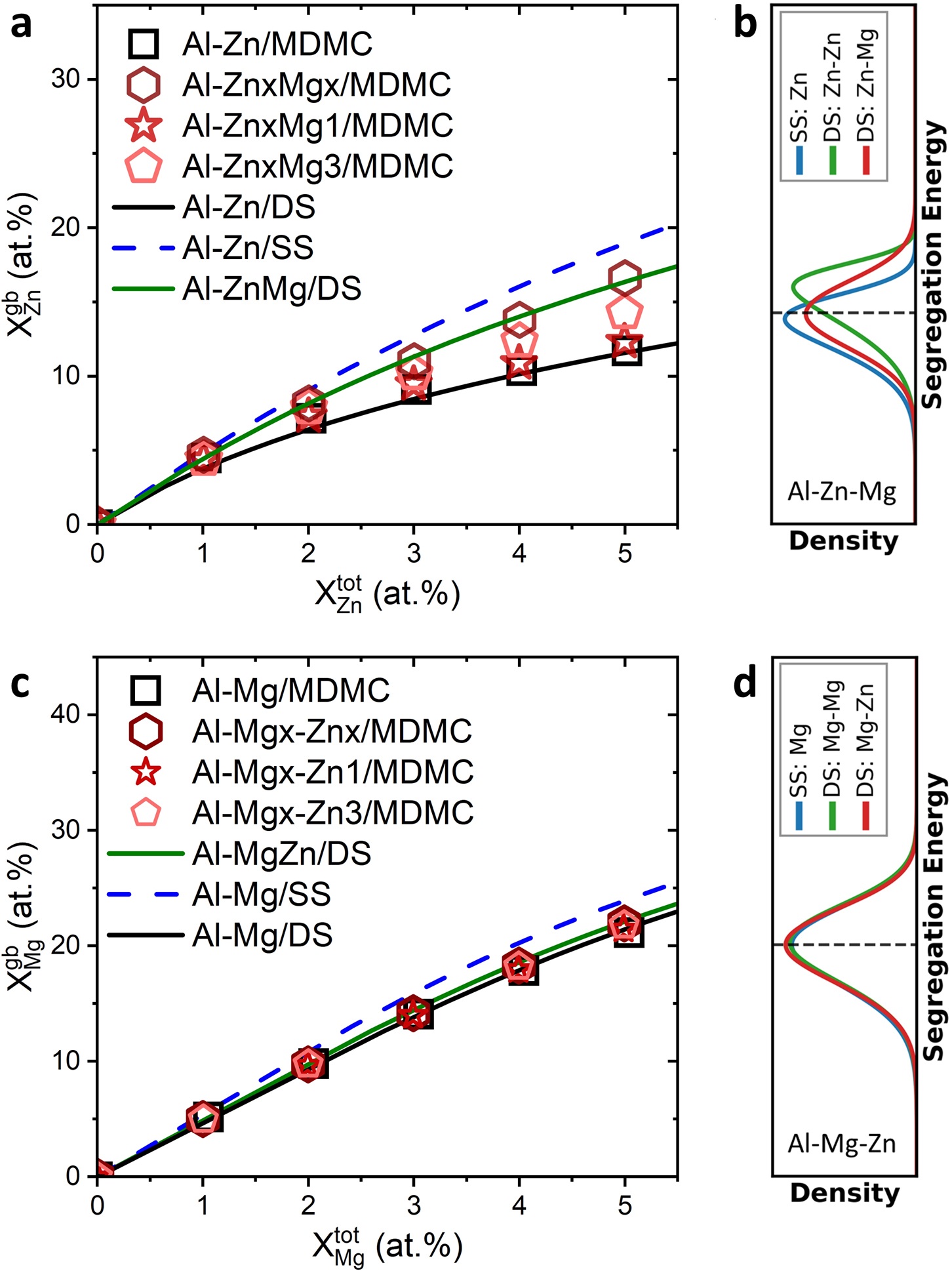}
		\caption{(a) and (c) demonstrate segregation prediction of Zn and Mg in the Al-(Mg,Zn) ternary system. (b) and (d) correspond spectra for Zn and Mg, respectively. The segregation energies were determined using a small NC Al sample with dimensions of 12×12×12 $nm^3$, containing five randomly oriented grains with the average grain size of $\sim$6.8 nm. A larger NC Al sample with dimensions of 15×15×15 $nm^3$ was employed for hybrid MD/MC simulations.}
		\label{Fig8}
	\end{figure}

	Applying the extended DS framework to other systems demonstrates its continued predictive capability for co-segregation behavior. For example, in the Al-(Mg,Zn) system, whose SS GB segregation energy correlation for Zn and Mg is $-0.9$, indicating no site competition between them. Moreover, Mg addition strongly shifts the DS spectra for Zn toward more negative values, suggesting a strong Zn-Mg attraction, as shown in Fig. \ref{Fig8}(b). Consequently, Mg promotes Zn segregation in Al. As shown in Fig. \ref{Fig8}(a), all hybrid MD/MC data points fall on or between the lower and upper bounds defined by the DS spectra for Zn-Zn and Zn-Mg, respectively. At a fixed global Zn concentration, increasing Mg content further enhances Zn segregation. 
	
	As for the Mg in the ternary system, the addition of Zn has little influence on the DS spectrum of Mg. Therefore, either the Mg-Mg or Mg-Zn DS spectrum can accurately capture the segregation behavior of Mg in the ternary system. As shown in Fig. \ref{Fig8}(c), all hybrid MD/MC simulation results lie within the narrow, nearly negligible region between the lower and upper bounds, indicating that Zn has a minimal effect on Mg segregation. This further confirms that Mg segregation is largely governed by its intrinsic behavior, with weak heteroatomic interactions playing an insignificant role. Accordingly, the present study focuses on predicting the co-segregation behavior of solutes in systems with multiple solute speceis where they exhibit weak intrinsic segregation tendencies, such as Al in the Mg-(Al,Cu) and Gd in the Mg-(Gd,Pb) ternaries shown in Fig. \ref{Fig5}.
	
	Apart from the Al-(Mg,Zn) system, the extended DS framework also successfully captures the co-segregation behavior in the Ni-(Cu,Pd) system, even in the presence of pronounced site competition between Cu and Pd, as shown in Fig. S7 in SM (i). The corresponding segregation spectra and prediction results are also provided in the Fig. S8, see SM (i). 
	
	These results demonstrate that site competition is an important factor influencing co-segregation behavior, however, it alone is insufficient to determine whether different solute species will co-segregate. In contrast, heteroatomic interactions play a more critical role in governing co-segregation behavior in host-rich alloys. Overall, these findings suggest that the extended DS segregation framework remains robust across diverse base alloy systems, underscoring its general applicability and strong predictive capability for complex co-segregation phenomena in alloys with multiple solute species.

	\subsection{Limitations and scope if the current framework}\label{sec6.4}
	
	The extended DS framework should be regarded as an effective extension of the spectral segregation model proposed by Malik and Schuh \cite{wagihSpectrumGrainBoundary2019}, rather than a fundamentally new thermodynamic theory. Its primary objective is to expand the applicability of the spectral framework by incorporating pairwise solute interactions while maintaining the simplicity and computational efficiency of the original formulation. Similar to the SS framework, the DS model describes segregation through an effective energy spectrum and predicts statistical average segregation behavior rather than site-specific occupancies. By enriching the segregation energy space with pairwise interacting configurations, the extended DS framework provides an effective thermodynamic representation of solute-solute interactions without explicitly resolving the full many-body statistical mechanics of segregation.
	
	As with any spectral approach, the current framework relies on several simplifying assumptions. The current formulation primarily captures pairwise solute interactions and assumes a fixed segregation energy spectrum for a given GB structure, such that structural evolution and higher-order interactions are not explicitly considered. In addition, the segregation spectra are constructed from 0 K molecular statics calculations. The contribution of entropy is not included, which may lead to deviations between the predicted and actual segregation behavior at elevated temperatures when solute-dependent vibrational effects are significant \cite{lejcekEntropyMattersGrain2021a}. Despite this approximation, segregation energy spectra obtained from 0 K calculations have been widely adopted and have shown good predictive capability for equilibrium GB segregation in many alloy systems. Additional hybrid MD/MC simulations of the Mg-Gd-Al system were performed to examine the predictive capability of the extended DS framework at elevated temperatures. The MD/MC results remain within the prediction bounds of the extended DS framework across the investigated temperature range (SM (i), Fig. S9), indicating that the framework can still describe the segregation behavior of Gd in this system at elevated temperatures.
	
	Previous studies have shown that GB characters \cite{maiSegregationTransitionMetals2022b,barrRoleGrainBoundary2021}, grain size \cite{tuchindaGrainSizeDependencies2022}, and triple junctions \cite{tuchindaTripleJunctionSolute2023} can significantly influence solute segregation. In this study, the smaller nanocrystalline (NC13) sample was used to construct the segregation energy spectra, while a slightly larger sample (NC16) was employed for the hybrid MD/MC simulations. The grains in both samples are randomly oriented, and therefore the contributions of GB characters and triple junctions are not explicitly distinguished or analyzed in this work. 
	
	To further assess the robustness of the extended DS framework with respect to the underlying microstructure, hybrid MD/MC simulations were also performed for the Mg-Gd-Al system using a larger nanocrystalline sample with an edge length of ~20 nm and an average grain size of ~10 nm. As shown in SM (i), Fig. S10, the predicted segregation bounds remain in good agreement with the hybrid MD/MC results, indicating that the extended DS framework continues to accurately bracket the segregation behavior of Gd in the ternary system using a larger NC microstructure. Although this additional validation does not constitute a systematic convergence study, it provides supporting evidence that the framework is robust with respect to moderate variations in the underlying NC microstructure.
	
	It is worth noting that the reliability of these prediction results strongly depends on the robustness of the interatomic potential. Some deviations from realistic material behavior may still occur, even though the NEP89 potential exhibits near \textit{ab initio} accuracy across a broad element space. However, the observed spectrum-sampling self-consistency indicates that the extended DS segregation framework is not only internally consistent but also physically meaningful in capturing the essential trends of solute co-segregation. This consistency between independently sampled configurations and the corresponding segregation spectra suggests that the framework is able to preserve the key statistical and energetic features governing site competition and solute-solute interactions, even in complex alloy environments. Once co-segregation is qualitatively established in a ternary alloy system, the corresponding solute concentrations at GBs can be then quantitatively bracketed by solving the DS model using the corresponding DS segregation energy spectra, which intrinsically incorporate solute-solute interactions.
	
	The extended DS framework is intended for co-segregation prediction in host-rich solid-solution alloys. It employs a dilute host reference state and pairwise segregation energy spectra to predict co-segregation behavior. Therefore, it is not directly applicable to multi-principal-element alloys, where a unique dilute host reference state is no longer well defined. In addition, secondary phase formation and segregation-induced complexion transitions are beyond the scope of this framework. Future developments should focus on establishing a rigorous thermodynamic formulation capable of simultaneously determining the equilibrium segregation behaviors of all solute species while retaining the simplicity and computational efficiency of existing segregation models. Further validation through first-principles calculations and experimental benchmarks, together with the incorporation of enthalpic and entropic contributions and temperature- and pressure-dependent effects, would further enhance the predictive capability and applicability of the future framework.
	
	\section{Conclusion}\label{sec7}
	
	In summary, we developed an extended DS segregation framework for predicting co-segregation in multicomponent alloys by explicitly accounting for solute-solute interactions through two-solute configurations, including both homoatomic and heteroatomic contributions. A machine-learning workflow was established to predict the pairwise segregation energies, which were then described using a skew-normal distribution. Application of the extended DS framework to Mg-based ternary and quaternary alloys shows that the spectral information enables determination of co-segregation states, followed by quantitative prediction of solute GB concentrations. The key conclusions are summarized as follows:
	
	\begin{itemize}
		\item The combined SOAP-based structural descriptors provide a comprehensive representation of LAEs in the presence of multiple solute species. Based on these descriptors, the trained XGBoost models achieve high accuracy in predicting pairwise segregation energies across diverse Mg-based ternary systems, demonstrating strong predictive capability and transferability.
		\item Solute-solute interactions can be captured from DS spectral shifts, where leftward and rightward shifts indicate attraction and repulsion, respectively. This interpretation is applicable to both homoatomic and heteroatomic interactions. Extending to GB segregation prediction, the right-most DS spectrum and left-most DS spectrum defines the lower and upper segregation limits for the corresponding solute species. This has been successfully applied to various Mg-based systems, where co-segregation occurs. The predicted lower and upper bounds show excellent agreement with hybrid MD/MC simulation results in these systems.
		\item In some cases, synergistic co-segregation still occurs despite pronounced site competition, indicating that heteroatomic interactions play a more dominant role than site competition in governing co-segregation states in these systems. 
		\item In systems where co-segregation is unfavorable due to their strong site competition or pronounced heteroatomic repulsion, a strategy can be applied to promote co-segregation by introducing an additional solute that interacts attractively with both primary solutes. Applying to the Mg-(Al,Zn) system, where a high Zn-to-Al ratio drives Al depletion from GBs, the introduced co-segregation mediator, such as Ca or Ni, effectively modulates the local heteroatomic interaction environments, enabling synergistic co-segregation of all three species, even in the presence of strong site competition between. 
		\item The extended DS framework is not limited to Mg-based systems. When applied to other alloy systems, such as Al-(Mg,Zn) and Ni-(Cu,Pd), it accurately predicts co-segregation behavior in excellent agreement with hybrid MD/MC simulations, demonstrating its robustness and broad applicability to multicomponent alloys.
	\end{itemize}
	
	We conclude that heteroatomic attraction is a necessary condition for synergistic co-segregation of different solute species at GBs. In certain cases, these heteroatomic attractions are sufficiently strong to overcome the inhibitory effects of site competition on co-segregation. Additionally, we have demonstrated the extensibility of the DS framework, which facilitates its application to systems containing multiple solute species. These findings advance our understanding of the critical role of solute-solute interactions in governing and predicting co-segregation behavior and provide new insight into the rational design of complex interfacial chemistries in alloys with multiple solute species.

	\textbf{CRediT authorship contribution statement}
	
	\textbf{Zuoyong Zhang}: Writing – review $\&$ editing, Writing – original draft, Visualization, Validation, Software, Methodology, Investigation, Formal analysis, Data curation, Conceptualization. \textbf{Chuang Deng}: Writing – review $\&$ editing, Supervision, Software, Resources, Project administration, Methodology, Funding acquisition, Conceptualization.
	
	\textbf{Declaration of Competing Interest }
	
	The authors declare that they have no known competing financial interests or personal relationships that could influence the work reported in this paper.
	
	\textbf{Acknowledgement}
	
	This research was supported by the NSERC Discovery Grant (RGPIN-2025-05559), Canada, and the use of computing resources provided by the Digital Research Alliance of Canada. \textbf{Z.Z.} also acknowledges the financial support from the University of Manitoba Graduate Fellowship (UMGF). During the preparation of this manuscript, the authors used ChatGPT to improve its readability. The authors carefully reviewed and edited the manuscript following the use of this tool and took full responsibility for the content of the publication.

	
	\bibliographystyle{elsarticle-num}
	\bibliography{reference}

@article{aksoyMachineLearningFramework2024,
  title = {A Machine Learning Framework for the Prediction of Grain Boundary Segregation in Chemically Complex Environments},
  author = {Aksoy, Doruk and Luo, Jian and Cao, Penghui and Rupert, Timothy J},
  year = 2024,
  journal = {Modelling and Simulation in Materials Science and Engineering},
  volume = {32},
  number = {6},
  pages = {065011},
  doi = {https://doi.org/10.1088/1361-651X/ad585f},
  urldate = {2025-10-21},
  language = {en}
}

@article{basuRoleAtomicScale2016,
  title = {The Role of Atomic Scale Segregation in Designing Highly Ductile Magnesium Alloys},
  author = {Basu, I. and Pradeep, K.G. and Mie{\ss}en, C. and {Barrales-Mora}, L.A. and {Al-Samman}, T.},
  year = 2016,
  journal = {Acta Materialia},
  volume = {116},
  pages = {77--94},
  doi = {https://doi.org/10.1016/j.actamat.2016.06.024},
  urldate = {2026-04-01},
  language = {en}
}

@article{cantwellGrainBoundaryComplexion2020,
  title = {Grain {{Boundary Complexion Transitions}}},
  author = {Cantwell, Patrick R. and Frolov, Timofey and Rupert, Timothy J. and Krause, Amanda R. and Marvel, Christopher J. and Rohrer, Gregory S. and Rickman, Jeffrey M. and Harmer, Martin P.},
  year = 2020,
  journal = {Annual Review of Materials Research},
  volume = {50},
  number = {1},
  pages = {465--492},
  doi = {https://doi.org/10.1146/annurev-matsci-081619-114055},
  urldate = {2025-10-21},
  language = {en}
}

@article{dasBayesianOptimizationGrainBoundary2025,
  title = {Bayesian {{Optimization}} of {{Grain-Boundary Segregation}} in {{High-Entropy Alloys}}},
  author = {Das, Shimanta and Oyeniran, Noah and Walter, Joshua and Gesch, Aidan and Hu, Chongze},
  year = 2025,
  journal = {npj Computational Materials},
  volume = {11},
  number = {1},
  pages = {371},
  doi = {https://doi.org/10.1038/s41524-025-01850-9},
  urldate = {2025-12-14},
  language = {en}
}

@article{mouhib2022MSEA,
  title = {Synergistic Effects of Solutes on Active Deformation Modes, Grain Boundary Segregation and Texture Evolution in {{Mg-Gd-Zn}} Alloys},
  author = {Mouhib, F. and Pei, R. and Erol, B. and Sheng, F. and {Korte-Kerzel}, S. and {Al-Samman}, T.},
  year = 2022,
  journal = {Materials Science and Engineering: A},
  volume = {847},
  pages = {143348},
  doi = {https://doi.org/10.1016/j.msea.2022.143348},
  urldate = {2025-10-15},
  language = {en}
}

@article{dillonComplexionNewConcept2007,
  title = {Complexion: {{A}} New Concept for Kinetic Engineering in Materials Science},
  shorttitle = {Complexion},
  author = {Dillon, Shen J. and Tang, Ming and Carter, W. Craig and Harmer, Martin P.},
  year = 2007,
  journal = {Acta Materialia},
  volume = {55},
  number = {18},
  pages = {6208--6218},
  doi = {https://doi.org/10.1016/j.actamat.2007.07.029},
  urldate = {2024-12-16},
  copyright = {https://www.elsevier.com/tdm/userlicense/1.0/},
  language = {en}
}

@article{grigorianCriticalCoolingRates2021,
  title = {Critical Cooling Rates for Amorphous-to-Ordered Complexion Transitions in {{Cu-rich}} Nanocrystalline Alloys},
  author = {Grigorian, Charlette M. and Rupert, Timothy J.},
  year = 2021,
  journal = {Acta Materialia},
  volume = {206},
  pages = {116650},
  doi = {https://doi.org/10.1016/j.actamat.2021.116650},
  urldate = {2025-10-21},
  language = {en}
}

@article{grigorianThickAmorphousComplexion2019,
  title = {Thick Amorphous Complexion Formation and Extreme Thermal Stability in Ternary Nanocrystalline {{Cu-Zr-Hf}} Alloys},
  author = {Grigorian, Charlette M. and Rupert, Timothy J.},
  year = 2019,
  journal = {Acta Materialia},
  volume = {179},
  pages = {172--182},
  doi = {https://doi.org/10.1016/j.actamat.2019.08.031},
  urldate = {2025-10-21},
  language = {en}
}

@article{guttmannThermodynamicsInteractiveCosegregation1982,
  title = {The Thermodynamics of Interactive Co-Segregation of Phosphorus and Alloying Elements in Iron and Temper-Brittle Steels},
  author = {Guttmann, M. and Dumoulin, {\relax Ph}. and Wayman, M.},
  year = 1982,
  journal = {Metallurgical Transactions A},
  volume = {13},
  number = {10},
  pages = {1693--1711},
  doi = {https://doi.org/10.1007/BF02647825},
  urldate = {2026-03-14},
  copyright = {https://www.springernature.com/gp/researchers/text-and-data-mining},
  language = {en}
}

@article{hessongModulationStructuralShortrange2026,
  title = {Modulation of Structural Short-Range Order Due to Chemical Patterning in Multi-Component Amorphous Interfacial Complexions},
  author = {Hessong, Esther C. and Zhang, Zhengyu and Lei, Tianjiao and Xu, Mingjie and Aoki, Toshihiro and Rupert, Timothy J.},
  year = 2026,
  journal = {Acta Materialia},
  pages = {122108},
  doi = {https://doi.org/10.1016/j.actamat.2026.122108},
  urldate = {2026-03-28},
  language = {en}
}

@article{hessongThickerAmorphousGrain,
  title = {Thicker Amorphous Grain Boundary Complexions Reduce Plastic Strain Localization in Nanocrystalline {{Cu-Zr}}},
  author = {Hessong, Esther C and Li, Nan and Fensin, Saryu and Boyce, Brad L and Rupert, Timothy J},
  year = 2026,
  doi = {https://doi.org/10.48550/arXiv.2601.18059},
  language = {en}
}

@article{hondrosSegregationInterfaces1977,
  title = {Segregation to Interfaces},
  author = {Hondros, E. D. and Seah, M. P.},
  year = 1977,
  journal = {International Metals Reviews},
  volume = {22},
  number = {1},
  pages = {262--301},
  publisher = {Taylor \& Francis},
  doi = {https://doi.org/10.1179/imtr.1977.22.1.262},
  urldate = {2026-03-31}
}

@article{huDatadrivenPredictionGrain2022,
  title = {Data-Driven Prediction of Grain Boundary Segregation and Disordering in High-Entropy Alloys in a {{5D}} Space},
  author = {Hu, Chongze and Luo, Jian},
  year = 2022,
  journal = {Materials Horizons},
  volume = {9},
  number = {3},
  pages = {1023--1035},
  doi = {https://doi.org/10.1039/D1MH01204E},
  urldate = {2025-08-21},
  language = {en}
}

@article{leiBinaryNanocrystallineAlloys2023,
  title = {Binary Nanocrystalline Alloys with Strong Glass Forming Interfacial Regions: {{Complexion}} Stability, Segregation Competition, and Diffusion Pathways},
  shorttitle = {Binary Nanocrystalline Alloys with Strong Glass Forming Interfacial Regions},
  author = {Lei, Tianjiao and Hessong, Esther C. and Gianola, Daniel S. and Rupert, Timothy J.},
  year = 2023,
  journal = {Materials Characterization},
  volume = {206},
  pages = {113415},
  doi = {https://doi.org/10.1016/j.matchar.2023.113415},
  urldate = {2025-10-21},
  language = {en}
}

@article{leiBulkNanocrystallineAlloys2021,
  title = {Bulk Nanocrystalline {{Al}} Alloys with Hierarchical Reinforcement Structures via Grain Boundary Segregation and Complexion Formation},
  author = {Lei, Tianjiao and Shin, Jungho and Gianola, Daniel S. and Rupert, Timothy J.},
  year = 2021,
  journal = {Acta Materialia},
  volume = {221},
  pages = {117394},
  doi = {https://doi.org/10.1016/j.actamat.2021.117394},
  urldate = {2025-10-21},
  language = {en}
}

@article{leiBulkNanocrystallineMg2025,
  title = {Bulk Nanocrystalline {{Al}}--{{Mg}}--{{Y}} Alloys with Amorphous Grain Boundary Complexions Display High Strength and Compressive Plasticity},
  author = {Lei, Tianjiao and Hessong, Esther C. and Fields, Brandon and Thiraux, Raphael Pierre and Gianola, Daniel S. and Rupert, Timothy J.},
  year = 2025,
  journal = {Journal of Materials Science},
  volume = {60},
  number = {39},
  pages = {18486--18501},
  doi = {https://doi.org/10.1007/s10853-025-11328-0},
  urldate = {2025-10-21},
  language = {en}
}

@article{matsonBondFocusedLocalAtomic2024,
  title = {A ``{{Bond-Focused}}'' {{Local Atomic Environment Representation}} for a {{High Throughput Solute Interaction Spectrum Analysis}}},
  author = {Matson, Thomas P. and Schuh, Christopher A.},
  year = 2024,
  journal = {Acta Materialia},
  pages = {120275},
  doi = {https://doi.org/10.1016/j.actamat.2024.120275},
  urldate = {2024-08-12},
  language = {en}
}

@book{mcleanGrainBoundariesMetalsM1957,
  title = {Grain {{Boundaries}} in {{Metals}},[{{M}}] {{Oxford}}},
  author = {McLean, Donald},
  year = 1957,
  publisher = {Clarendon Press}
}

@article{mengAchievingExtraordinaryThermal2022,
  title = {Achieving Extraordinary Thermal Stability of Fine-Grained Structure in a Dilute Magnesium Alloy},
  author = {Meng, Zhao-Yuan and Wang, Cheng and Hua, Zhen-Ming and Zha, Min and Wang, Hui-Yuan},
  year = 2022,
  journal = {Materials Research Letters},
  volume = {10},
  number = {12},
  pages = {797--804},
  doi = {https://doi.org/10.1080/21663831.2022.2106797},
  urldate = {2025-12-20},
  language = {en}
}

@article{mulfordTemperEmbrittlementNiCr1976,
  title = {Temper Embrittlement of {{Ni-Cr Steels}} by Phosphorus},
  author = {Mulford, R. A. and Mcmahon, C. J. and Pope, D. P. and Feng, H. C.},
  year = 1976,
  journal = {Metallurgical Transactions A},
  volume = {7},
  number = {8},
  pages = {1183--1195},
  doi = {https://doi.org/10.1007/BF02656602},
  urldate = {2026-03-13},
  copyright = {https://www.springernature.com/gp/researchers/text-and-data-mining},
  language = {en}
}

@article{niePeriodicSegregationSolute2013,
  title = {Periodic {{Segregation}} of {{Solute Atoms}} in {{Fully Coherent Twin Boundaries}}},
  author = {Nie, J. F. and Zhu, Y. M. and Liu, J. Z. and Fang, X. Y.},
  year = 2013,
  journal = {Science},
  volume = {340},
  number = {6135},
  pages = {957--960},
  doi = {https://doi.org/10.1126/science.1229369},
  urldate = {2023-02-19},
  language = {en}
}

@article{peiGrainBoundaryCosegregation2021,
  title = {Grain Boundary Co-Segregation in Magnesium Alloys with Multiple Substitutional Elements},
  author = {Pei, Risheng and Zou, Yongchun and Wei, Daqing and {Al-Samman}, Talal},
  year = 2021,
  journal = {Acta Materialia},
  volume = {208},
  pages = {116749},
  doi = {https://doi.org/10.1016/j.actamat.2021.116749},
  urldate = {2024-08-26},
  language = {en}
}

@article{peiSoluteCosegregationMechanisms2026,
  title = {Solute Co-Segregation Mechanisms at Low-Angle Grain Boundaries in Magnesium: {{A}} Combined Atomic-Scale Experimental and Modeling Study},
  shorttitle = {Solute Co-Segregation Mechanisms at Low-Angle Grain Boundaries in Magnesium},
  author = {Pei, Risheng and Petrazoller, Jo{\'e} and Atila, Achraf and Arnoldi, Simon and Xiao, Lei and Liu, Xiaoqing and Wang, Hexin and {Korte-Kerzel}, Sandra and Berbenni, St{\'e}phane and Richeton, Thiebaud and Gu{\'e}nol{\'e}, Julien and Xie, Zhuocheng and {Al-Samman}, Talal},
  year = 2026,
  journal = {Acta Materialia},
  volume = {306},
  pages = {121947},
  doi = {https://doi.org/10.1016/j.actamat.2026.121947},
  urldate = {2026-02-02},
  language = {en}
}

@article{peiSynergisticEffectCa2022,
  title = {Synergistic Effect of {{Y}} and {{Ca}} Addition on the Texture Modification in {{AZ31B}} Magnesium Alloy},
  author = {Pei, Risheng and Zou, Yongchun and Zubair, Muhammad and Wei, Daqing and {Al-Samman}, Talal},
  year = 2022,
  journal = {Acta Materialia},
  volume = {233},
  pages = {117990},
  doi = {https://doi.org/10.1016/j.actamat.2022.117990},
  urldate = {2026-02-02},
  language = {en}
}

@article{qianImprovedTensileForming2026,
  title = {Improved Tensile and Forming Properties by the Enhanced Strain Coordination at Grain Boundary with Alloying Elements Co-Segregation in {{Mg-Zn-Ca}} Alloys},
  author = {Qian, Xiaoying and Dong, Zhihua and Luo, Qiang and Li, Junzhuo and Luo, Yulun and Lei, Bin and Zhang, Ang and Jiang, Bin},
  year = 2026,
  journal = {Journal of Alloys and Compounds},
  volume = {1050},
  pages = {185527},
  doi = {https://doi.org/10.1016/j.jallcom.2025.185527},
  urldate = {2025-12-13},
  language = {en}
}

@article{qianInfluenceAlloyingElement2022,
  title = {Influence of Alloying Element Segregation at Grain Boundary on the Microstructure and Mechanical Properties of {{Mg-Zn}} Alloy},
  author = {Qian, Xiaoying and Dong, Zhihua and Jiang, Bin and Lei, Bin and Yang, Huabao and He, Chao and Liu, Lintao and Wang, Cuihong and Yuan, Ming and Yang, Hong and Yang, Baoqing and Zheng, Changyong and Pan, Fusheng},
  year = 2022,
  journal = {Materials \& Design},
  volume = {224},
  pages = {111322},
  doi = {https://doi.org/10.1016/j.matdes.2022.111322},
  urldate = {2026-03-30},
  language = {en}
}

@article{schulerMaterialsSelectionRules2017,
  title = {Materials Selection Rules for Amorphous Complexion Formation in Binary Metallic Alloys},
  author = {Schuler, Jennifer D. and Rupert, Timothy J.},
  year = 2017,
  journal = {Acta Materialia},
  volume = {140},
  pages = {196--205},
  doi = {https://doi.org/10.1016/j.actamat.2017.08.042},
  urldate = {2025-10-21},
  language = {en}
}

@article{seahGrainBoundarySegregation1973,
  title = {Grain Boundary Segregation},
  author = {Seah, M. P. and Hondros, E. D.},
  year = 1973,
  journal = {Proceedings of the Royal Society of London. A. Mathematical and Physical Sciences},
  volume = {335},
  number = {1601},
  pages = {191--212},
  doi = {https://doi.org/10.1098/rspa.1973.0121},
  urldate = {2026-03-13}
}

@article{tuchindaComputedEntropySpectra2024,
  title = {Computed Entropy Spectra for Grain Boundary Segregation in Polycrystals},
  author = {Tuchinda, Nutth and Schuh, Christopher A.},
  year = 2024,
  journal = {npj Computational Materials},
  volume = {10},
  number = {1},
  pages = {72},
  doi = {https://doi.org/10.1038/s41524-024-01260-3},
  urldate = {2024-04-15},
  language = {en}
}

@article{tuchindaGrainSizeDependencies2022,
  title = {Grain Size Dependencies of Intergranular Solute Segregation in Nanocrystalline Materials},
  author = {Tuchinda, Nutth and Schuh, Christopher A.},
  year = 2022,
  journal = {Acta Materialia},
  volume = {226},
  pages = {117614},
  doi = {https://doi.org/10.1016/j.actamat.2021.117614},
  urldate = {2023-03-02},
  language = {en}
}

@article{tuchindaVibrationalEntropySpectra2023,
  title = {The Vibrational Entropy Spectra of Grain Boundary Segregation in Polycrystals},
  author = {Tuchinda, Nutth and Schuh, Christopher A.},
  year = 2023,
  journal = {Acta Materialia},
  volume = {245},
  pages = {118630},
  doi = {https://doi.org/10.1016/j.actamat.2022.118630},
  urldate = {2023-01-25},
  language = {en}
}

@article{wagihDesigningCooperativeGrain2025,
  title = {Designing for Cooperative Grain Boundary Segregation in Multicomponent Alloys},
  author = {Wagih, Malik and Naunheim, Yannick and Lei, Tianjiao and Schuh, Christopher A.},
  year = 2025,
  journal = {Proceedings of the National Academy of Sciences},
  volume = {122},
  number = {38},
  pages = {e2511930122},
  publisher = {Proceedings of the National Academy of Sciences},
  doi = {https://doi.org/10.1073/pnas.2511930122},
  urldate = {2026-04-02}
}

@article{wagihGrainBoundarySegregation2020,
  title = {Grain Boundary Segregation beyond the Dilute Limit: {{Separating}} the Two Contributions of Site Spectrality and Solute Interactions},
  shorttitle = {Grain Boundary Segregation beyond the Dilute Limit},
  author = {Wagih, Malik and Schuh, Christopher A.},
  year = 2020,
  journal = {Acta Materialia},
  volume = {199},
  pages = {63--72},
  doi = {https://doi.org/10.1016/j.actamat.2020.08.022},
  urldate = {2022-02-20},
  language = {en}
}

@article{wagihLearningGrainBoundary2020,
  title = {Learning Grain Boundary Segregation Energy Spectra in Polycrystals},
  author = {Wagih, Malik and Larsen, Peter M. and Schuh, Christopher A.},
  year = 2020,
  journal = {Nature Communications},
  volume = {11},
  number = {1},
  pages = {6376},
  doi = {https://doi.org/10.1038/s41467-020-20083-6},
  urldate = {2022-02-27},
  language = {en}
}

@article{wagihLearningGrainBoundarySegregation2022,
  title = {Learning {{Grain-Boundary Segregation}}: {{From First Principles}} to {{Polycrystals}}},
  shorttitle = {Learning {{Grain-Boundary Segregation}}},
  author = {Wagih, Malik and Schuh, Christopher A.},
  year = 2022,
  journal = {Physical Review Letters},
  volume = {129},
  number = {4},
  pages = {046102},
  doi = {https://doi.org/10.1103/PhysRevLett.129.046102},
  urldate = {2022-07-21},
  language = {en}
}

@article{wagihSpectrumGrainBoundary2019,
  title = {Spectrum of Grain Boundary Segregation Energies in a Polycrystal},
  author = {Wagih, Malik and Schuh, Christopher A.},
  year = 2019,
  journal = {Acta Materialia},
  volume = {181},
  pages = {228--237},
  doi = {https://doi.org/10.1016/j.actamat.2019.09.034},
  urldate = {2022-02-20},
  language = {en}
}

@article{wangCALPHADIntegratedGrain2023,
  title = {{{CALPHAD}} Integrated Grain Boundary Co-Segregation Design: {{Towards}} Safe High-Entropy Alloys},
  shorttitle = {{{CALPHAD}} Integrated Grain Boundary Co-Segregation Design},
  author = {Wang, Lei and Darvishi Kamachali, Reza},
  year = 2023,
  journal = {Journal of Alloys and Compounds},
  volume = {933},
  pages = {167717},
  doi = {https://doi.org/10.1016/j.jallcom.2022.167717},
  urldate = {2024-08-26},
  language = {en}
}

@article{xiaoCurrentProgressSolute2026,
  title = {Current Progress on Solute Segregation at Interfaces in {{Mg}} Alloys},
  author = {Xiao, Lirong and Chen, Xinjian and Gao, Bo and Dong, Xinxin and Pan, Zaifan and Zhou, Hao},
  year = 2026,
  journal = {Materials Research Letters},
  pages = {1--20},
  doi = {https://doi.org/10.1080/21663831.2026.2619451},
  urldate = {2026-01-28},
  language = {en}
}

@article{xingSoluteInteractionEffects2018,
  title = {Solute Interaction Effects on Grain Boundary Segregation in Ternary Alloys},
  author = {Xing, Wenting and Kalidindi, Arvind R. and Amram, Dor and Schuh, Christopher A.},
  year = 2018,
  journal = {Acta Materialia},
  volume = {161},
  pages = {285--294},
  doi = {https://doi.org/10.1016/j.actamat.2018.09.005},
  urldate = {2023-10-27},
  language = {en}
}

@article{zengTextureEvolutionStatic2016,
  title = {Texture Evolution during Static Recrystallization of Cold-Rolled Magnesium Alloys},
  author = {Zeng, Z.R. and Zhu, Y.M. and Xu, S.W. and Bian, M.Z. and Davies, C.H.J. and Birbilis, N. and Nie, J.F.},
  year = 2016,
  journal = {Acta Materialia},
  volume = {105},
  pages = {479--494},
  doi = {https://doi.org/10.1016/j.actamat.2015.12.045},
  urldate = {2026-03-27},
  language = {en}
}

@article{zhangDesigningHighMechanical2024,
  title = {Towards Designing High Mechanical Performance Low-Alloyed Wrought Magnesium Alloys via Grain Boundary Segregation Strategy: {{A}} Review},
  shorttitle = {Towards Designing High Mechanical Performance Low-Alloyed Wrought Magnesium Alloys via Grain Boundary Segregation Strategy},
  author = {Zhang, Zhi and Xie, Jinshu and Zhang, Jinghuai and Yang, Xu-Sheng and Wu, Ruizhi},
  year = 2024,
  journal = {Journal of Magnesium and Alloys},
  pages = {S2213956724001208},
  doi = {https://doi.org/10.1016/j.jma.2024.03.016},
  urldate = {2024-04-15},
  language = {en}
}

@article{zhangDisconnectionFormationSegregationinduced2026,
  title = {Disconnection Formation via Segregation-Induced Grain Boundary Phase Transitions},
  author = {Zhang, Zuoyong and Deng, Chuang},
  year = 2026,
  journal = {Acta Materialia},
  volume = {309},
  pages = {122077},
  doi = {https://doi.org/10.1016/j.actamat.2026.122077},
  urldate = {2026-03-07},
  language = {en}
}

@article{zhangGrainBoundaryInterstitial2025,
  title = {Grain Boundary Interstitial Segregation in Substitutional Binary Alloys},
  author = {Zhang, Zuoyong and Deng, Chuang},
  year = 2025,
  journal = {Acta Materialia},
  volume = {291},
  pages = {121019},
  publisher = {Elsevier BV},
  doi = {https://doi.org/10.1016/j.actamat.2025.121019},
  urldate = {2025-04-30},
  copyright = {https://www.elsevier.com/tdm/userlicense/1.0/},
  language = {en}
}

@article{zhangGrainBoundarySegregation2024,
  title = {Grain Boundary Segregation Prediction with a Dual-Solute Model},
  author = {Zhang, Zuoyong and Deng, Chuang},
  year = 2024,
  journal = {Physical Review Materials},
  volume = {8},
  number = {10},
  pages = {103605},
  doi = {https://doi.org/10.1103/PhysRevMaterials.8.103605},
  urldate = {2025-06-01},
  language = {en}
}

@article{zhangHydrostaticPressureinducedTransition2023,
  title = {Hydrostatic Pressure-Induced Transition in Grain Boundary Segregation Tendency in Nanocrystalline Metals},
  author = {Zhang, Zuoyong and Deng, Chuang},
  year = 2023,
  journal = {Scripta Materialia},
  volume = {234},
  pages = {115576},
  doi = {https://doi.org/10.1016/j.scriptamat.2023.115576},
  urldate = {2024-02-04},
  language = {en}
}

@article{zhaoDirectObservationImpact2019,
  title = {Direct Observation and Impact of Co-Segregated Atoms in Magnesium Having Multiple Alloying Elements},
  author = {Zhao, Xiaojun and Chen, Houwen and Wilson, Nick and Liu, Qing and Nie, Jian-Feng},
  year = 2019,
  journal = {Nature Communications},
  volume = {10},
  number = {1},
  pages = {3243},
  doi = {https://doi.org/10.1038/s41467-019-10921-7},
  urldate = {2025-09-29},
  language = {en}
}

@article{fang2022machine,
  title = {Machine Learning Accelerates the Materials Discovery},
  author = {Fang, Jiheng and Xie, Ming and He, Xingqun and Zhang, Jiming and Hu, Jieqiong and Chen, Yongtai and Yang, Youcai and Jin, Qinglin},
  year = 2022,
  journal = {Materials Today Communications},
  volume = {33},
  pages = {104900},
  doi = {https://doi.org/10.1016/j.mtcomm.2022.104900},
  urldate = {2026-04-03},
  language = {en}
}

@article{nenningerLocalAtomicEnvironment2023,
	title = {Local Atomic Environment Analysis of Short and Long-Range Solute-Solute Interactions in a Symmetric Tilt Grain Boundary},
	author = {Nenninger, Tara and Sansoz, Frederic},
	year = 2023,
	journal = {Scripta Materialia},
	volume = {222},
	pages = {115045},
	doi = {https://doi.org/10.1016/j.scriptamat.2022.115045},
	urldate = {2023-01-31},
	language = {en}
}

@article{nenningerSoluteClusteringPolycrystals2025,
	title = {Solute Clustering in Polycrystals: {{Unveiling}} the Interplay of Grain Boundary Junction and Long-Range Solute Attraction Effects},
	shorttitle = {Solute Clustering in Polycrystals},
	author = {Nenninger, Tara and Sansoz, Frederic},
	year = 2025,
	journal = {Acta Materialia},
	volume = {290},
	pages = {120946},
	doi = {https://doi.org/10.1016/j.actamat.2025.120946},
	urldate = {2025-10-21},
	language = {en}
}

@article{nikitinSizedependentAttractionCu2025,
	title = {Size-Dependent Attraction of {{Cu}} Solutes to Clusters Formed at {{Ag}} Grain Boundaries},
	author = {Nikitin, Pavel and Sansoz, Frederic},
	year = 2025,
	journal = {Physical Review Materials},
	volume = {9},
	number = {9},
	pages = {093603},
	doi = {https://doi.org/10.1103/t7qb-g8n6},
	urldate = {2025-10-21},
	language = {en}
}

@article{guttmannEquilibriumSegregationTernary1975,
	title = {Equilibrium Segregation in a Ternary Solution: {{A}} Model for Temper Embrittlement},
	shorttitle = {Equilibrium Segregation in a Ternary Solution},
	author = {Guttmann, M.},
	year = 1975,
	journal = {Surface Science},
	volume = {53},
	number = {1},
	pages = {213--227},
	doi = {https://doi.org/10.1016/0039-6028(75)90125-9},
	urldate = {2026-03-13},
	copyright = {https://www.elsevier.com/tdm/userlicense/1.0/},
	language = {en}
}

@article{trelewiczGrainBoundarySegregation2009,
	title = {Grain Boundary Segregation and Thermodynamically Stable Binary Nanocrystalline Alloys},
	author = {Trelewicz, Jason R. and Schuh, Christopher A.},
	year = 2009,
	journal = {Physical Review B},
	volume = {79},
	number = {9},
	pages = {094112},
	doi = {https://doi.org/10.1103/PhysRevB.79.094112},
	urldate = {2023-01-31},
	language = {en}
}

@article{liuMaterialsDiscoveryDesign2017,
	title = {Materials Discovery and Design Using Machine Learning},
	author = {Liu, Yue and Zhao, Tianlu and Ju, Wangwei and Shi, Siqi},
	year = 2017,
	journal = {Journal of Materiomics},
	volume = {3},
	number = {3},
	pages = {159--177},
	doi = {https://doi.org/10.1016/j.jmat.2017.08.002},
	urldate = {2026-04-03},
	language = {en}
}

@article{raccugliaMachinelearningassistedMaterialsDiscovery2016,
	title = {Machine-Learning-Assisted Materials Discovery Using Failed Experiments},
	author = {Raccuglia, Paul and Elbert, Katherine C. and Adler, Philip D. F. and Falk, Casey and Wenny, Malia B. and Mollo, Aurelio and Zeller, Matthias and Friedler, Sorelle A. and Schrier, Joshua and Norquist, Alexander J.},
	year = 2016,
	journal = {Nature},
	volume = {533},
	number = {7601},
	pages = {73--76},
	doi = {https://doi.org/10.1038/nature17439},
	urldate = {2026-04-03},
	language = {en}
}

@article{bartok2013mlprb,
	title = {On Representing Chemical Environments},
	author = {Bart{\'o}k, Albert P. and Kondor, Risi and Cs{\'a}nyi, G{\'a}bor},
	year = 2013,
	journal = {Physical Review B},
	volume = {87},
	number = {18},
	pages = {184115},
	doi = {https://doi.org/10.1103/PhysRevB.87.184115},
	urldate = {2025-01-05},
	copyright = {http://link.aps.org/licenses/aps-default-license},
	language = {en}
}

@article{laakso2023dscribe,
	title = {Updates to the {{DScribe}} Library: {{New}} Descriptors and Derivatives},
	shorttitle = {Updates to the {{DScribe}} Library},
	author = {Laakso, Jarno and Himanen, Lauri and Homm, Henrietta and Morooka, Eiaki V. and J{\"a}ger, Marc O. J. and Todorovi{\'c}, Milica and Rinke, Patrick},
	year = 2023,
	journal = {The Journal of Chemical Physics},
	volume = {158},
	number = {23},
	pages = {234802},
	doi = {https://doi.org/10.1063/5.0151031},
	urldate = {2025-01-06},
	language = {en}
}

@article{pedregosa2011scikit,
	title = {Scikit-Learn: {{Machine}} Learning in {{Python}}},
	author = {Pedregosa, Fabian and Varoquaux, Ga{\"e}l and Gramfort, Alexandre and Michel, Vincent and Thirion, Bertrand and Grisel, Olivier and Blondel, Mathieu and Prettenhofer, Peter and Weiss, Ron and Dubourg, Vincent},
	year = 2011,
	journal = {the Journal of machine Learning research},
	volume = {12},
	pages = {2825--2830},
	publisher = {JMLR. org},
	isbn = {1532-4435}
}

@article{xie2025predsegmg,
	title = {Predicting Grain Boundary Segregation in Magnesium Alloys: {{An}} Atomistically Informed Machine Learning Approach},
	shorttitle = {Predicting Grain Boundary Segregation in Magnesium Alloys},
	author = {Xie, Zhuocheng and Atila, Achraf and Gu{\'e}nol{\'e}, Julien and {Korte-Kerzel}, Sandra and {Al-Samman}, Talal and Kerzel, Ulrich},
	year = 2025,
	journal = {Journal of Magnesium and Alloys},
	publisher = {Elsevier BV},
	doi = {https://doi.org/10.1016/j.jma.2025.03.021},
	urldate = {2025-07-15},
	copyright = {https://www.elsevier.com/tdm/userlicense/1.0/},
	language = {en}
}

@article{pal2021spectrumvol,
	title = {The Spectrum of Atomic Excess Free Volume in Grain Boundaries},
	author = {Pal, Snehanshu and Reddy, K. Vijay and Yu, Tingting and Xiao, Jianwei and Deng, Chuang},
	year = 2021,
	journal = {Journal of Materials Science},
	volume = {56},
	number = {19},
	pages = {11511--11528},
	doi = {https://doi.org/10.1007/s10853-021-06028-4},
	urldate = {2022-02-28},
	language = {en}
}

@inproceedings{chen2016xgboost,
	title = {{{XGBoost}}: {{A Scalable Tree Boosting System}}},
	shorttitle = {{{XGBoost}}},
	booktitle = {Proceedings of the 22nd {{ACM SIGKDD International Conference}} on {{Knowledge Discovery}} and {{Data Mining}}},
	author = {Chen, Tianqi and Guestrin, Carlos},
	year = 2016,
	pages = {785--794},
	publisher = {ACM},
	address = {San Francisco California USA},
	doi = {https://doi.org/10.1145/2939672.2939785},
	urldate = {2026-01-11},
	isbn = {978-1-4503-4232-2},
	language = {en}
}

@article{yuan2025multitaskml,
	title = {Multi-Task Learning of Solute Segregation Energy across Multiple Alloy Systems},
	author = {Yuan, Liang and Ma, Zongyi and Pan, Zhiliang},
	year = 2025,
	journal = {Computational Materials Science},
	volume = {253},
	pages = {113846},
	publisher = {Elsevier BV},
	doi = {https://doi.org/10.1016/j.commatsci.2025.113846},
	urldate = {2025-07-14},
	copyright = {https://www.elsevier.com/tdm/userlicense/1.0/},
	language = {en}
}

@article{plimpton1995lammps,
	title = {Fast {{Parallel Algorithms}} for {{Short-Range Molecular Dynamics}}},
	author = {Plimpton, Steve},
	year = 1995,
	journal = {Journal of Computational Physics},
	volume = {117},
	number = {1},
	pages = {1--19},
	doi = {https://doi.org/10.1006/jcph.1995.1039}
}

@article{thompson2022lammps,
	title = {{{LAMMPS}} - a Flexible Simulation Tool for Particle-Based Materials Modeling at the Atomic, Meso, and Continuum Scales},
	author = {Thompson, Aidan P. and Aktulga, H. Metin and Berger, Richard and Bolintineanu, Dan S. and Brown, W. Michael and Crozier, Paul S. and {in 't Veld}, Pieter J. and Kohlmeyer, Axel and Moore, Stan G. and Nguyen, Trung Dac and Shan, Ray and Stevens, Mark J. and Tranchida, Julien and Trott, Christian and Plimpton, Steven J.},
	year = 2022,
	journal = {Computer Physics Communications},
	volume = {271},
	pages = {108171},
	doi = {https://doi.org/10.1016/j.cpc.2021.108171},
	urldate = {2022-09-22},
	language = {en}
}

@article{stukowski2022ovito,
	title = {Automated Identification and Indexing of Dislocations in Crystal Interfaces},
	author = {Stukowski, Alexander and Bulatov, Vasily V and Arsenlis, Athanasios},
	year = 2012,
	journal = {Modelling and Simulation in Materials Science and Engineering},
	volume = {20},
	number = {8},
	pages = {085007},
	doi = {https://doi.org/10.1088/0965-0393/20/8/085007},
	urldate = {2022-10-09},
	language = {en}
}

@article{larsen2016acna,
	title = {Robust Structural Identification via Polyhedral Template Matching},
	author = {Larsen, Peter Mahler and Schmidt, S{\o}ren and Schi{\o}tz, Jakob},
	year = 2016,
	journal = {Modelling and Simulation in Materials Science and Engineering},
	volume = {24},
	number = {5},
	pages = {055007},
	doi = {https://doi.org/10.1088/0965-0393/24/5/055007},
	urldate = {2022-10-07},
	language = {en}
}

@misc{liang2025nep89,
	title = {{{NEP89}}: {{Universal}} Neuroevolution Potential for Inorganic and Organic Materials across 89 Elements},
	shorttitle = {{{NEP89}}},
	author = {Liang, Ting and Xu, Ke and Lindgren, Eric and Chen, Zherui and Zhao, Rui and Liu, Jiahui and Tang, Benrui and Zhang, Bohan and Wang, Yanzhou and Song, Keke and Ying, Penghua and Dong, Haikuan and Chen, Shunda and Erhart, Paul and Fan, Zheyong and {Ala-Nissila}, Tapio and Xu, Jianbin},
	year = 2025,
	number = {arXiv:2504.21286},
	eprint = {2504.21286},
	primaryclass = {cond-mat},
	publisher = {arXiv},
	doi = {https://doi.org/10.48550/arXiv.2504.21286},
	urldate = {2025-05-02},
	archiveprefix = {arXiv}
}

@article{hirel2015atomsk,
	title = {Atomsk: {{A}} Tool for Manipulating and Converting Atomic Data Files},
	shorttitle = {Atomsk},
	author = {Hirel, Pierre},
	year = 2015,
	journal = {Computer Physics Communications},
	volume = {197},
	pages = {212--219},
	doi = {https://doi.org/10.1016/j.cpc.2015.07.012},
	urldate = {2022-10-07},
	language = {en}
}

@article{berendsen1984mdcoupling,
	title = {Molecular Dynamics with Coupling to an External Bath},
	author = {Berendsen, H. J. C. and Postma, J. P. M. and {van Gunsteren}, W. F. and DiNola, A. and Haak, J. R.},
	year = 1984,
	journal = {The Journal of Chemical Physics},
	volume = {81},
	number = {8},
	pages = {3684--3690},
	doi = {https://doi.org/10.1063/1.448118},
	urldate = {2023-01-09},
	language = {en}
}

@article{parrinello1981poly,
	title = {Polymorphic Transitions in Single Crystals: {{A}} New Molecular Dynamics Method},
	shorttitle = {Polymorphic Transitions in Single Crystals},
	author = {Parrinello, M. and Rahman, A.},
	year = 1981,
	journal = {Journal of Applied Physics},
	volume = {52},
	number = {12},
	pages = {7182--7190},
	doi = {https://doi.org/10.1063/1.328693},
	urldate = {2022-04-03},
	language = {en}
}

@article{fanGPUMDPackage2022,
	title = {{{GPUMD}}: {{A}} Package for Constructing Accurate Machine-Learned Potentials and Performing Highly Efficient Atomistic Simulations},
	shorttitle = {{{GPUMD}}},
	author = {Fan, Zheyong and Wang, Yanzhou and Ying, Penghua and Song, Keke and Wang, Junjie and Wang, Yong and Zeng, Zezhu and Xu, Ke and Lindgren, Eric and Rahm, J. Magnus and Gabourie, Alexander J. and Liu, Jiahui and Dong, Haikuan and Wu, Jianyang and Chen, Yue and Zhong, Zheng and Sun, Jian and Erhart, Paul and Su, Yanjing and {Ala-Nissila}, Tapio},
	year = 2022,
	journal = {The Journal of Chemical Physics},
	volume = {157},
	number = {11},
	pages = {114801},
	doi = {https://doi.org/10.1063/5.0106617},
	urldate = {2026-04-08},
	language = {en}
}

@article{xu2025gpumd4.0,
	title = {{{GPUMD}} 4.0: {{A}} High-performance Molecular Dynamics Package for Versatile Materials Simulations with Machine-learned Potentials},
	shorttitle = {{{GPUMD}} 4.0},
	author = {Xu, Ke and Bu, Hekai and Pan, Shuning and Lindgren, Eric and Wu, Yongchao and Wang, Yong and Liu, Jiahui and Song, Keke and Xu, Bin and Li, Yifan and Hainer, Tobias and Svensson, Lucas and Wiktor, Julia and Zhao, Rui and Huang, Hongfu and Qian, Cheng and Zhang, Shuo and Zeng, Zezhu and Zhang, Bohan and Tang, Benrui and Xiao, Yang and Yan, Zihan and Shi, Jiuyang and Liang, Zhixin and Wang, Junjie and Liang, Ting and Cao, Shuo and Wang, Yanzhou and Ying, Penghua and Xu, Nan and Chen, Chengbing and Zhang, Yuwen and Chen, Zherui and Wu, Xin and Jiang, Wenwu and Berger, Esme and Li, Yanlong and Chen, Shunda and Gabourie, Alexander J. and Dong, Haikuan and Xiong, Shiyun and Wei, Ning and Chen, Yue and Xu, Jianbin and Ding, Feng and Sun, Zhimei and Ala-Nissila, Tapio and Harju, Ari and Zheng, Jincheng and Guan, Pengfei and Erhart, Paul and Sun, Jian and Ouyang, Wengen and Su, Yanjing and Fan, Zheyong},
	year = 2025,
	journal = {Materials Genome Engineering Advances},
	volume = {3},
	number = {3},
	pages = {e70028},
	doi = {https://doi.org/10.1002/mgea.70028},
	urldate = {2025-12-22},
	language = {en}
}

@article{sadigh2012vcsgc,
	title = {Scalable Parallel {{Monte Carlo}} Algorithm for Atomistic Simulations of Precipitation in Alloys},
	author = {Sadigh, Babak and Erhart, Paul and Stukowski, Alexander and Caro, Alfredo and Martinez, Enrique and {Zepeda-Ruiz}, Luis},
	year = 2012,
	journal = {Physical Review B},
	volume = {85},
	number = {18},
	pages = {184203},
	doi = {https://doi.org/10.1103/PhysRevB.85.184203},
	urldate = {2022-01-06},
	language = {en}
}

@article{bitzekStructuralRelaxationMade2006,
	title = {Structural {{Relaxation Made Simple}}},
	author = {Bitzek, Erik and Koskinen, Pekka and G{\"a}hler, Franz and Moseler, Michael and Gumbsch, Peter},
	year = 2006,
	journal = {Physical Review Letters},
	volume = {97},
	number = {17},
	pages = {170201},
	doi = {https://doi.org/10.1103/PhysRevLett.97.170201},
	urldate = {2026-04-09},
	copyright = {http://link.aps.org/licenses/aps-default-license},
	language = {en}
}

@article{guenole2020fire,
	title = {Assessment and Optimization of the Fast Inertial Relaxation Engine (Fire) for Energy Minimization in Atomistic Simulations and Its Implementation in Lammps},
	author = {Gu{\'e}nol{\'e}, Julien and N{\"o}hring, Wolfram G. and Vaid, Aviral and Houll{\'e}, Fr{\'e}d{\'e}ric and Xie, Zhuocheng and Prakash, Aruna and Bitzek, Erik},
	year = 2020,
	journal = {Computational Materials Science},
	volume = {175},
	pages = {109584},
	doi = {https://doi.org/10.1016/j.commatsci.2020.109584},
	urldate = {2022-12-21},
	language = {en}
}

@article{azzaliniClass1985,
	title = {A Class of Distributions Which Includes the Normal Ones},
	author = {Azzalini, Adelchi},
	year = 1985,
	journal = {Scandinavian journal of statistics},
	pages = {171--178},
	publisher = {JSTOR},
	isbn = {0303-6898}
}

@article{leaSiteCompetitionSurface1975,
  title = {Site Competition in Surface Segregation},
  author = {Lea, Colin and Seah, M. P.},
  year = 1975,
  journal = {Surface Science},
  volume = {53},
  number = {1},
  pages = {272--285},
  doi = {https://doi.org/10.1016/0039-6028(75)90129-6},
  urldate = {2026-04-14}
}

@article{xue2025utilizing,
  title = {Utilizing Atomic-Scale Grain Boundary Segregation: A New Way for Constructing the Lamellar Heterogeneous Magnesium Alloy},
  shorttitle = {Utilizing Atomic-Scale Grain Boundary Segregation},
  author = {Xue, Hansong and Zhou, Yang and Wei, Zengjun and Zhang, Ming and Li, Jun and Liu, Ying and She, Jia and Hu, Jia and Jiang, Bin},
  year = 2025,
  journal = {Materials Research Letters},
  pages = {1--8},
  doi = {https://doi.org/10.1080/21663831.2025.2544775},
  urldate = {2025-09-22},
  language = {en}
}

@article{zhouEffectAgInterfacial2015,
  title = {Effect of {{Ag}} on Interfacial Segregation in {{Mg}}--{{Gd}}--{{Y}}--({{Ag}})--{{Zr}} Alloy},
  author = {Zhou, H. and Cheng, G.M. and Ma, X.L. and Xu, W.Z. and Mathaudhu, S.N. and Wang, Q.D. and Zhu, Y.T.},
  year = 2015,
  journal = {Acta Materialia},
  volume = {95},
  pages = {20--29},
  doi = {https://doi.org/110.1016/j.actamat.2015.05.020},
  urldate = {2026-04-17},
  language = {en}
}

@article{jang2022acomparative,
  title = {A Comparative Study on Grain Boundary Segregation and Solute Clustering in {{Mg-Al-Zn}} and {{Mg-Zn-Ca}} Alloys},
  author = {Jang, Hyo-Sun and Seol, Donghyuk and Lee, Byeong-Joo},
  year = 2022,
  journal = {Journal of Alloys and Compounds},
  volume = {894},
  pages = {162539},
  doi = {https://doi.org/10.1016/j.jallcom.2021.162539},
  urldate = {2025-10-15},
  language = {en}
}

@article{linSynergistic2023a,
  title = {Synergistic Effect of Multi-Element Co-Segregation on Mechanical Properties of {{Mg}} 10 1 - 2 Twin Grain Boundary},
  author = {Lin, Jiarui and Han, Hui and Tian, Yanzhong and Pang, Xueyong and Qin, Gaowu},
  year = 2023,
  journal = {Modelling and Simulation in Materials Science and Engineering},
  volume = {31},
  number = {6},
  pages = {065009},
  doi = {https://doi.org/10.1088/1361-651X/ace0d2},
  urldate = {2026-04-16},
  language = {en}
}

@article{zeng2026MSEA,
  title = {Tuning Grain Boundary Cohesion and Plasticity in {{Mg}} Alloys by Grain Boundary Segregation and Deformation Mechanism Manipulation},
  author = {Zeng, Huaqiang and Shi, Dongfeng and Yang, Biaobiao and Shi, Chenying and Zhang, Jin},
  year = 2026,
  journal = {Materials Science and Engineering: A},
  volume = {956},
  pages = {149910},
  doi = {https://doi.org/10.1016/j.msea.2026.149910},
  urldate = {2026-03-07},
  language = {en}
}

@article{nakata2022scripta,
  title = {Role of Grain Boundary Segregation on Microstructural Development in Basal-Textured {{Mg-Al-Zn}} Alloy Sheet},
  author = {Nakata, T. and Li, Z.H. and Sasaki, T.T. and Hono, K. and Kamado, S.},
  year = 2022,
  journal = {Scripta Materialia},
  volume = {218},
  pages = {114828},
   doi = {https://doi.org/10.1016/j.scriptamat.2022.114828},
  urldate = {2026-01-20},
  language = {en}
}

@article{bian2021solute,
  title = {Solute Segregation Assisted Grain Boundary Precipitation and Its Impact to Ductility of a Precipitation-Hardenable Magnesium Alloy},
  author = {Bian, Mingzhe and Huang, Xinsheng and Chino, Yasumasa},
  year = 2021,
  journal = {Materials Science and Engineering: A},
  volume = {819},
  pages = {141481},
  doi = {https://doi.org/10.1016/j.msea.2021.141481},
  urldate = {2026-02-02},
  language = {en}
}

@article{bian2022Improving,
  title = {Improving the Mechanical and Corrosion Properties of Pure Magnesium by Parts-per-Million-Level Alloying},
  author = {Bian, Mingzhe and Nakatsugawa, Isao and Matsuoka, Yusuke and Huang, Xinsheng and Tsukada, Yuhki and Koyama, Toshiyuki and Chino, Yasumasa},
  year = 2022,
  journal = {Acta Materialia},
  volume = {241},
  pages = {118393},
  doi = {https://doi.org/10.1016/j.actamat.2022.118393},
  urldate = {2026-04-17},
  language = {en}
}

@article{yang2026age-hardening,
  title = {Age-Hardening Response Limited by Solute Clustering in an Ultrafine-Grained {{Mg}}--{{Gd}}--{{Ag}} Alloy},
  author = {Yang, Zhenquan and Ma, Aibin and Jiang, Jinghua and Xu, Bingqian and Sun, Jiapeng},
  year = 2026,
  journal = {Journal of Alloys and Compounds},
  volume = {1052},
  pages = {186117},
  doi = {https://doi.org/10.1016/j.jallcom.2026.186117},
  urldate = {2026-01-13},
  language = {en}
}

@article{yang2025toward,
  title = {Towards Age-Hardening Ability Enhancement and High Strength in {{Mg}}--{{Gd}}--{{Ag}} Alloy by Balancing Grain Refinement and Weakening of Dynamic Precipitation},
  author = {Yang, Zhenquan and Ma, Aibin and Xu, Bingqian and Wang, Guowei and Jiang, Jinghua and Sun, Jiapeng},
  year = 2025,
  journal = {Journal of Magnesium and Alloys},
  volume = {13},
  number = {4},
  pages = {1699--1720},
  doi = {https://doi.org/10.1016/j.jma.2024.06.032},
  urldate = {2026-01-14},
  language = {en}
}

@article{qian2026jmst,
  title = {Mechanism of Alloying Element Co-Segregation at the Grain Boundary and Its Influence on Mechanical Properties in {{Mg-Zn-Ca}} Alloy},
  author = {Qian, Xiaoying and Dong, Zhihua and Jiang, Bin and Zheng, Zhiying and Zhang, Ang and Li, Changle and Vitos, Levente},
  year = 2026,
  journal = {Journal of Materials Science \& Technology},
  volume = {267},
  pages = {132--145},
  doi = {https://doi.org/10.1016/j.jmst.2025.12.016},
  urldate = {2026-03-14},
  language = {en}
}

@article{zhang2022significantly,
  title = {Significantly Enhanced Grain Boundary {{Zn}} and {{Ca}} Co-Segregation of Dilute {{Mg}} Alloy via Trace {{Sm}} Addition},
  author = {Zhang, Zhi and Zhang, Jinghuai and Xie, Jinshu and Liu, Shujuan and He, Yuying and Wang, Ru and Fang, Daqing and Fu, Wei and Jiao, Yunlei and Wu, Ruizhi},
  year = 2022,
  journal = {Materials Science and Engineering: A},
  volume = {831},
  pages = {142259},
  doi = {https://doi.org/10.1016/j.msea.2021.142259},
  urldate = {2025-07-30},
  language = {en}
}

@article{maEffectExtrusionParameters2025,
  title = {Effect of Extrusion Parameters on Microstructure and Corrosion Behavior of {{Mg-Gd-Ni-Ca}} Alloy},
  author = {Ma, Kai and Wang, Jingfeng and Wang, Liqing and Wang, Kui},
  year = 2025,
  journal = {Journal of Alloys and Compounds},
  volume = {1033},
  pages = {181281},
  doi = {https://doi.org/10.1016/j.jallcom.2025.181281},
  urldate = {2026-04-17},
  language = {en}
}

@article{matsonAtomisticAssessmentSoluteSolute2021,
  title = {Atomistic Assessment of Solute-Solute Interactions during Grain Boundary Segregation},
  author = {Matson, Thomas P. and Schuh, Christopher A.},
  year = 2021,
  journal = {Nanomaterials},
  volume = {11},
  number = {9},
  pages = {2360},
  doi = {https://doi.org/10.3390/nano11092360},
  urldate = {2022-10-08},
  language = {en}
}

@article{scheiberImpactSolutesoluteInteractions2018c,
  title = {Impact of Solute-Solute Interactions on Grain Boundary Segregation and Cohesion in Molybdenum},
  author = {Scheiber, Daniel and Romaner, Lorenz and Pippan, Reinhard and Puschnig, Peter},
  year = 2018,
  journal = {Physical Review Materials},
  volume = {2},
  number = {9},
  pages = {093609},
  doi = {https://doi.org/10.1103/PhysRevMaterials.2.093609},
  urldate = {2026-07-17},
  language = {en}
}

@article{maiSegregationTransitionMetals2022b,
  title = {The Segregation of Transition Metals to Iron Grain Boundaries and Their Effects on Cohesion},
  author = {Mai, Han Lin and Cui, Xiang-Yuan and Scheiber, Daniel and Romaner, Lorenz and Ringer, Simon P.},
  year = 2022,
  journal = {Acta Materialia},
  volume = {231},
  pages = {117902},
  doi = {https://doi.org/10.1016/j.actamat.2022.117902},
  urldate = {2026-07-17},
  language = {en}
}

@article{liDMD2011,
  title = {Diffusive Molecular Dynamics and Its Application to Nanoindentation and Sintering},
  author = {Li, Ju and Sarkar, Sanket and Cox, William T. and Lenosky, Thomas J. and Bitzek, Erik and Wang, Yunzhi},
  year = 2011,
  journal = {Physical Review B},
  volume = {84},
  number = {5},
  pages = {054103},
  doi = {https://doi.org/10.1103/PhysRevB.84.054103},
  urldate = {2024-08-19},
  copyright = {http://link.aps.org/licenses/aps-default-license},
  language = {en}
}

@article{najafabadiFiniteTemperatureStructure1990a,
  title = {Finite Temperature Structure and Thermodynamics of the {{Au $\Sigma$5}} (001) Twist Boundary},
  author = {Najafabadi, R. and Srolovitz, D. J. and LeSar, R.},
  year = 1990,
  journal = {Journal of Materials Research},
  doi = {https://doi.org/10.1557/jmr.1990.2663},
  urldate = {2026-07-15}
}

@article{dontsovaSoluteSegregationKinetics2015,
  title = {Solute Segregation Kinetics and Dislocation Depinning in a Binary Alloy},
  author = {Dontsova, E. and Rottler, J. and Sinclair, C. W.},
  year = 2015,
  journal = {Physical Review B},
  volume = {91},
  number = {22},
  pages = {224103},
  doi = {https://doi.org/10.1103/PhysRevB.91.224103},
  urldate = {2024-08-19},
  copyright = {http://link.aps.org/licenses/aps-default-license},
  language = {en}
}

@article{mengGrainBoundaryMisorientationdependent2026,
  title = {Grain Boundary Misorientation-Dependent {{Ca}} Solute Segregation Enabling Superior Strength-Formability Synergy of a Low-Alloyed {{Mg-Al-Sn-Ca}} Sheet},
  author = {Meng, Yun-Peng and Hua, Zhen-Ming and Chen, Jia-Rui and Jia, Hai-Long and Jin, Shen-Bao and Wang, Cheng and Zha, Min and Wang, Hui-Yuan},
  year = 2026,
  journal = {Materials Research Letters},
  volume = {0},
  number = {0},
  pages = {1--10},
  publisher = {Taylor \& Francis},
  doi = {https://doi.org/10.1080/21663831.2026.2684786},
  urldate = {2026-06-09}
}

@article{lejcekEntropyMattersGrain2021a,
  title = {Entropy Matters in Grain Boundary Segregation},
  author = {Lej{\v c}ek, P. and Hofmann, S. and V{\v s}iansk{\'a}, M. and {\v S}ob, M.},
  year = 2021,
  journal = {Acta Materialia},
  volume = {206},
  pages = {116597},
  doi = {https://doi.org/10.1016/j.actamat.2020.116597},
  urldate = {2026-07-17}
}

@article{barrRoleGrainBoundary2021,
  title = {The Role of Grain Boundary Character in Solute Segregation and Thermal Stability of Nanocrystalline {{Pt}}--{{Au}}},
  author = {Barr, Christopher M. and Foiles, Stephen M. and Alkayyali, Malek and Mahmood, Yasir and Price, Patrick M. and Adams, David P. and Boyce, Brad L. and Abdeljawad, Fadi and Hattar, Khalid},
  year = 2021,
  journal = {Nanoscale},
  volume = {13},
  number = {6},
  pages = {3552--3563},
  doi = {https://doi.org/10.1039/D0NR07180C},
  urldate = {2026-07-19},
  language = {en}
}

@article{tuchindaTripleJunctionSolute2023,
  title = {Triple Junction Solute Segregation in {{Al-based}} Polycrystals},
  author = {Tuchinda, Nutth and Schuh, Christopher A.},
  year = 2023,
  journal = {Physical Review Materials},
  volume = {7},
  number = {2},
  pages = {023601},
  doi = {https://doi.org/10.1103/PhysRevMaterials.7.023601},
  urldate = {2023-02-18},
  language = {en}
}
	
	\appendix
	\clearpage
	\section{SM1}\label{SM1}
	\begin{center}
\textbf{\large Supplementary materials for\\[26pt]
		Predicting co-segregation in alloys with solute-solute interactions}
		
\large Zuoyong Zhang and Chuang Deng*

\small \textit{Department of Mechanical Engineering, University of Manitoba, Winnipeg, R3T 5V6 Manitoba, Canada}

\vspace{1cm}

*Corresponding author: Chuang.Deng@umanitoba.ca

\end{center}

\vspace{3cm}
\textbf{\large This file includes:}

\leftskip=1cm Supplementary Materials (i)

\leftskip=0cm \textbf{\large Other supplementary materials are included in:}

\leftskip=1cm Supplementary Materials (ii)\\

\leftskip=0cm

\vspace{2.5cm}

\textbf{\large Abstract}

This supplemental file includes all Supplementary Materials that are not included in Supplementary materials (ii).

\newpage

	\setcounter{table}{0}  
\renewcommand{\thetable}{S\arabic{table}}
\renewcommand{\arraystretch}{1.3}
\begin{table}
	
	\centering
	\caption{Hyperparameters used for machine-learning prediction of segregation energies. Moderate hyperparameter settings were employed, which already achieved excellent prediction performance. Therefore, additional hyperparameter optimization or cross-validation was not performed.}\label{Tab1}
	\begin{tabular}{cc}

		\toprule[1.5pt]
		\textbf{XGBoost} & \textbf{Hyperparameters}\\
		\midrule[1.0pt]
		Learning rate & 0.1\\
		Maximum tree depth & 6\\
		Subsample ratio & 0.8\\
		Column sampling ratio & 0.8\\
		Maximum boosting rounds & 1000\\
		Early stopping rounds & 50\\
		Random seed & 9\\
		\bottomrule[1.5pt]
		
	\end{tabular}

\end{table}

\setcounter{figure}{0}
\renewcommand{\thefigure}{S\arabic{figure}}
\begin{figure}
	\centering
	\includegraphics[width=0.9\textwidth]{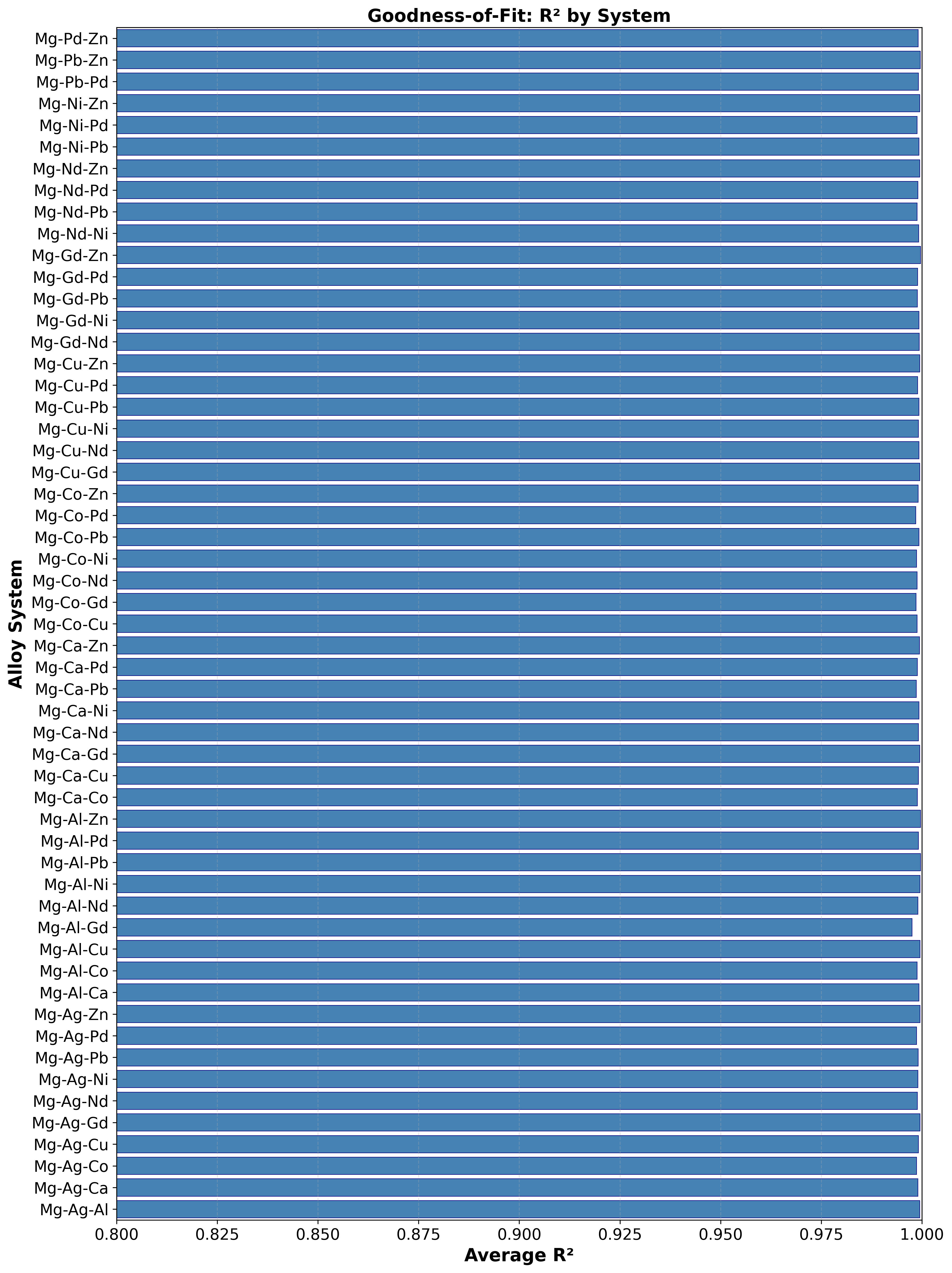}
	\caption{The skew-normal fit accuracy in $R^2$ of the DS spectra within the 55 Mg-based ternary systems}
	\label{figs1}
\end{figure}

\setcounter{figure}{1}
\renewcommand{\thefigure}{S\arabic{figure}}
\begin{figure}
	\centering
	\includegraphics[width=0.9\textwidth]{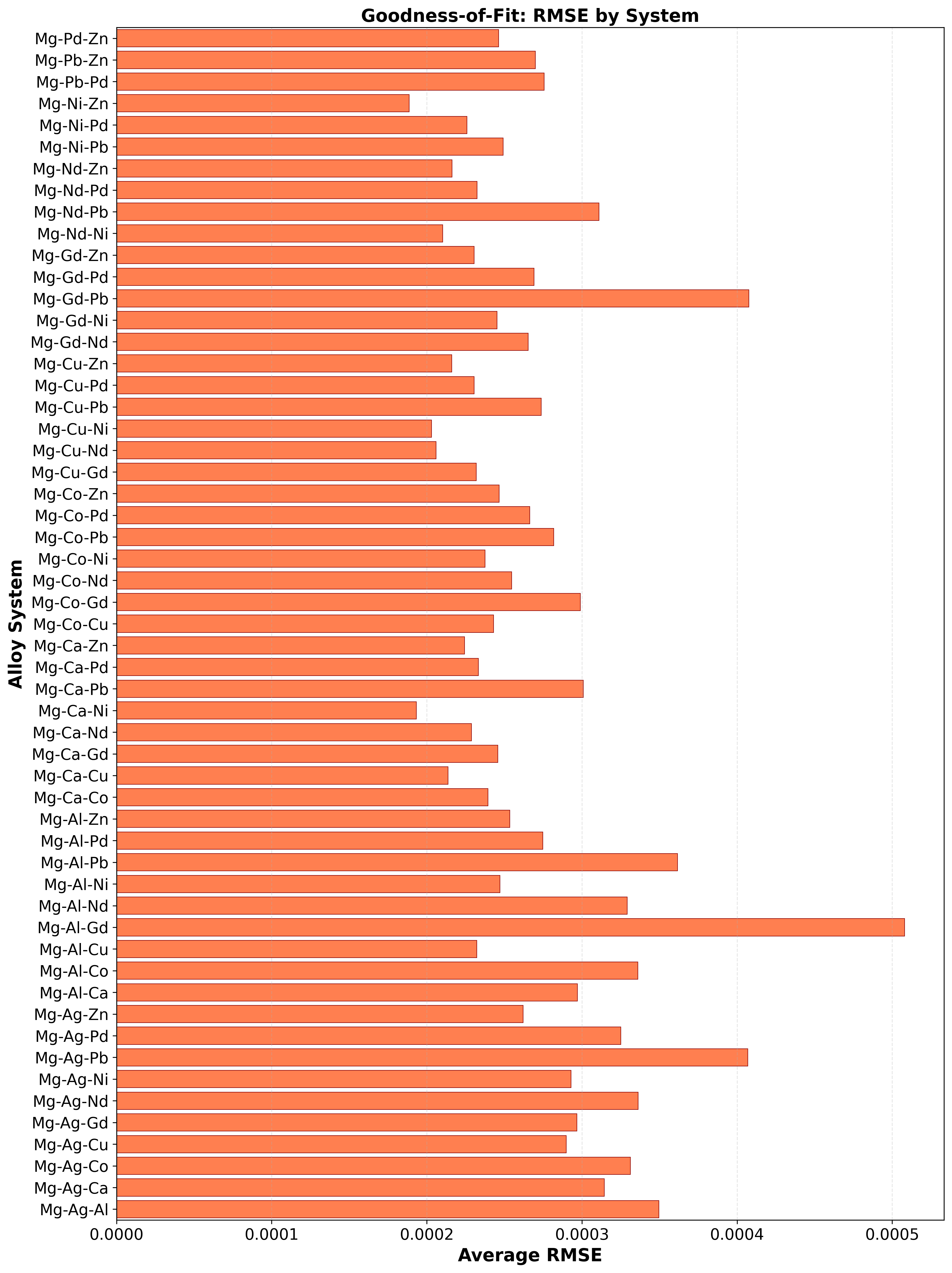}
	\caption{The skew-normal fit accuracy in root mean square error ($RMSE$) of the DS spectra within the 55 Mg-based ternary systems.}
	\label{figs2}
\end{figure}

\setcounter{figure}{2}
\renewcommand{\thefigure}{S\arabic{figure}}
\begin{figure}
	\centering
	\includegraphics[width=0.6\textwidth]{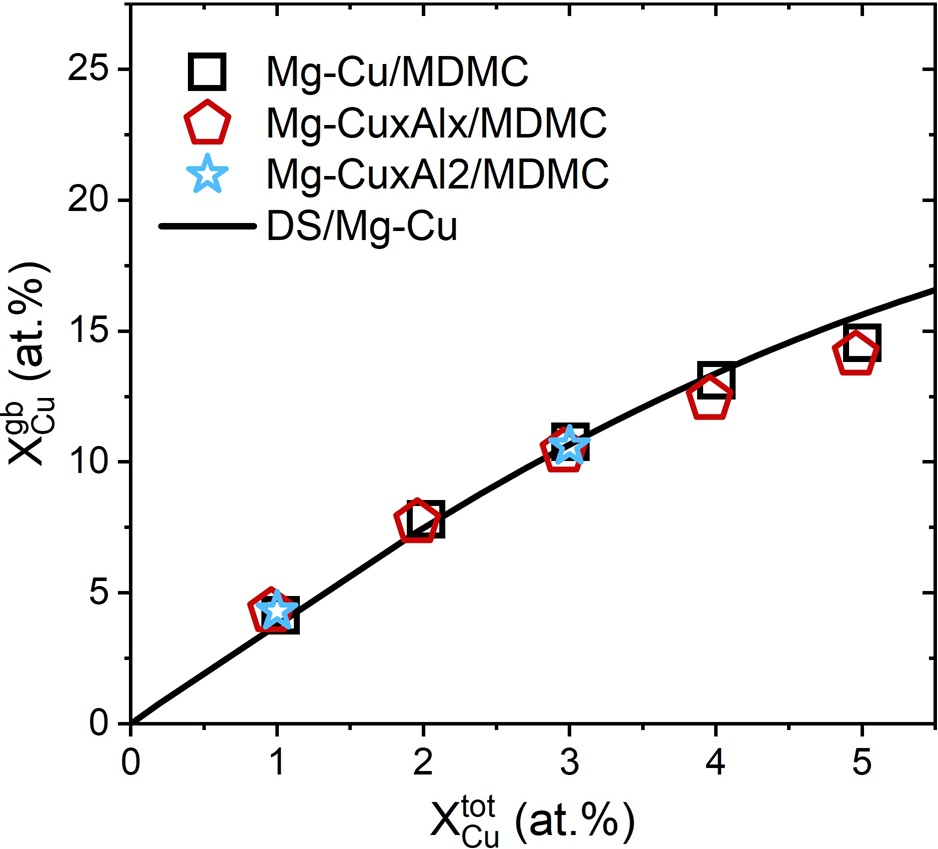}
	\caption{GB segregation prediction vs hybrid molecular dynamics (MD)/Monte Carlo (MC) data points for the Mg-Cu-Al system. The Cu segregation at grain boundaries (GBs) is indenpendent of the Al content due to its intrinsic strong segregation tendency.}
	\label{figs3}
\end{figure}

\setcounter{figure}{3}
\renewcommand{\thefigure}{S\arabic{figure}}
\begin{figure}
	\centering
	\includegraphics[width=1\textwidth]{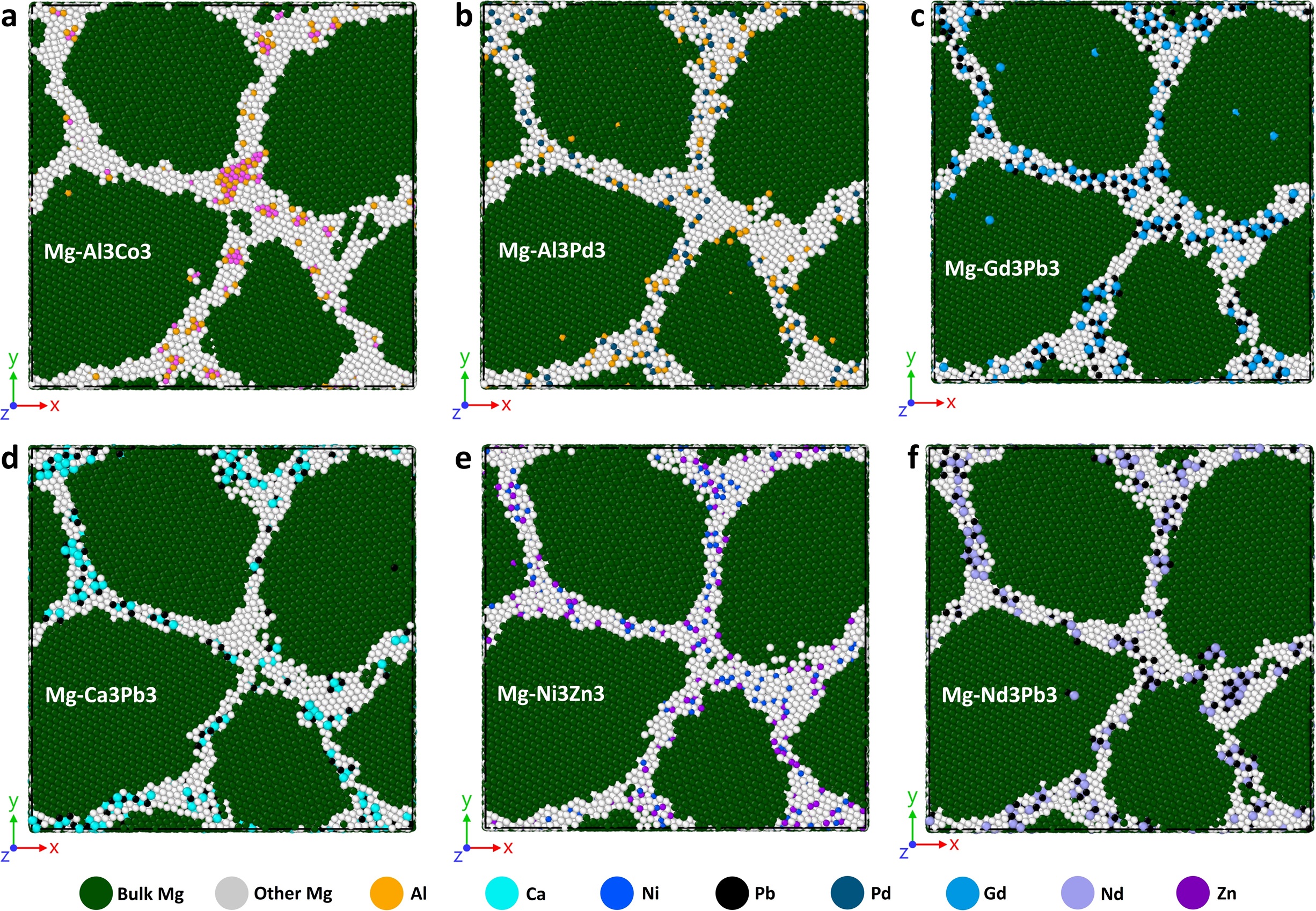}
	\caption{Hybrid MD/MC simulation results for the (a) Mg-(Al,Co), (b) Mg-(Al,Pd), (c) Mg-(Gd,Pb), (d) Mg-(Ca,Pb), (e) Mg-(Ni,Zn), and (f) Mg-(Nd,Pb) respectively. Synergistic co-segregation is observed in these systems despite the presence of site competition. In contrast, for systems without site competition, heteroatomic attraction readily induces synergistic co-segregation, and therefore it is not necessary to present additional hybrid MD/MC results here. For the Mg-(Al,Cu) and Mg-(Al,Ni) systems, their co-segregation behavior has been discussed in the main text.}
	\label{figs4}
\end{figure}

\setcounter{figure}{4}
\renewcommand{\thefigure}{S\arabic{figure}}
\begin{figure}
	\centering
	\includegraphics[width=1\textwidth]{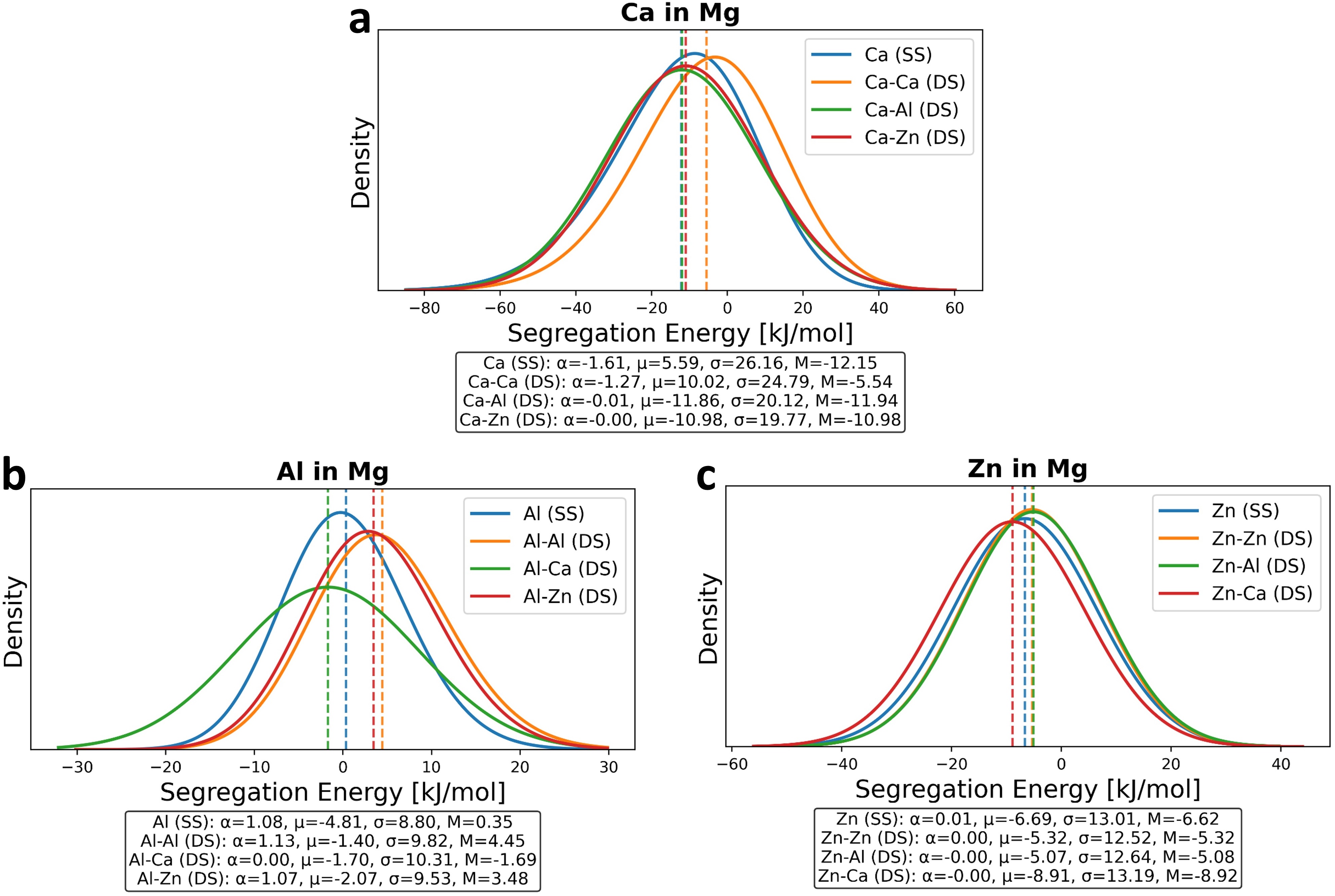}
	\caption{Dual-solute (DS) segregation energy spectrm set for (a) Ca, (b) Al, and (c) Zn, respectively, in the Mg-(Al,Zn,Ca) multicomponent system. The colored dashed vertical lines and the term M in the parameter box represent the ${mean}$ values for corresponding spectra.}
	\label{figs5}
\end{figure}

\setcounter{figure}{5}
\renewcommand{\thefigure}{S\arabic{figure}}
\begin{figure}
	\centering
	\includegraphics[width=1\textwidth]{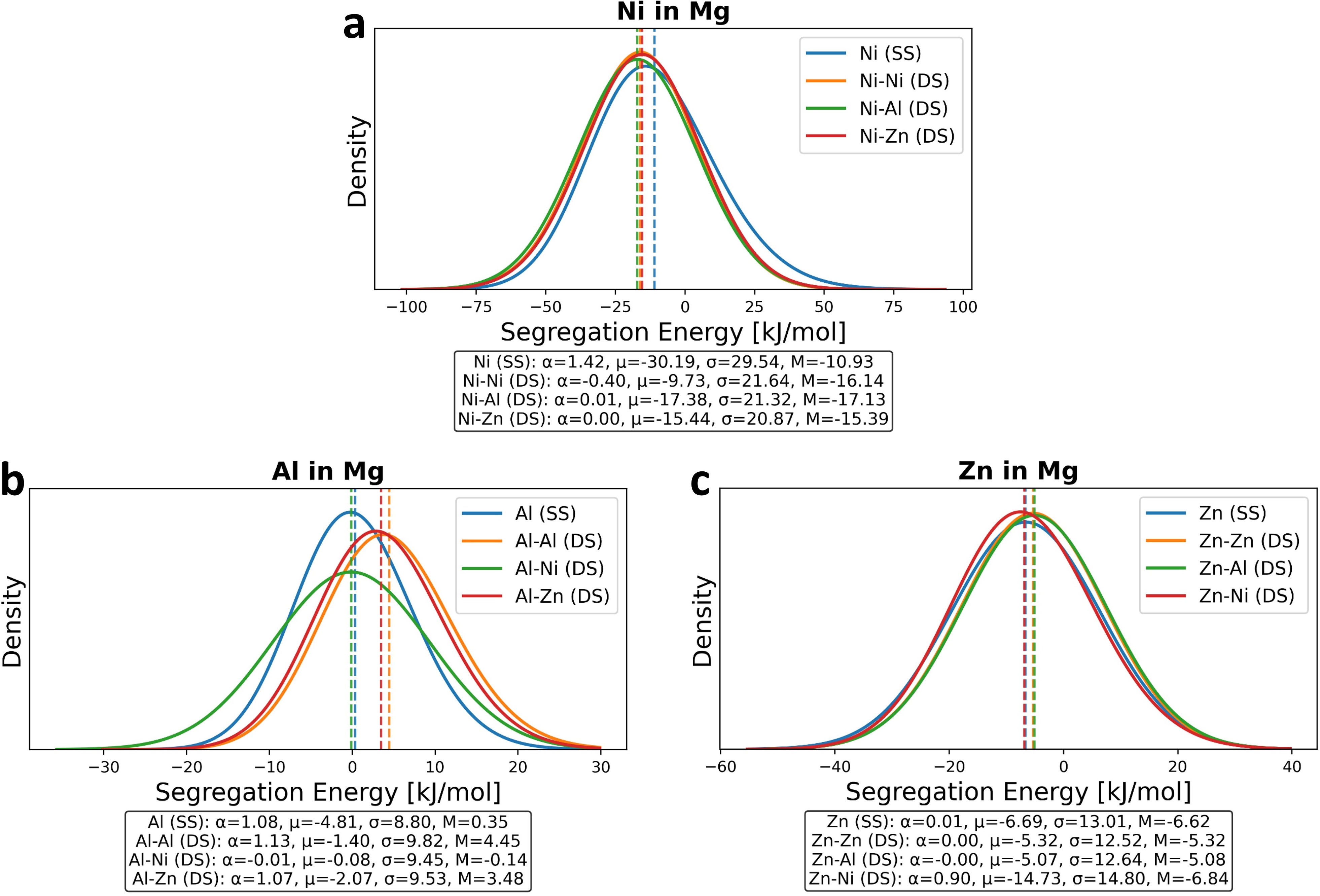}
	\caption{Dual-solute (DS) segregation energy spectrm set for (a) Ni, (b) Al, and (c) Zn, respectively, in the Mg-(Al,Zn,Ni) multicomponent system. The colored dashed vertical lines represent the ${mean}$ values for corresponding spectra.}
	\label{figs6}
\end{figure}

\setcounter{figure}{6}
\renewcommand{\thefigure}{S\arabic{figure}}
\begin{figure}
	\centering
	\includegraphics[width=0.4\textwidth]{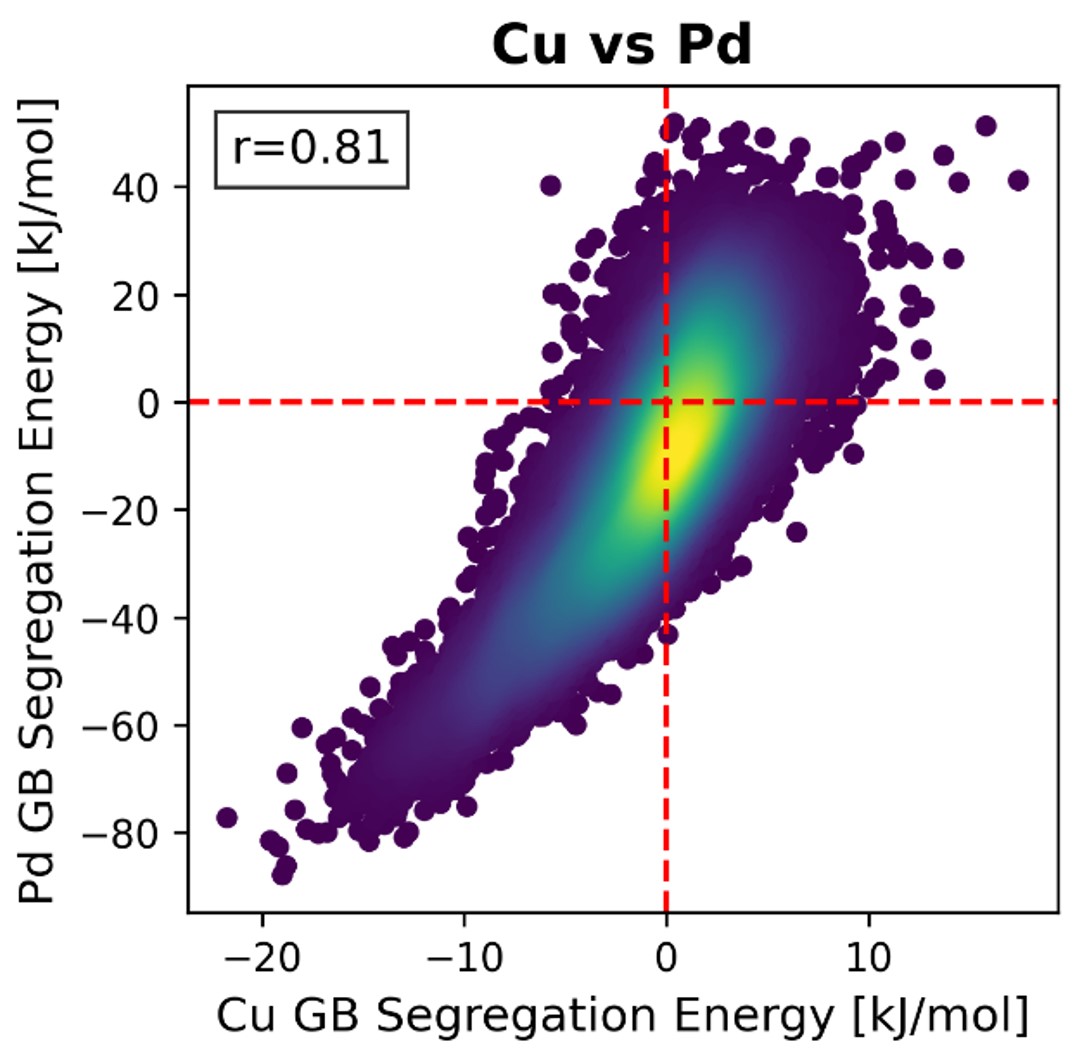}
	\caption{The site-wise GB segregation energy correlation plot for Cu and Pd in Ni. The Pearson's r value is 0.81, which indicates the presence of strong site competition between Cu and Pd.}
	\label{figs7}
\end{figure}

\setcounter{figure}{7}
\renewcommand{\thefigure}{S\arabic{figure}}
\begin{figure}
	\centering
	\includegraphics[width=1\textwidth]{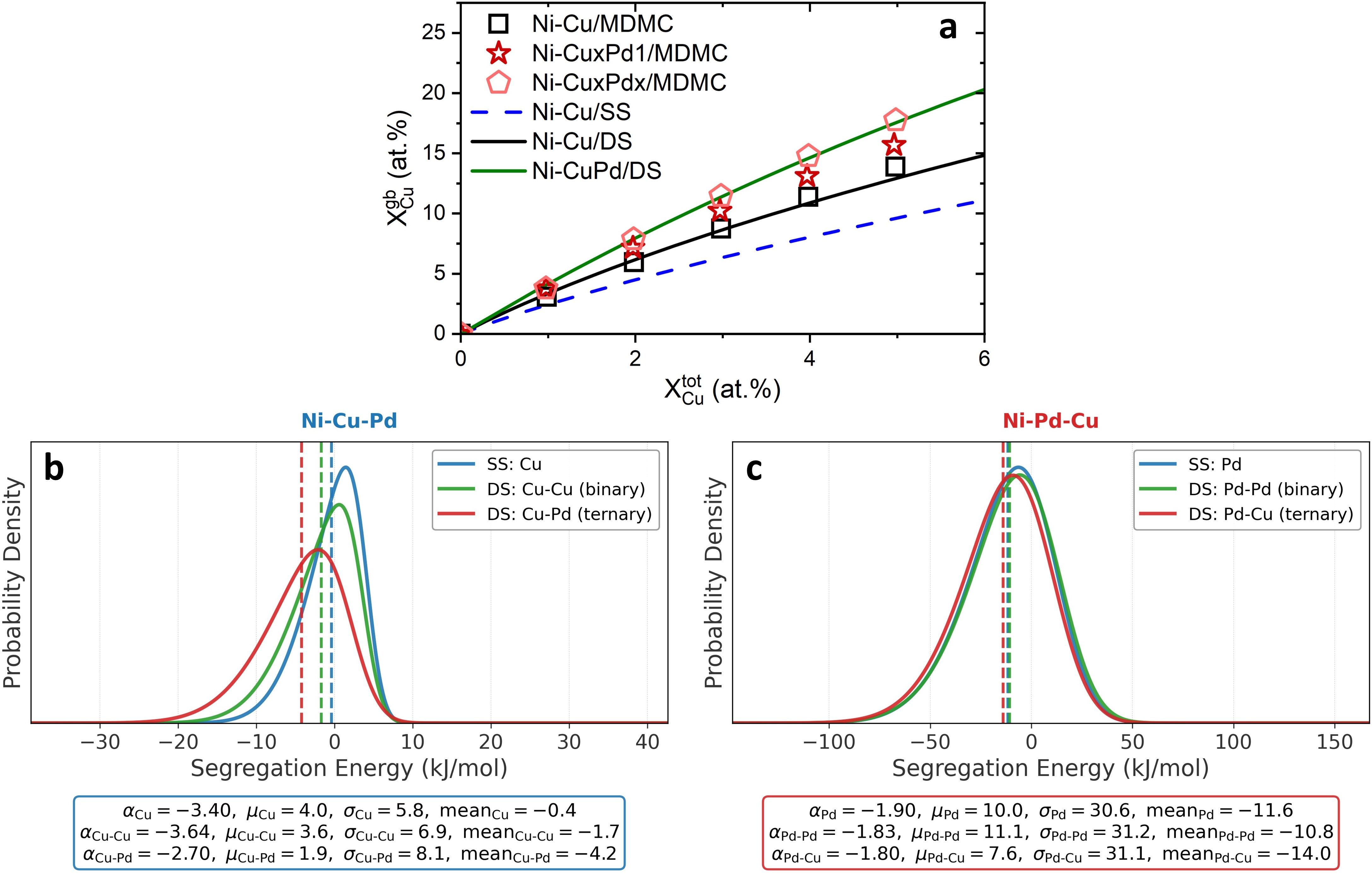}
	\caption{(a) GB segregation prediction for Cu in the Ni-(Cu,Pd) system. It can be seen that Cu segregation has been enhanced in the presence of Pd. (b) and (c) present the corresponding spectra for Ni-Cu-Pd and Ni-Pd-Cu, respectively. Synergistic co-segregation is observed attributed to the strong heteroatomic attraction between Cu and Pd in the Ni matrix regardless of the strong site competition between them.}
	\label{figs8}
\end{figure}

\setcounter{figure}{8}
\renewcommand{\thefigure}{S\arabic{figure}}
\begin{figure}
	\centering
	\includegraphics[width=0.6\textwidth]{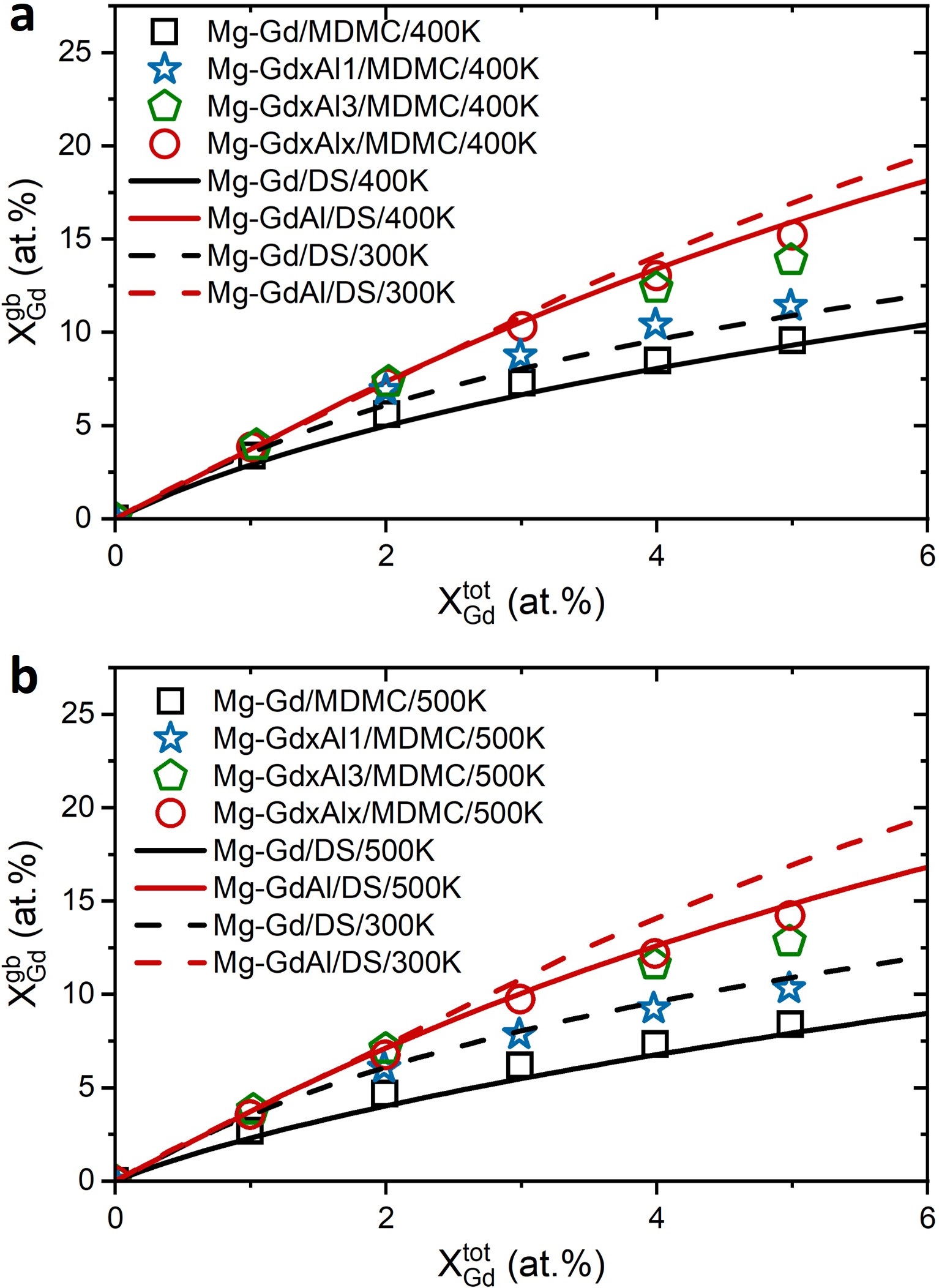}
	\caption{Hybrid MD/MC results compared with prediction bounds at (a) 400 K and (b) 500 K. Under both elevated-temperature conditions, all hybrid MD/MC results fall within the predicted lower and upper bounds. The segregation of Gd is monotonically enhanced with increasing global Al concentration. These results are consistent with those at 300 K in the main text. Furthermore, the segregation strength decreases with increasing temperature.}
	\label{figs9}
\end{figure}

\setcounter{figure}{9}
\renewcommand{\thefigure}{S\arabic{figure}}
\begin{figure}
	\centering
	\includegraphics[width=0.6\textwidth]{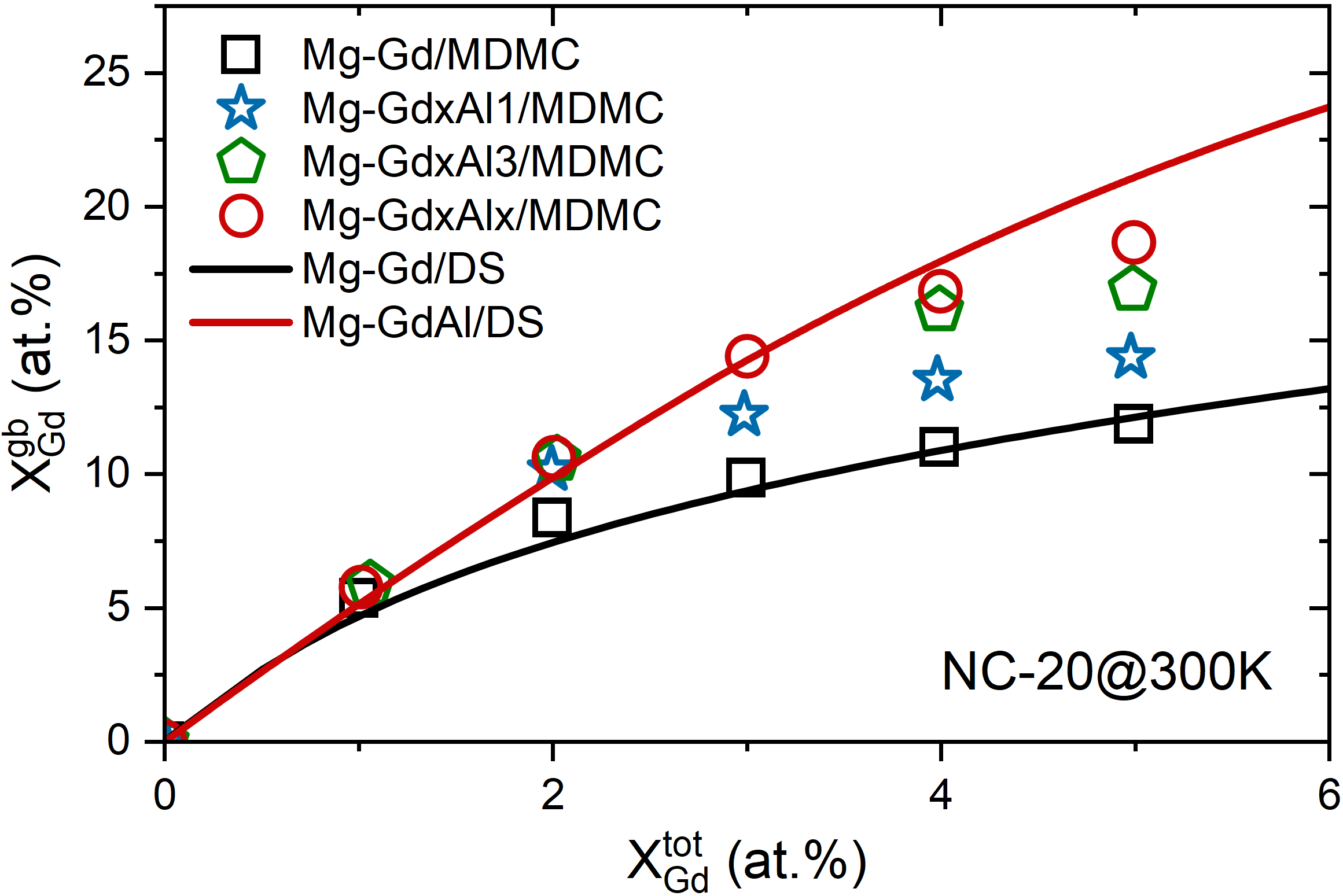}
	\caption{The results show that the predicted bounds successfully bracket the segregation behavior in the multicomponent system using a larger nanocrystalline sample, even though the segregation energy spectra were constructed from a much smaller nanocrystalline sample.}
	\label{figs10}
\end{figure}

	\clearpage
	\section{SM2}\label{SM2}
	\begin{center}
	\textbf{\large Supplementary materials for\\[26pt]
		Predicting co-segregation in alloys with solute-solute interactions}
	
	\large Zuoyong Zhang and Chuang Deng*

	\small \textit{Department of Mechanical Engineering, University of Manitoba, Winnipeg, R3T 5V6 Manitoba, Canada}

	\vspace{1cm}
	
	*Corresponding author: Chuang.Deng@umanitoba.ca
	
\end{center}

\vspace{3cm}
\textbf{\large This file includes:}

\leftskip=1cm Supplementary Materials (ii)

\leftskip=0cm \textbf{\large Other supplementary materials are included in:}

\leftskip=1cm Supplementary Materials (i)\\

\leftskip=0cm

\vspace{2.5cm}

\textbf{\large Abstract}

This supplemental file provides the complete spectral dataset (pages 2 to 15) for 55 Mg-based ternary systems with 11 solute species (Ag, Al, Ca, Co, Cu, Gd, Nd, Ni, Pb, Pd, and Zn), in which the single-solute (SS) grain boundary (GB) segregation energy spectra are directly obtained from molecular statics (MS) simulations, while the dual-solute (DS) spectra are derived from machine-learning predictions. The GB SS segregation energy correlations between different solute species in magnesium are also included here to serve as indicators for assessing whether site competition exists between them.

    \clearpage

	\begin{figure}
		\centering
		\includegraphics[width=1\textwidth]{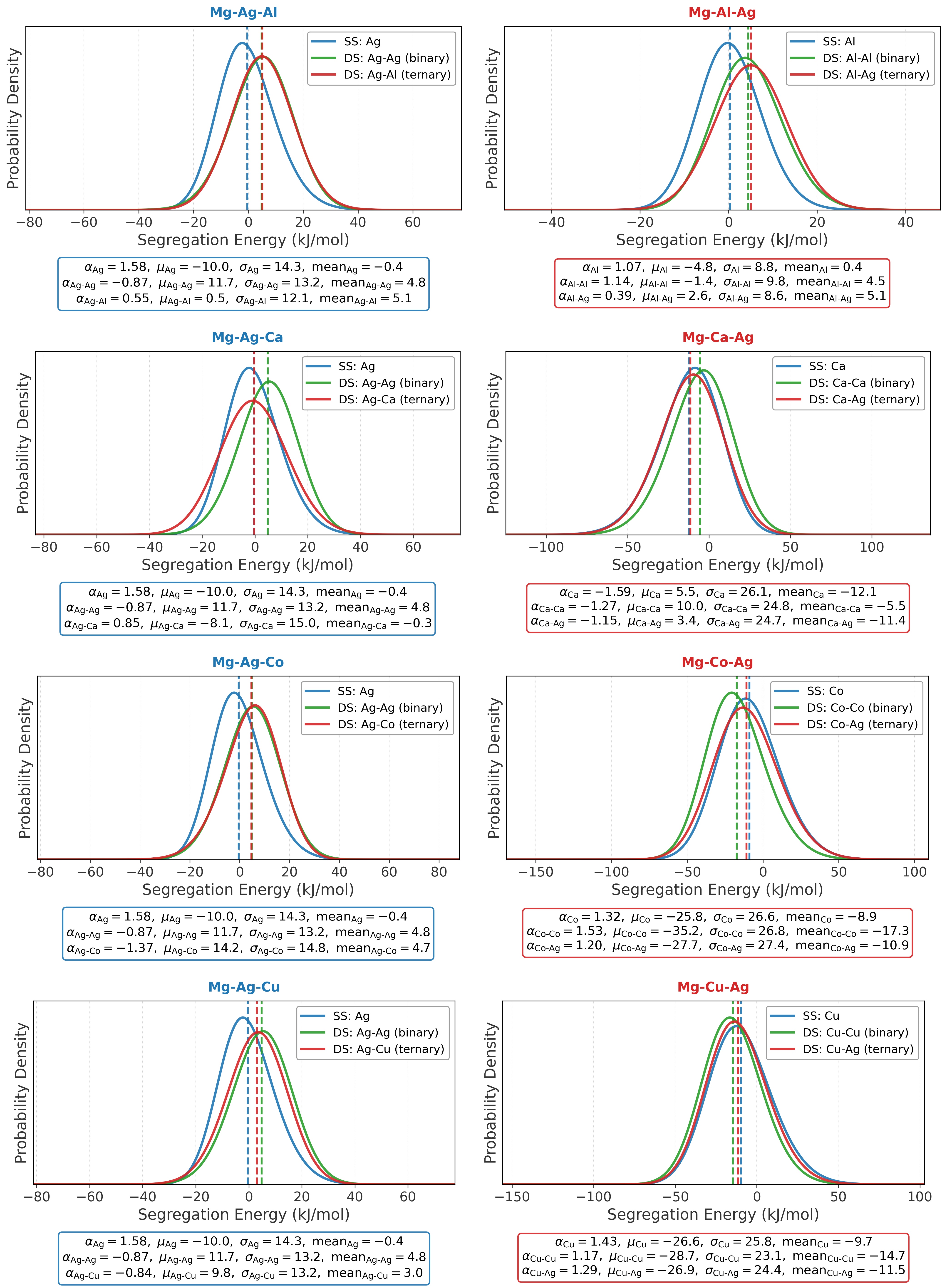}
	\end{figure}

	\begin{figure}
		\centering
		\includegraphics[width=1\textwidth]{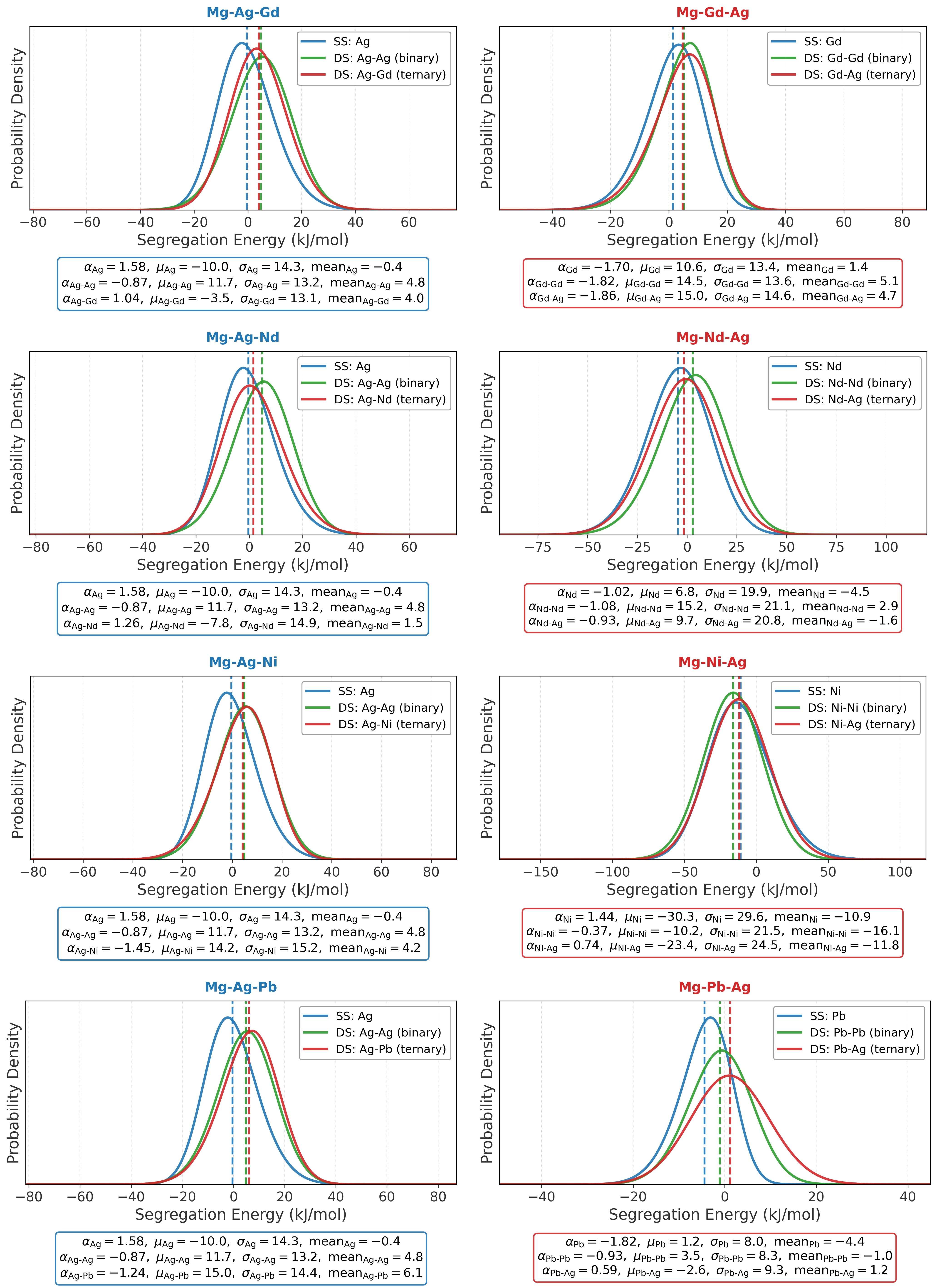}
	\end{figure}

	\begin{figure}
		\centering
		\includegraphics[width=1\textwidth]{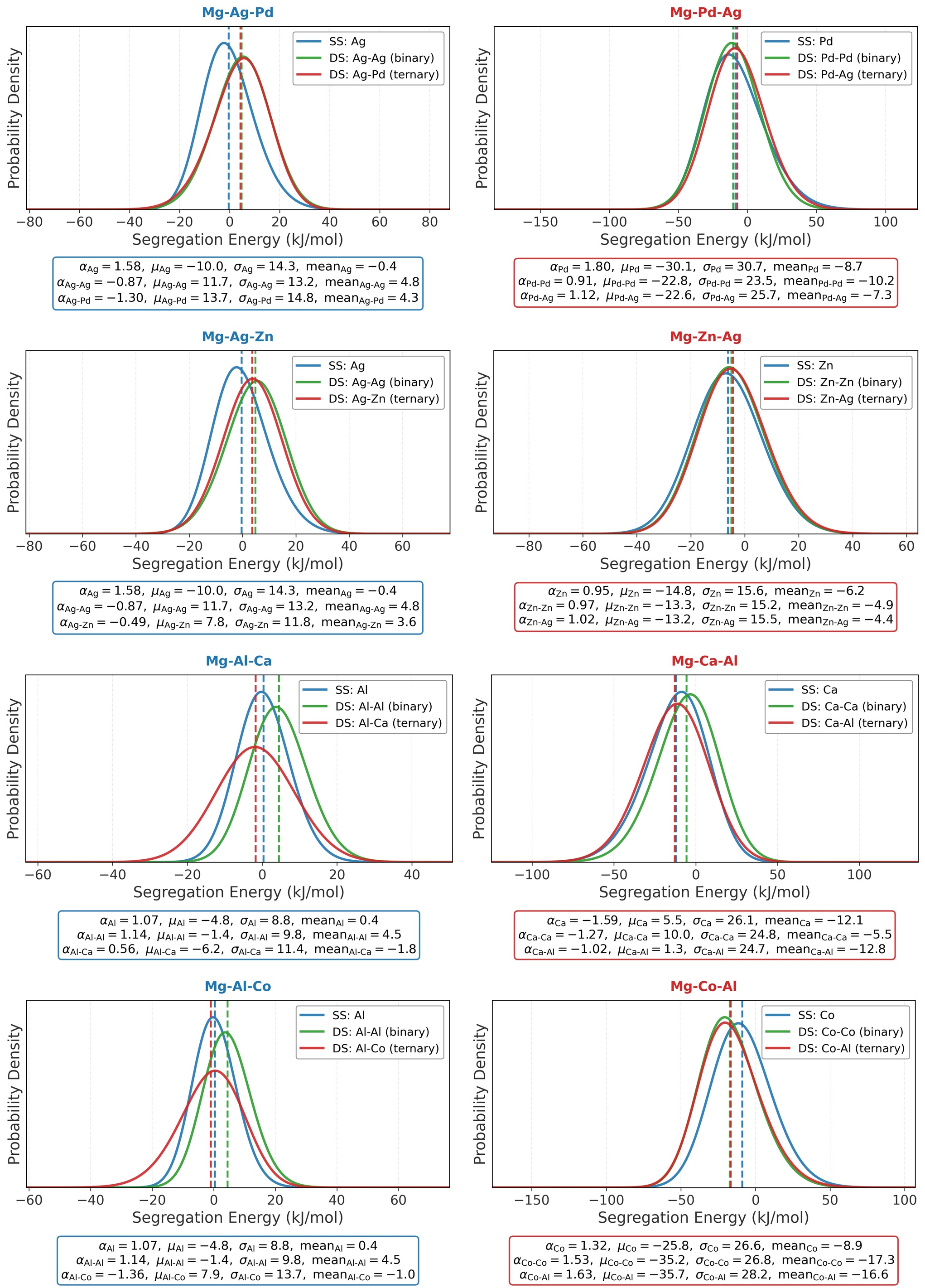}
	\end{figure}

	\begin{figure}
		\centering
		\includegraphics[width=1\textwidth]{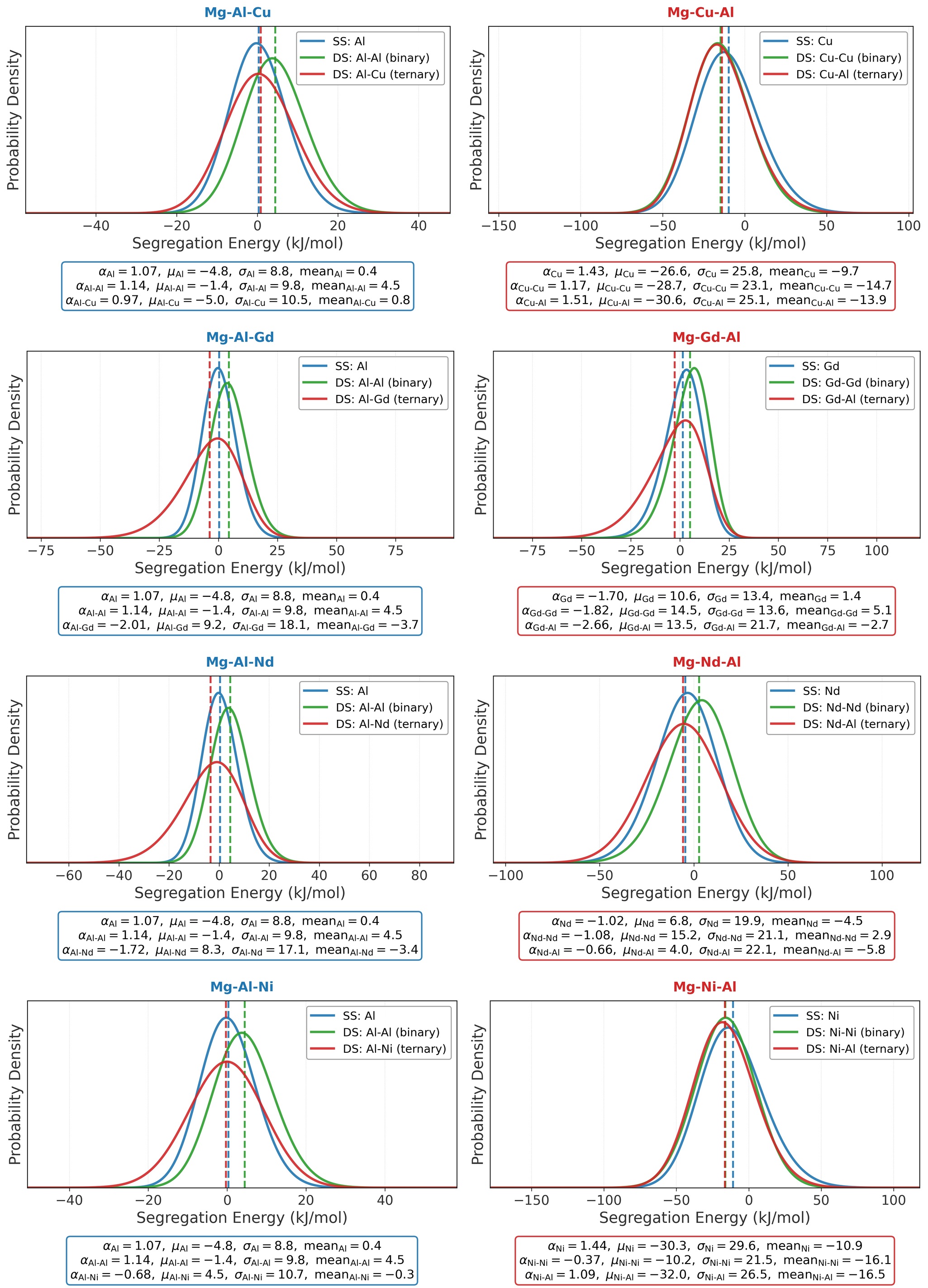}
	\end{figure}
	
	\begin{figure}
		\centering
		\includegraphics[width=1\textwidth]{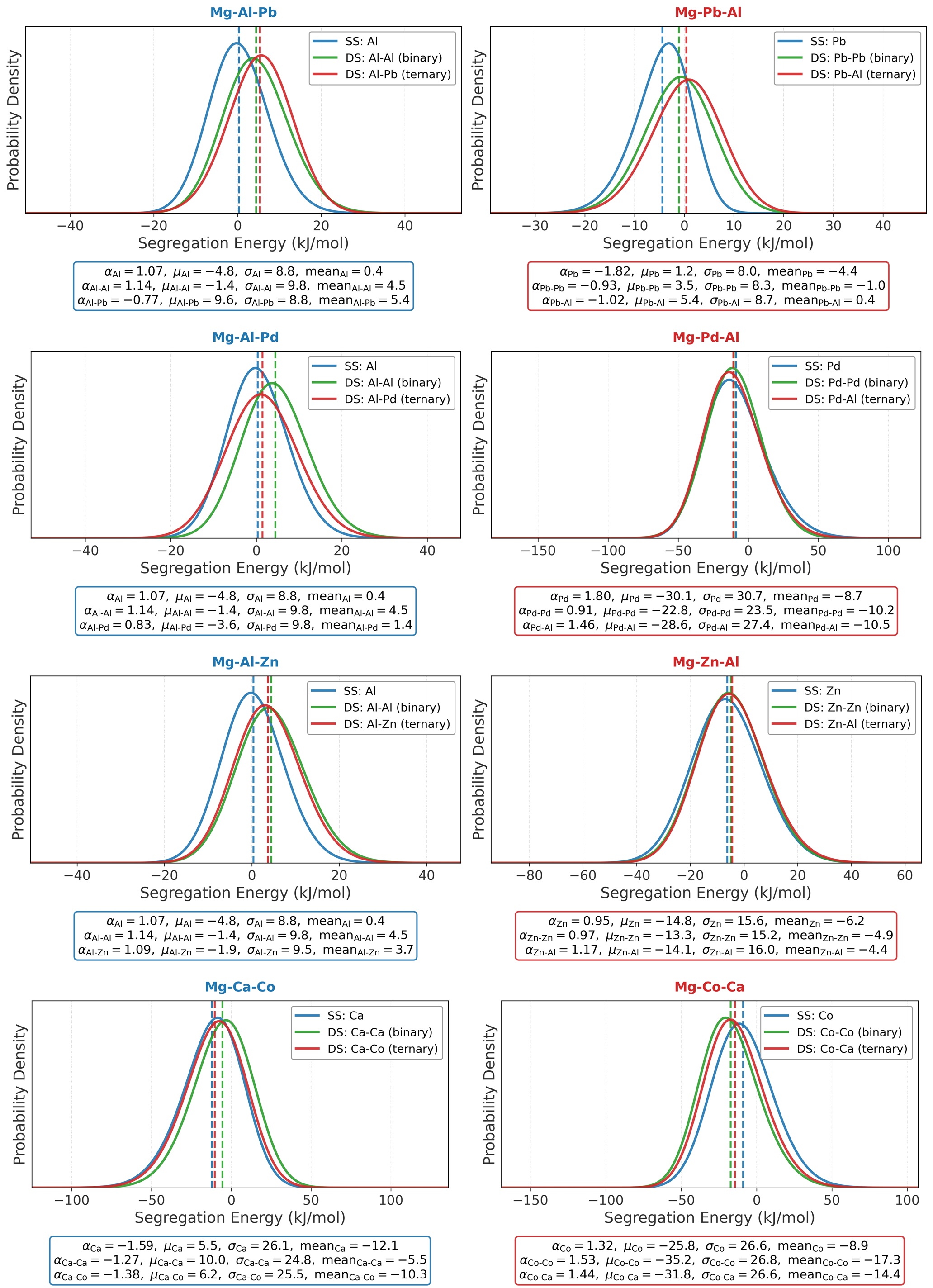}
	\end{figure}

	\begin{figure}
		\centering
		\includegraphics[width=1\textwidth]{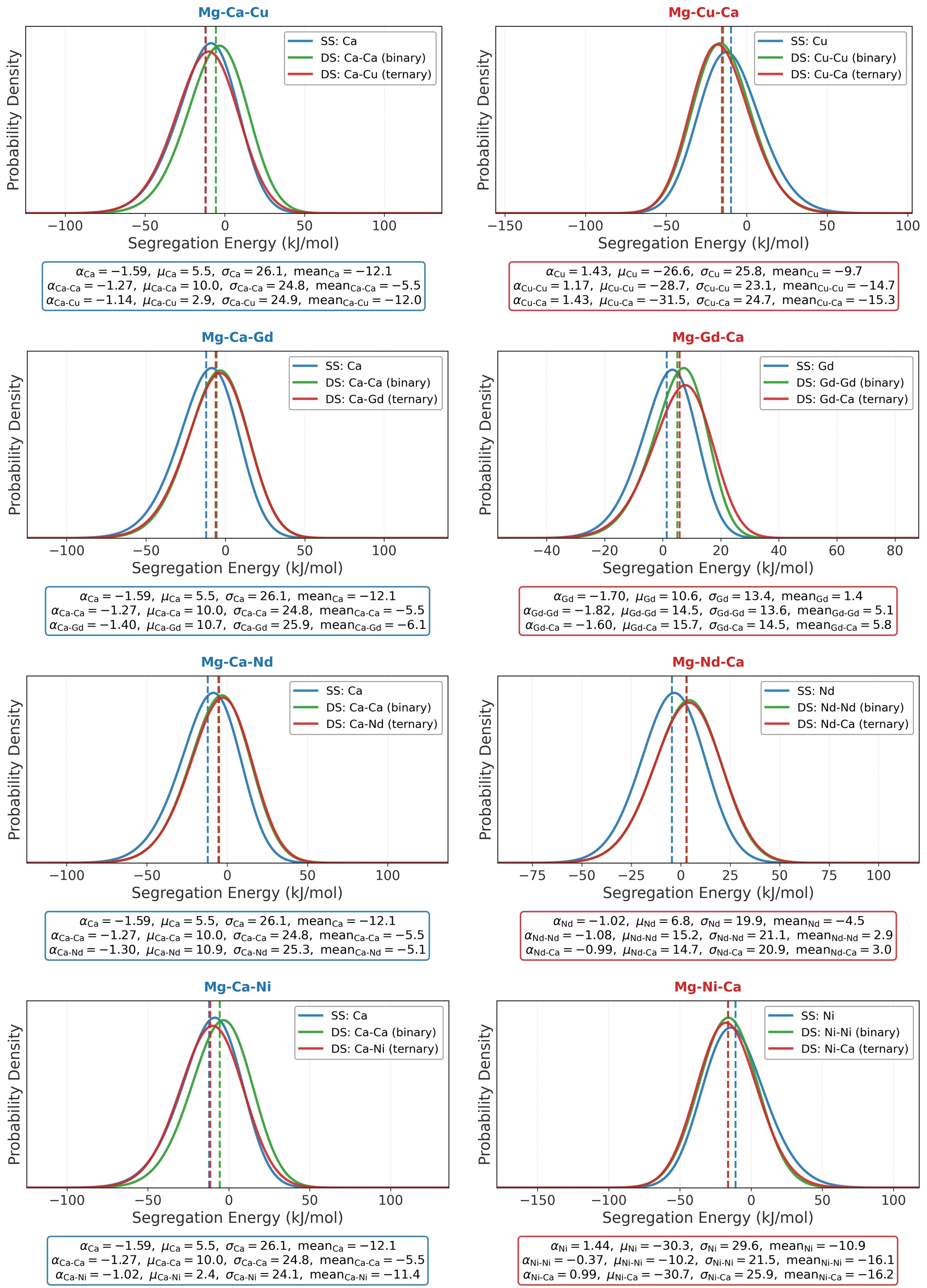}
	\end{figure}
	
	\begin{figure}
		\centering
		\includegraphics[width=1\textwidth]{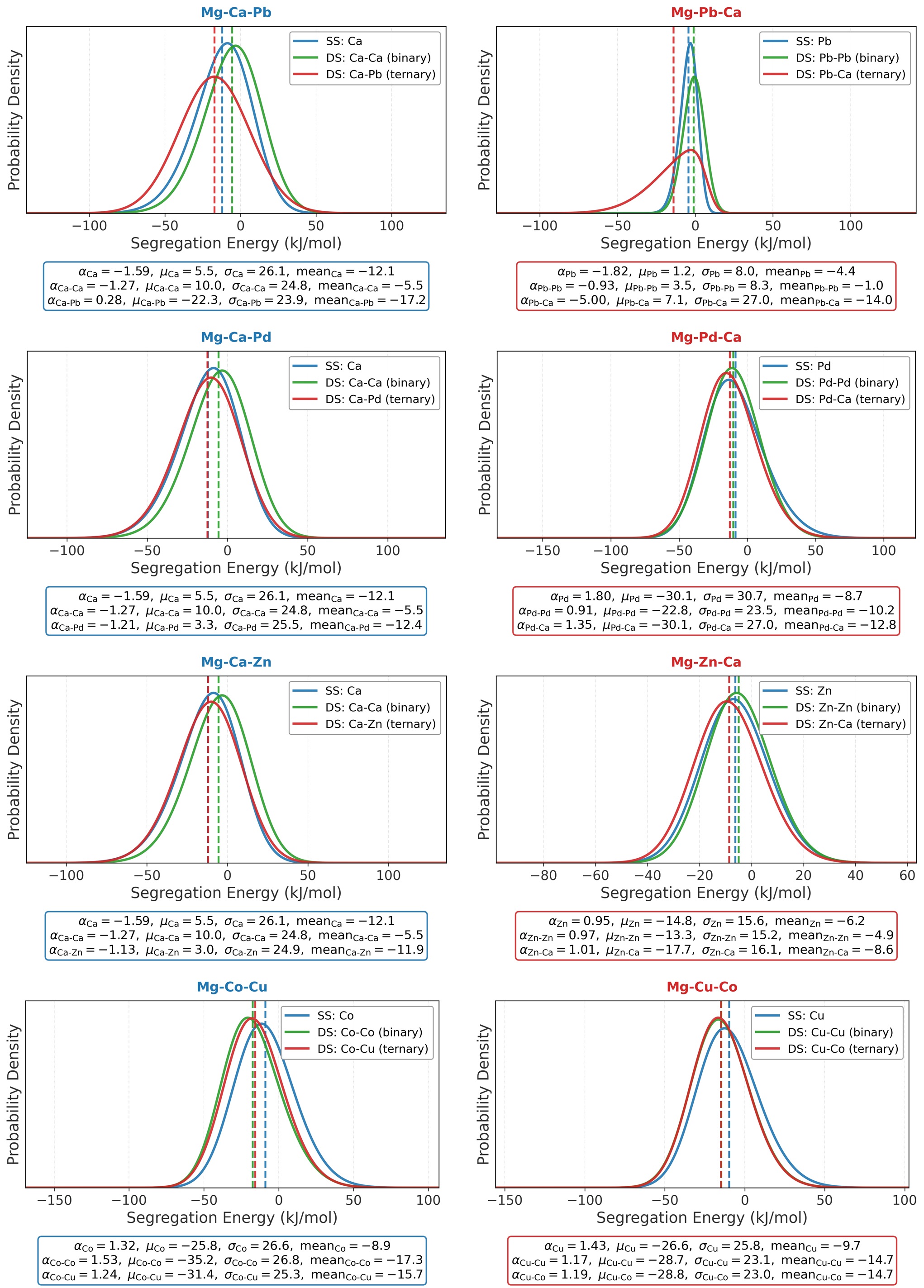}
	\end{figure}
	
	\begin{figure}
		\centering
		\includegraphics[width=1\textwidth]{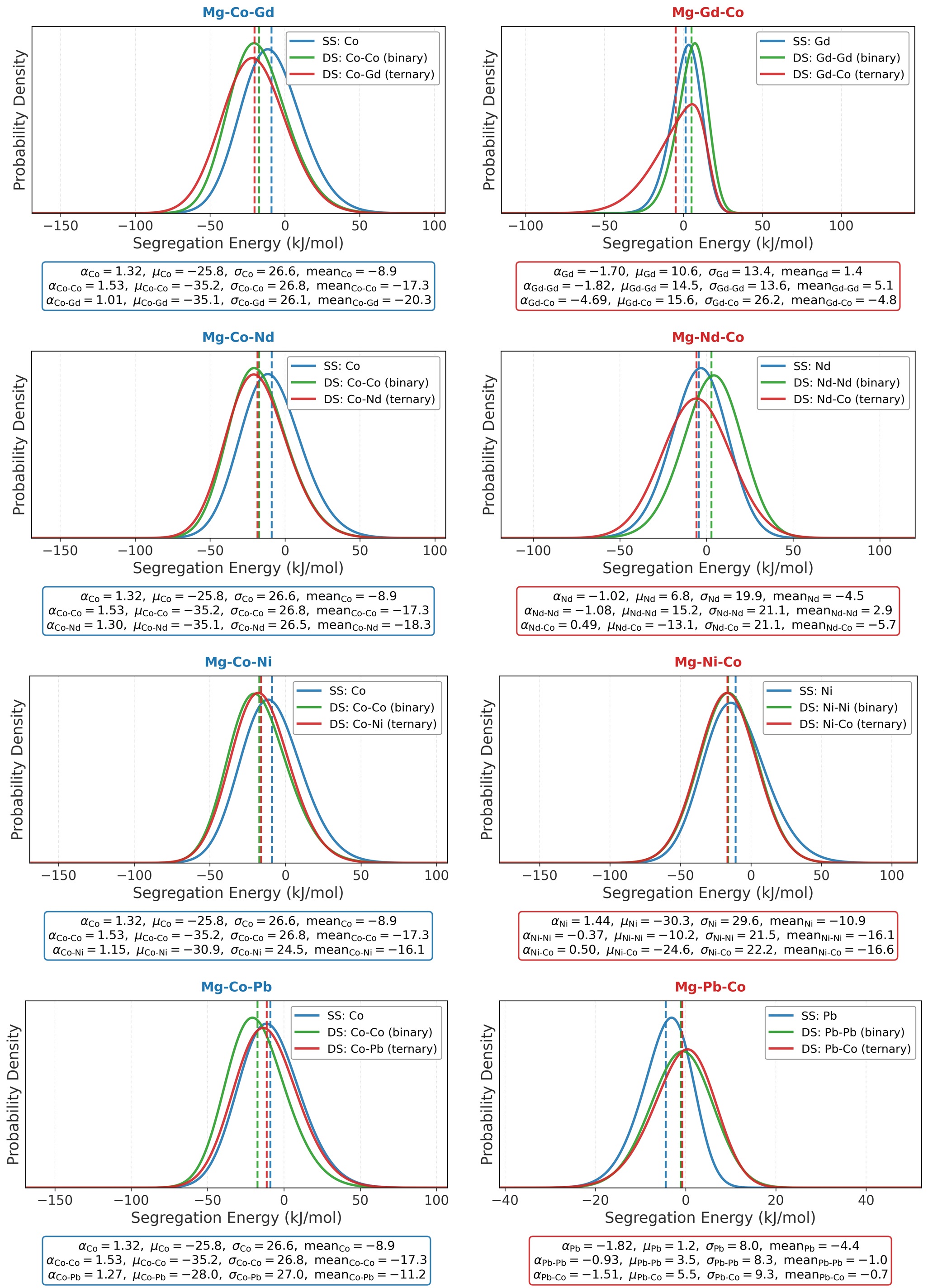}
	\end{figure}

	\begin{figure}
		\centering
		\includegraphics[width=1\textwidth]{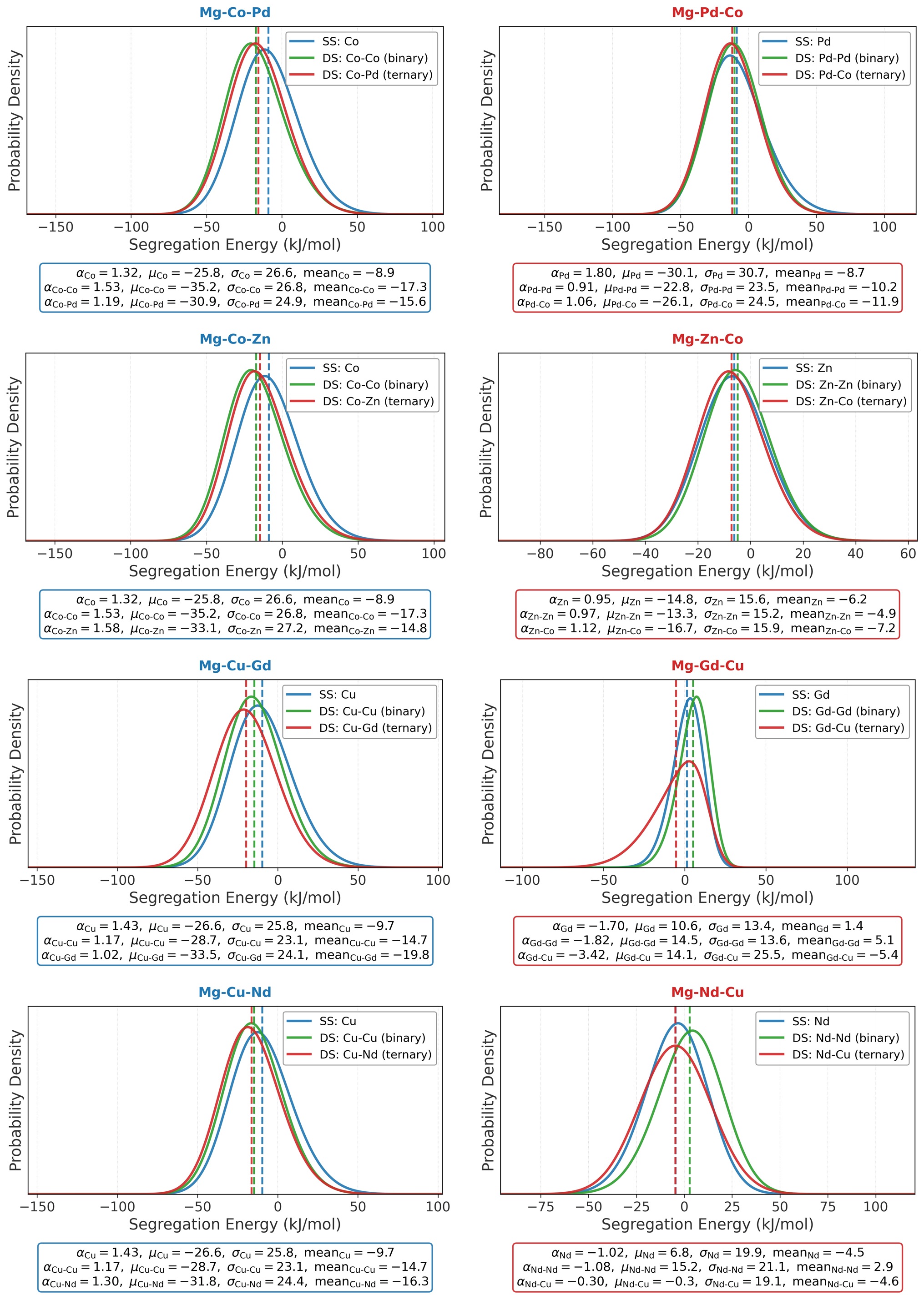}
	\end{figure}
	
	\begin{figure}
		\centering
		\includegraphics[width=1\textwidth]{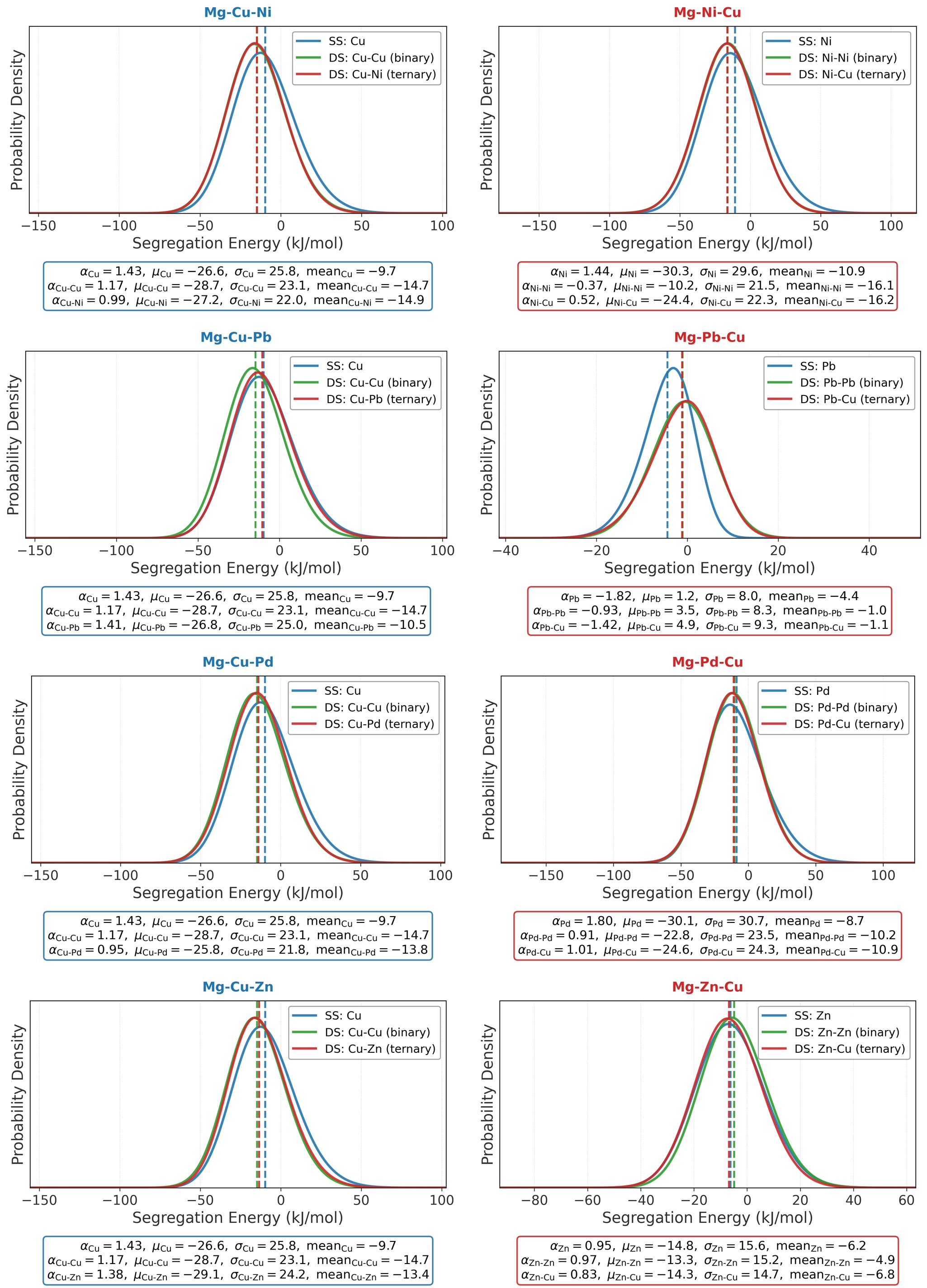}
	\end{figure}
	
	\begin{figure}
		\centering
		\includegraphics[width=1\textwidth]{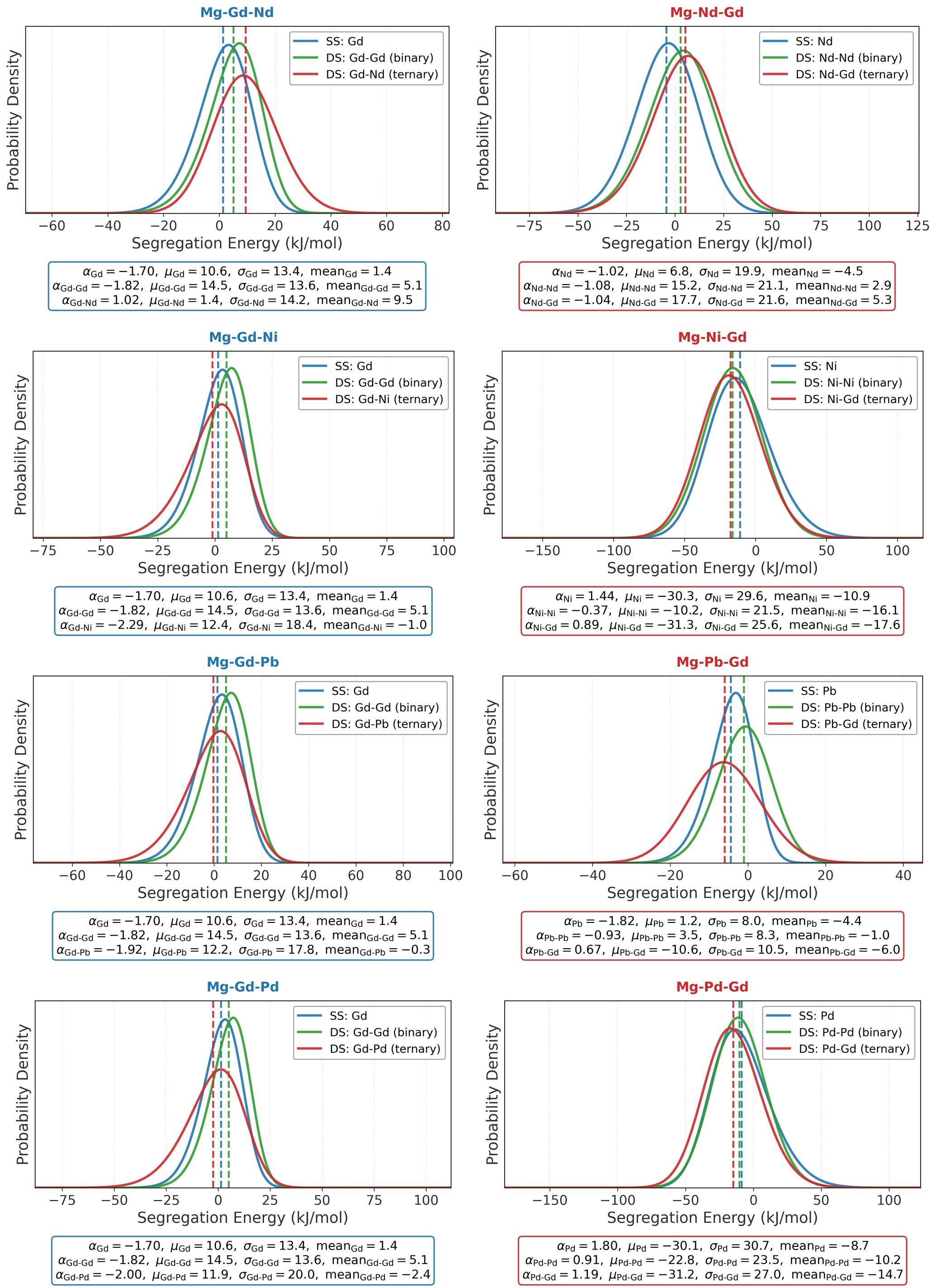}
	\end{figure}

	\begin{figure}
		\centering
		\includegraphics[width=1\textwidth]{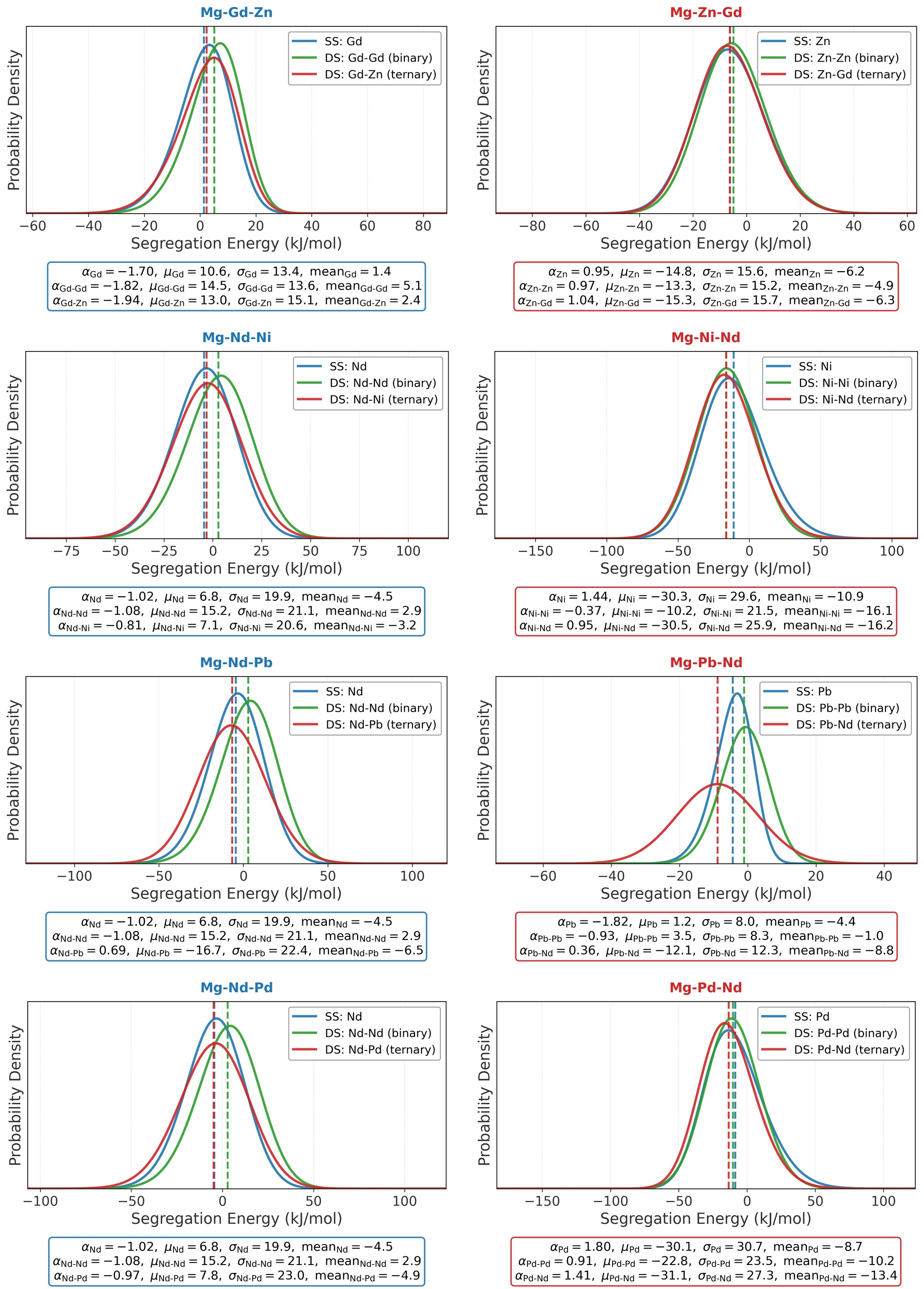}
	\end{figure}
	
	\begin{figure}
		\centering
		\includegraphics[width=1\textwidth]{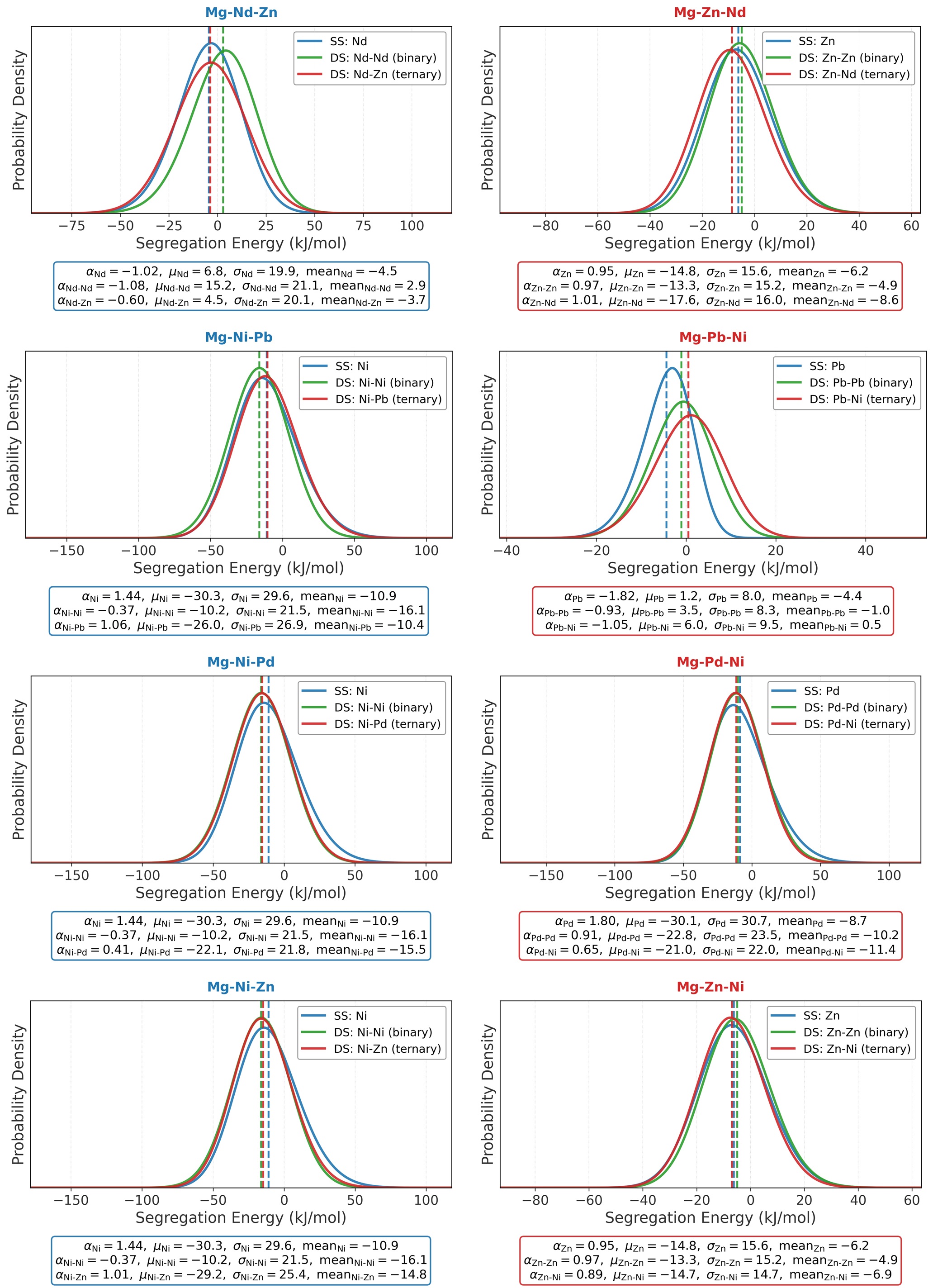}
	\end{figure}
	
	\begin{figure}
		\centering
		\includegraphics[width=1\textwidth]{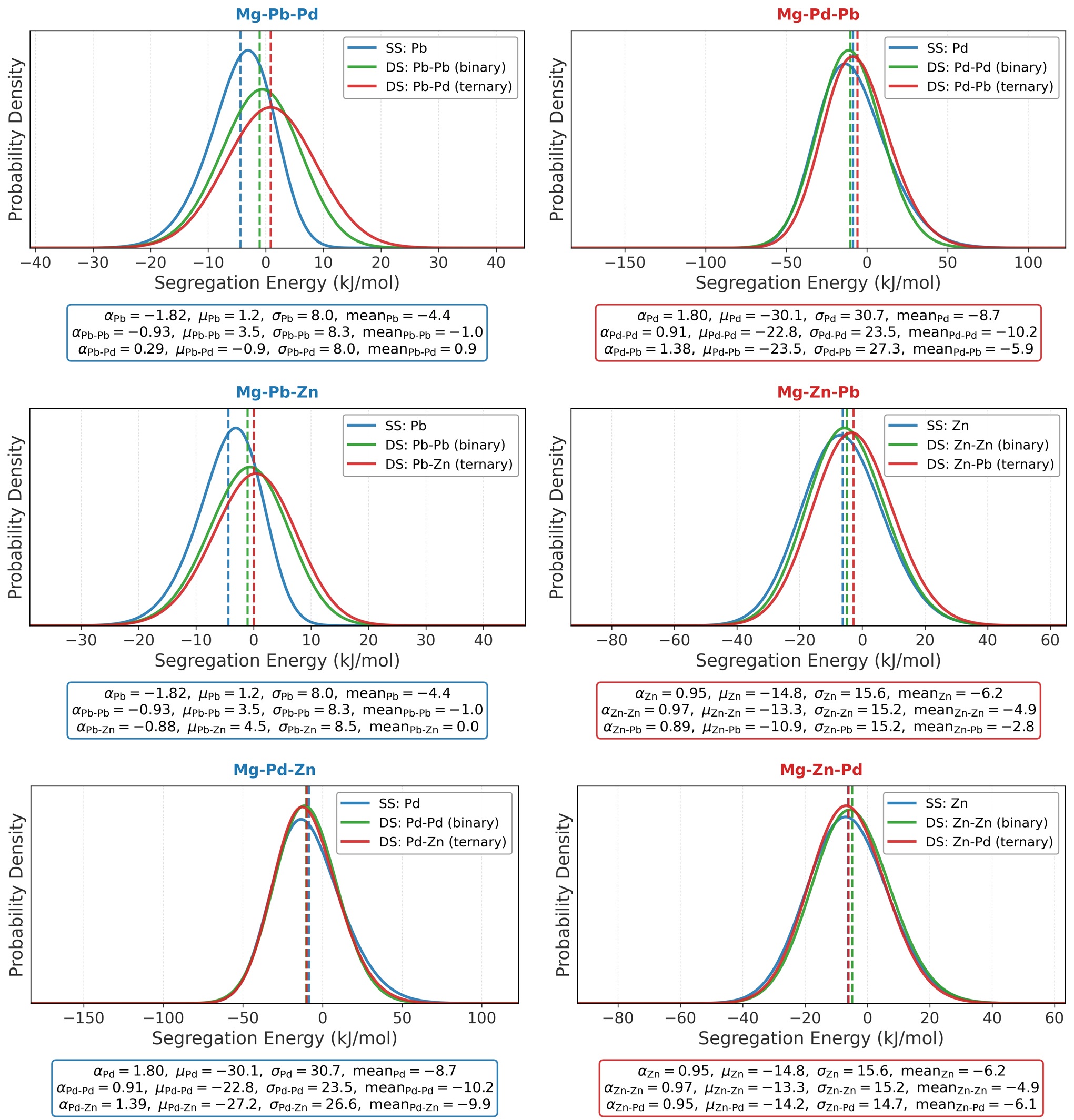}
	\end{figure}

	\begin{figure}
		\centering
		\includegraphics[width=1\textwidth]{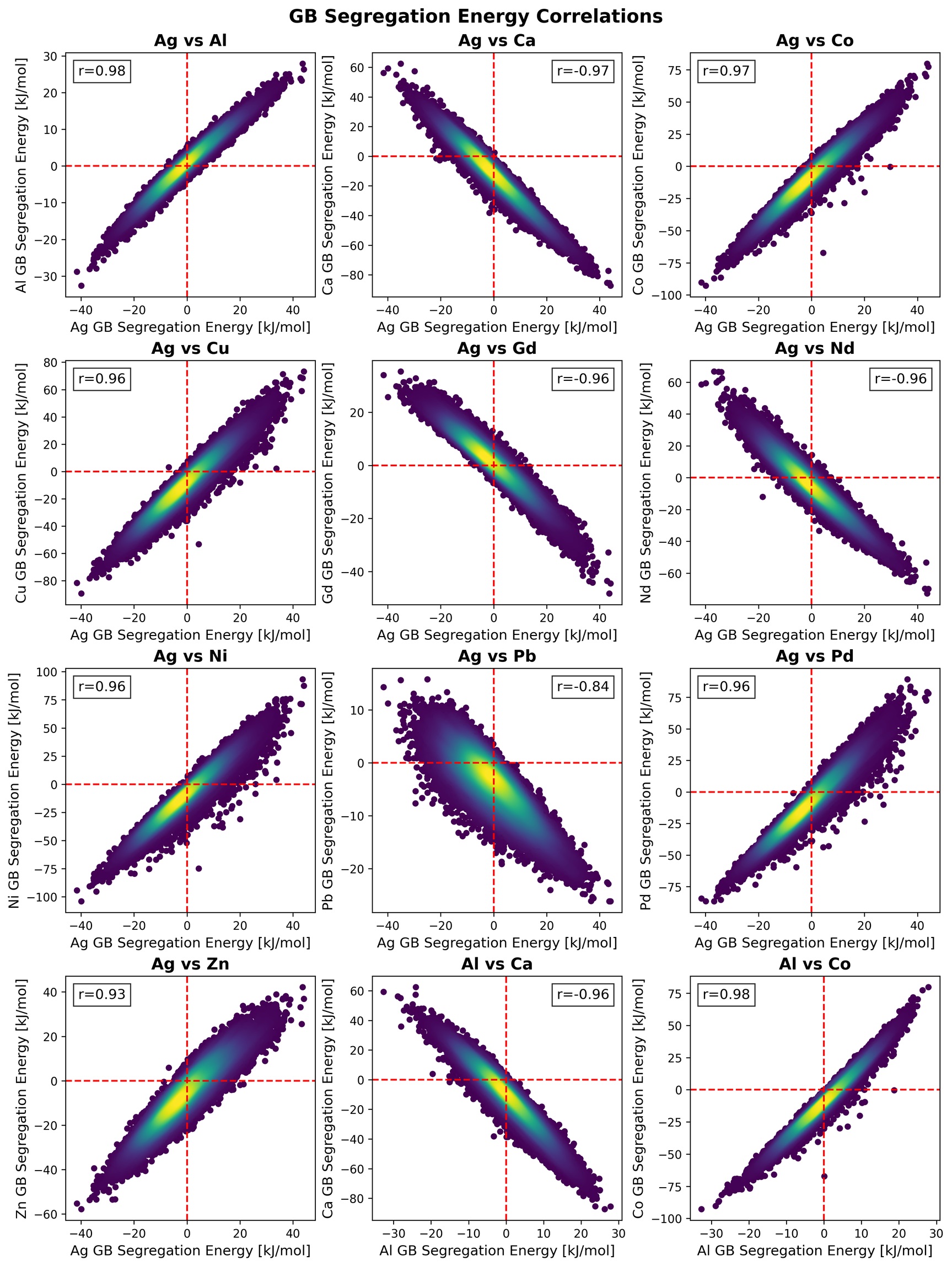}
	\end{figure}

	\begin{figure}
		\centering
		\includegraphics[width=1\textwidth]{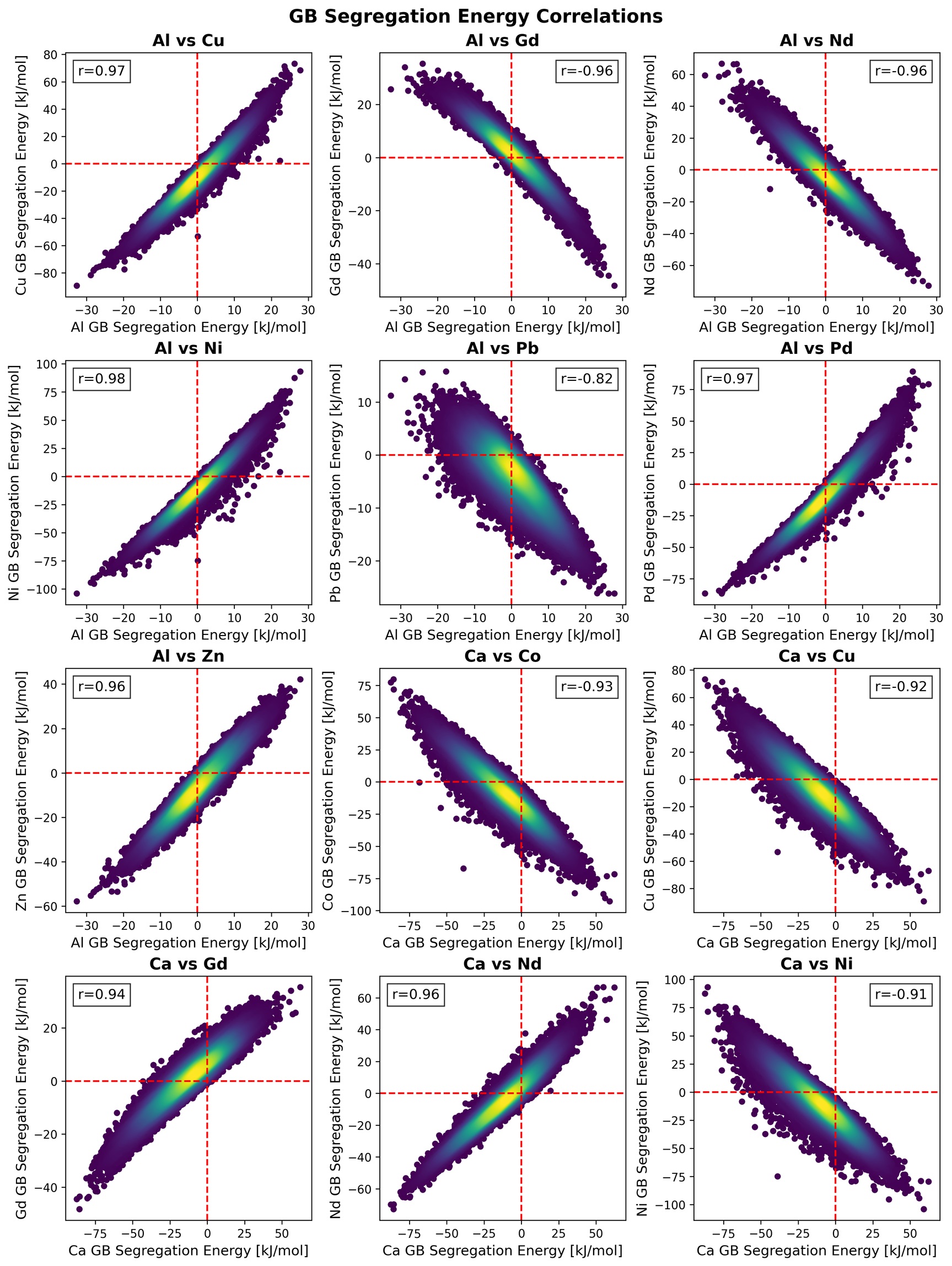}
	\end{figure}
	
	\begin{figure}
		\centering
		\includegraphics[width=1\textwidth]{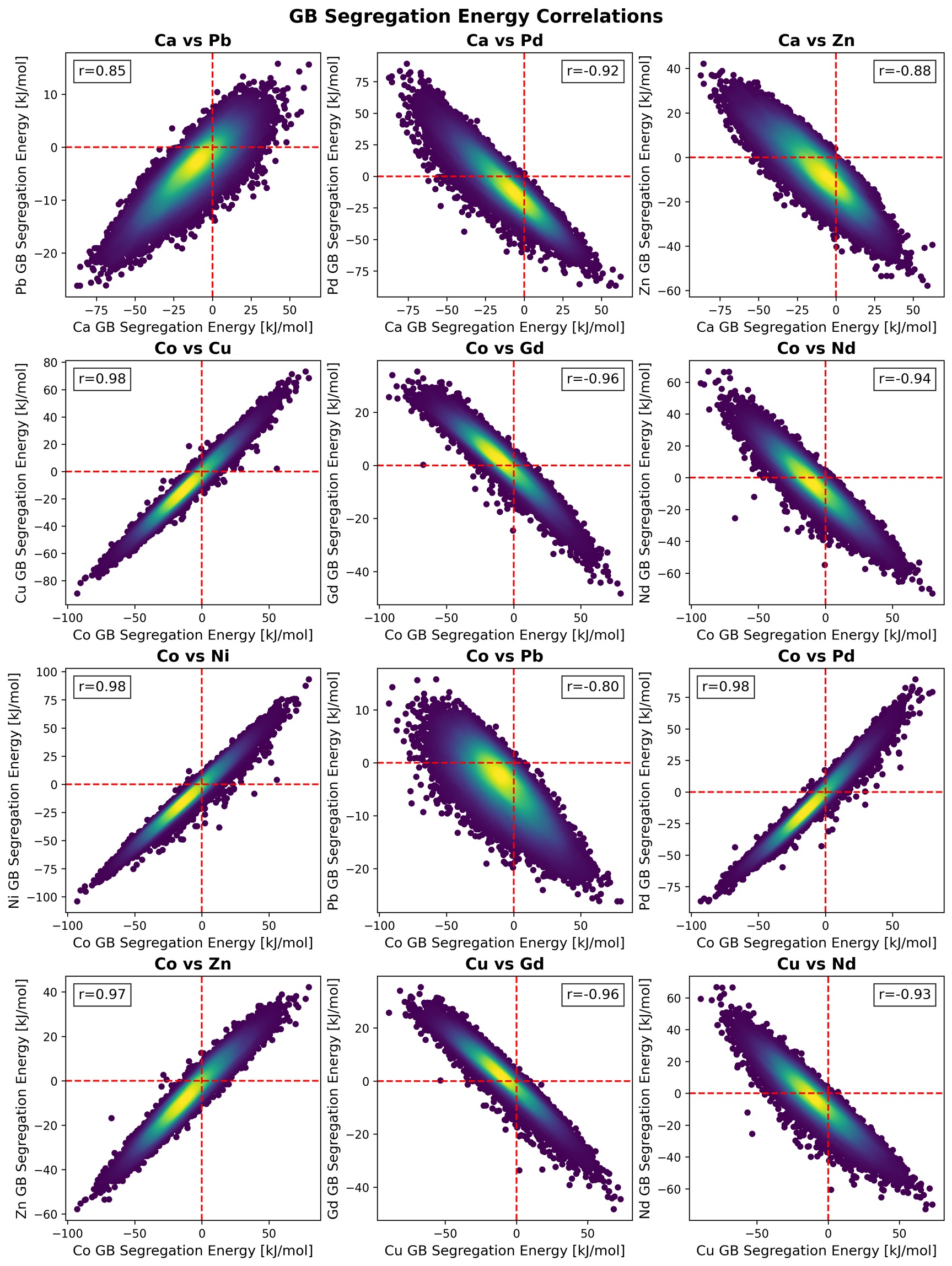}
		
	\end{figure}

	\begin{figure}
		\centering
		\includegraphics[width=1\textwidth]{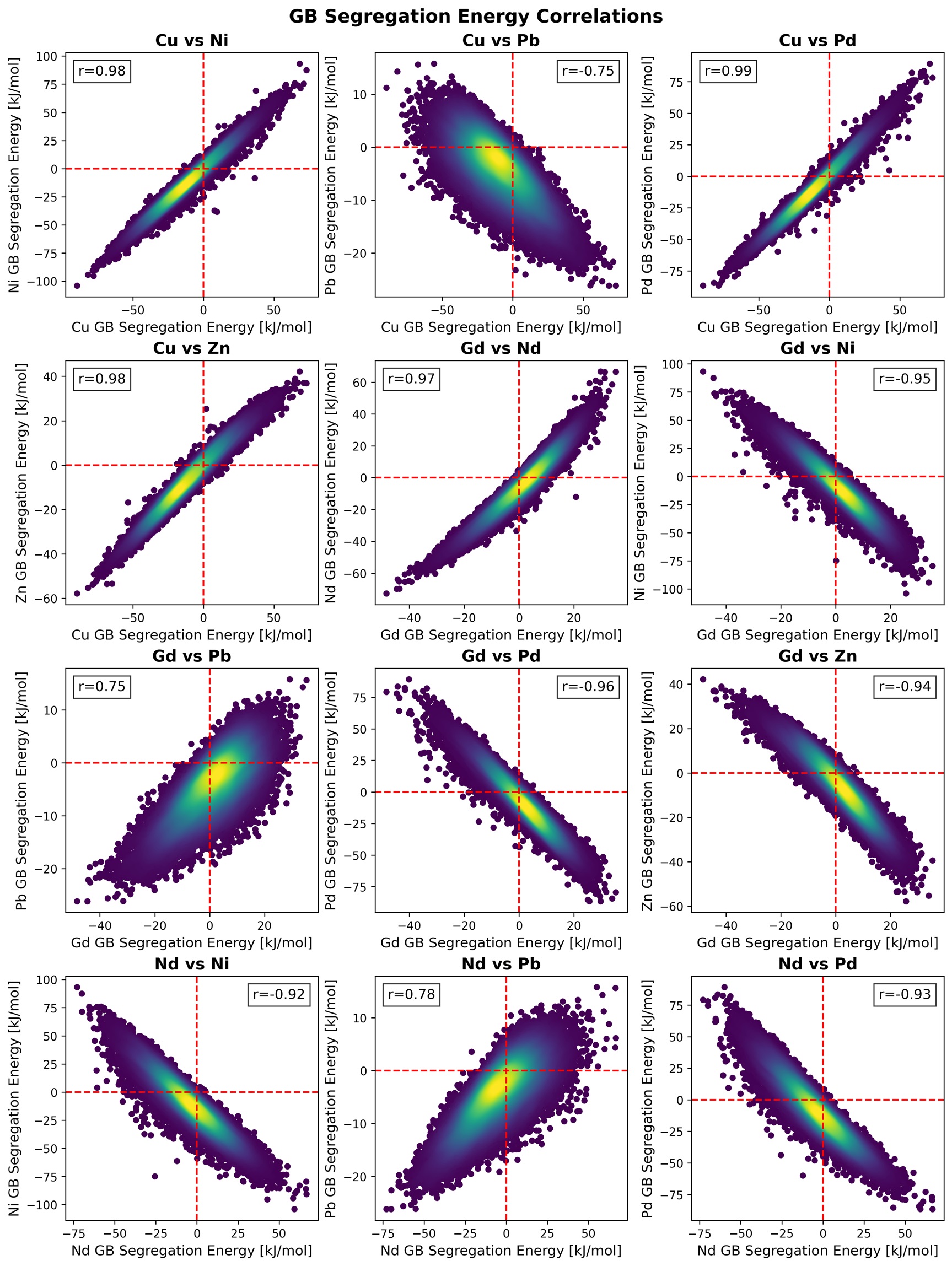}
		
	\end{figure}
	
	\begin{figure}
		\centering
		\includegraphics[width=1\textwidth]{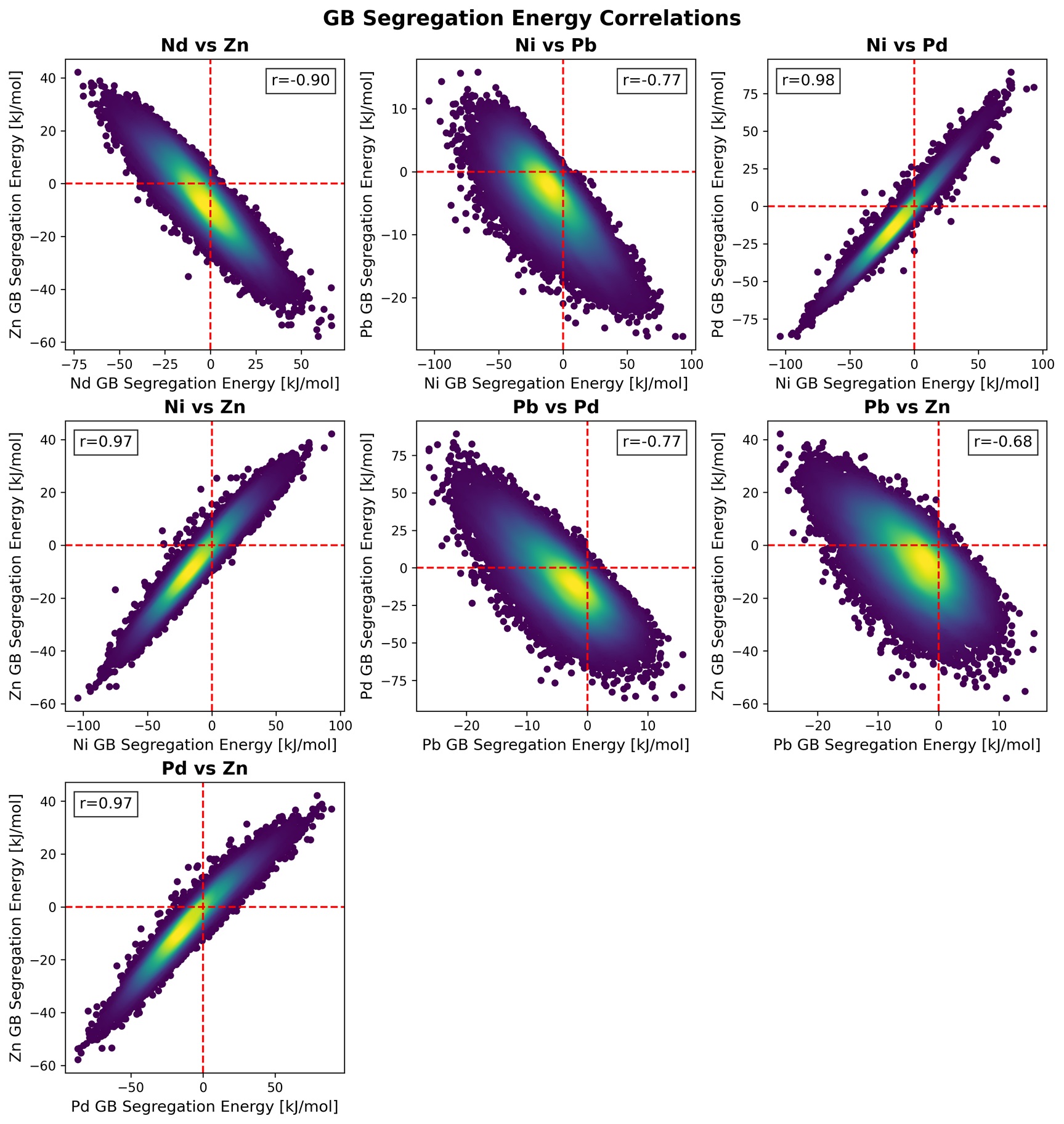}
		
	\end{figure}

\end{document}